\documentclass[english,aps,twocolumn,10pt,floatfix]{revtex4-1}
\usepackage[utf8]{inputenc}
\usepackage{amsmath}
\usepackage{amssymb}
\usepackage{array}
\usepackage{babel}
\usepackage{bbm}
\usepackage{booktabs}
\usepackage{color}
\usepackage{etoolbox}
\usepackage{graphicx}
\usepackage{hyperref,cleveref}
\usepackage{longtable}
\usepackage{microtype}
\usepackage{multirow}
\usepackage[para]{threeparttable}

\makeatletter
\providecommand{\tabularnewline}{\\}
\makeatother

\crefname{equation}{Eq.}{Eqs.}
\Crefname{equation}{Eq.}{Eqs.}
\crefname{figure}{Fig.}{Figs.}
\Crefname{figure}{Fig.}{Figs.}

\DeclareMathOperator{\tr}{tr}

\hypersetup{colorlinks=true, linkcolor=blue, urlcolor=blue, breaklinks=true}

\usepackage{appendix}
\usepackage{amsmath}
\usepackage{amssymb}
\usepackage{cleveref}
\usepackage{microtype}
\usepackage[resetlabels]{multibib}
\usepackage{titlesec}

\usepackage{multirow}

\usepackage{texmf/abbreviations}
\usepackage{texmf/formulautils}

\NewDocumentCommand{\Texp}{}{\mathcal{T}\mathrm{exp}}

\newcites{suppl}{Supplementary references}

\crefname{equation}{Eq.}{Eqs.}
\Crefname{equation}{Eq.}{Eqs.}

\crefname{figure}{Fig.}{Figs.}
\Crefname{figure}{Fig.}{Figs.}

\crefname{appsec}{suppl.\ sec.}{suppl.\ secs.}
\Crefname{appsec}{Suppl.\ sec.}{Suppl.\ secs.}

\crefname{appsubsubsec}{suppl.\ sec.}{suppl.\ secs.}
\Crefname{appsubsubsec}{Suppl.\ sec.}{Suppl.\ secs.}

\RenewDocumentCommand{\mconstXtra}{m}{
	\IfEqCase{#1}{%
		{pauli_x}{\hat{\sigma}_\mathrm{x}}%
		{pauli_y}{\hat{\sigma}_\mathrm{y}}%
		{pauli_z}{\hat{\sigma}_\mathrm{z}}%
	}%
	[\PackageError{}{Undefined math. constant:}{}]%
}

\NewDocumentCommand{\pconst}{m}{
	\IfEqCase{#1}{%
		{hbar}{\hbar}%
	}%
	[\PackageError{formulautils}{Undefined phys. constant:}{}]%
}

\NewDocumentCommand{\pquant}{m}{
	\IfEqCase{#1}{%
		{strict_cp_map}{\Phi}%
		{strict_cp_map_tf_ti}{\Phi(t_\mathrm{f},t_\mathrm{i})}%
		{cp_map}{\phi}%
		{cp_map_tf_ti}{\phi(t_\mathrm{f},t_\mathrm{i})}%
		{delta_rho_x}{\delta\hat{\rho}_\mathrm{x}}%
		{delta_rho_y}{\delta\hat{\rho}_\mathrm{y}}%
		{delta_rho_z}{\delta\hat{\rho}_\mathrm{z}}%
		{delta_rho_j}{\delta\hat{\rho}_j}%
		{delta_rho_k}{\delta\hat{\rho}_k}%
		{D_n}{D_{\vec{n}}} 
		{magic}{\mathcal{R}_\mathrm{Q}}%
		{exp_magic}{\bar{\mathcal{R}}_\mathrm{Q}}%
		{punish_destr}{P} 
		{rho_xyz}{\hat{\rho}_{x,y,z}}%
		{rho_0}{\hat{\rho}_0}%
		{rho_n}{\hat{\rho}_{\vec{n}}}%
		{ti}{t_\mathrm{i}}%
		{tf}{t_\mathrm{f}}%
		{time_step}{\Delta t}%
		{T_dec}{T_\mathrm{dec}} 
		{T_eff}{T_\mathrm{eff}} 
		{T_triv}{T_\mathrm{triv}} 
		{T_triv^cn}{T_\mathrm{triv}^\mathrm{(cn)}} 
	}%
	[\PackageError{}{Undefined phys. quantity:}{}]%
}

\NewDocumentCommand{\maintextref}{m}{
	\IfEqCase{#1}{%
		{fig:setup}{\abbr{fig} 1}%
		{fig:setup:agent_environment}{\abbr{fig} 1a}%
		{fig:setup:state_aware}{\abbr{fig} 1b}%
		{fig:setup:event_aware}{\abbr{fig} 1c}%
		{fig:training}{\abbr{fig} 3}%
		{fig:training:learning_curve}{\abbr{fig} 3a}%
		{fig:training:gate_seq}{\abbr{fig} 3b}%
		{fig:training:magic_vs_time}{\abbr{fig} 3c}%
		{fig:training:branching}{\abbr{fig} 3d}%
		{fig:analysis}{\abbr{fig} 4}%
		{fig:analysis:expected_msmts}{\abbr{fig} 4a}%
		{fig:analysis:surprising_msmt}{\abbr{fig} 4b}%
		{fig:analysis:tsne}{\abbr{fig} 4c}%
		{fig:variations}{\abbr{fig} 5}%
		{fig:variations:conn_seq}{\abbr{fig} 5a}%
		{fig:variations:conn_bar}{\abbr{fig} 5b}%
		{fig:variations:msmt_errors}{\abbr{fig} 5c}%
		{fig:corr_noise}{\abbr{fig} 6}%
		{fig:corr_noise:setup}{\abbr{fig} 6a}%
		{fig:corr_noise:improvement_two_ancillas}{\abbr{fig} 6b}%
		{fig:corr_noise:num_ancillas_bar}{\abbr{fig} 6c}%
		{fig:corr_noise:gate_seq}{\abbr{fig} 6d}%
		{fig:corr_noise:bfts_branching}{\abbr{fig} 6e}%
		{fig:decoding_and_event_aware}{\abbr{fig} 7}%
		{fig:decoding_and_event_aware:learning_curve_state_aware}{\abbr{fig} 7a}%
		{fig:decoding_and_event_aware:learning_curve_event_aware}{\abbr{fig} 7b}%
		{fig:decoding_and_event_aware:neuron_act}{\abbr{fig} 7c}%
	}%
	[\PackageError{}{Undefined maintext reference:}{}]%
}

\begin{document}

\title{Reinforcement Learning with Neural Networks for Quantum Feedback}

\author{Thomas Fösel}
\affiliation{Max Planck Institute for the Science of Light, Staudtstr. 2, 91058 Erlangen, Germany}

\author{Petru Tighineanu}
\affiliation{Max Planck Institute for the Science of Light, Staudtstr. 2, 91058 Erlangen, Germany}

\author{Talitha Weiss}
\affiliation{Max Planck Institute for the Science of Light, Staudtstr. 2, 91058 Erlangen, Germany}

\author{Florian Marquardt}
\affiliation{Max Planck Institute for the Science of Light, Staudtstr. 2, 91058 Erlangen, Germany}
\affiliation{Physics Department, University of Erlangen-Nuremberg, Staudtstr. 5, 91058 Erlangen, Germany}

\begin{abstract}
	Machine learning with artificial neural networks is revolutionizing
	science. The most advanced challenges require discovering answers
	autonomously. This is the domain of reinforcement learning, where
	control strategies are improved according to a reward function. The
	power of neural-network-based reinforcement learning has been highlighted
	by spectacular recent successes, such as playing Go, but its benefits
	for physics are yet to be demonstrated. Here, we show how a network-based
	``agent'' can discover complete quantum-error-correction strategies,
	protecting a collection of qubits against noise. These strategies
	require feedback adapted to measurement outcomes. Finding them from
	scratch, without human guidance, tailored to different hardware resources,
	is a formidable challenge due to the combinatorially large search
	space. To solve this, we develop two ideas: two-stage learning with
	teacher/student networks and a reward quantifying the capability to
	recover the quantum information stored in a multi-qubit system. Beyond
	its immediate impact on quantum computation, our work more generally
	demonstrates the promise of neural-network-based reinforcement learning
	in physics.
\end{abstract}

\maketitle

We are witnessing rapid progress in applications of artificial neural
networks (ANN) for tasks like image classification, speech recognition,
natural language processing, and many others \cite{lecun_deep_2015,russell_artificial_2016}.
Within physics, the examples emerging during the past two years range
across areas like statistical physics, quantum many-body systems,
and quantum error correction \cite{carleo_solving_2017,carrasquilla_machine_2017,van_nieuwenburg_learning_2017,torlai_neural_2017,august_using_2017,dunjko_machine_2017,baireuther_machine-learning-assisted_2017,krastanov_deep_2017,mehta_high-bias_2018}.
To date, most applications of neural networks employ supervised learning,
where a large collection of samples has to be provided together with
the correct labeling.

However, inspired by the long-term vision of artificial scientific
discovery \cite{king_automation_2009,schmidt_distilling_2009}, one
is led to search for more powerful techniques that explore solutions
to a given task autonomously. Reinforcement learning (RL) is a general
approach of this kind \cite{russell_artificial_2016}, where an ``agent''
interacts with an ``environment''. The agent's ``policy'', i.\,e.\ the
choice of actions in response to the environment's evolution, is updated
to increase some reward. The power of this method, when combined with
ANNs, was demonstrated convincingly through learning to play games
beyond human expertise \cite{mnih_human-level_2015,silver_mastering_2017}.
In physics, RL \emph{without} neural networks has been introduced
recently, for example to study qubit control \cite{chen_fidelity-based_2014,bukov_machine_2017}
and invent quantum optics experiments \cite{melnikov_active_2018}.
Moving to neural-network-based RL promises access to the vast variety
of techniques currently being developed for ANNs.

\begin{figure}
	\includegraphics[width=0.9\columnwidth]{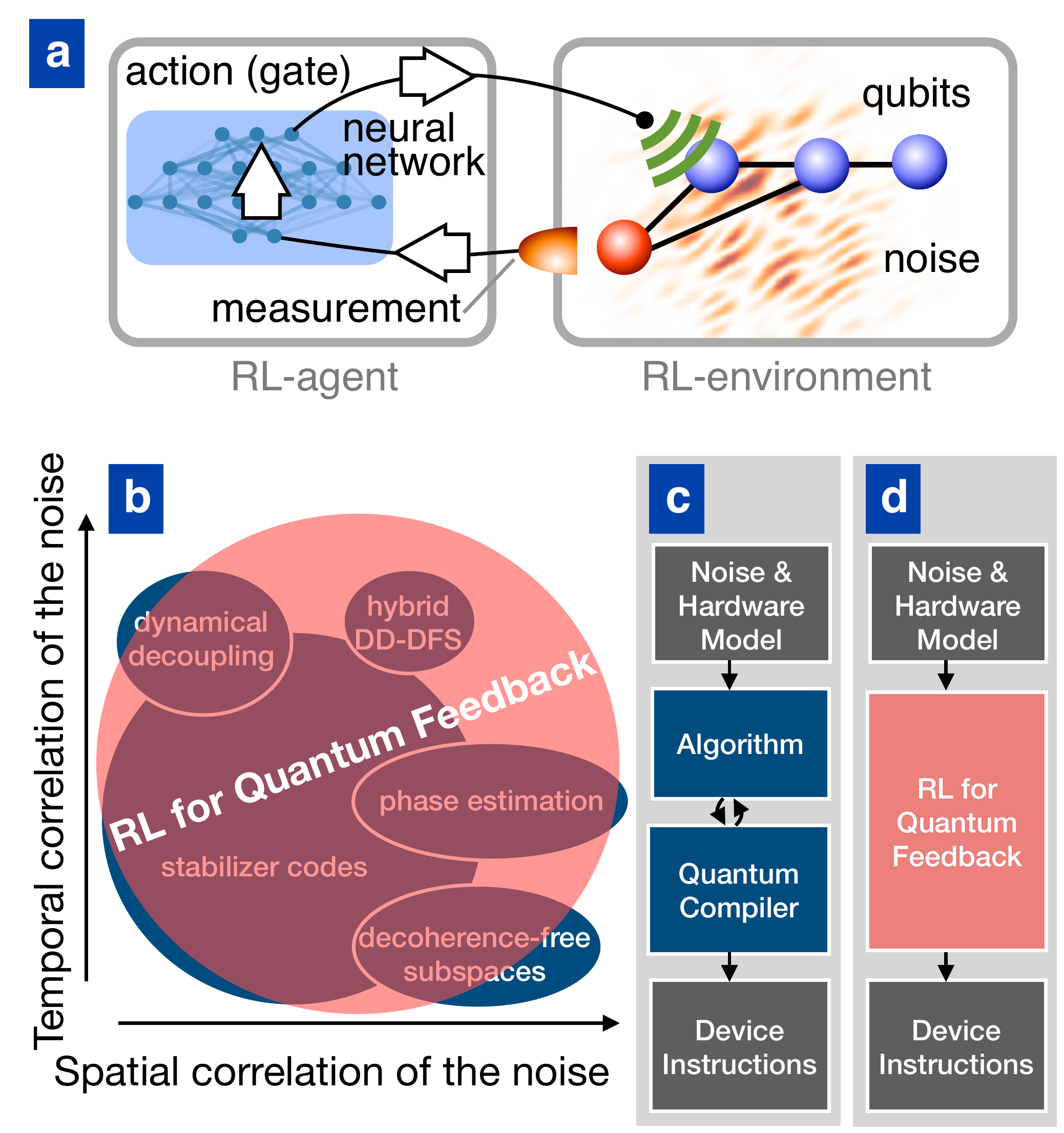}
	
	\caption{\label{fig:SetupAndGenerality}(color) (a) The general setting of
	this work: A few-qubit quantum device with a neural-network-based
	controller whose task is to protect the quantum memory residing in
	this device against noise. Reinforcement learning (RL) lets the controller
	(``RL-agent'') discover on its own how to best choose gate sequences,
	perform measurements, and react to measurement results, by interacting
	with the quantum device (``RL-environment''). (b) visualizes the
	flexibility of our approach (schematic). Depending on the type of
	noise and hardware setting, different approaches are optimal (DD,
	dynamical decoupling; DFS, decoherence-free subspace). By contrast,
	the RL approach is designed to automatically discover the best strategy,
	adapted to the situation. In (c) we show the conventional procedure
	to select some QEC algorithm and then produce hardware-adapted device
	instructions (possibly re-iterating until an optimal choice is found).
	We compare this to our approach (d) that takes care of all these steps
	at once and provides QEC strategies fully adapted to the concrete
	specifications of the quantum device.}
\end{figure}

In this work, we introduce network-based RL in physics (\cref{fig:SetupAndGenerality})
and illustrate its versatility in the domain of quantum feedback.
Specifically, we devise a unified, fully autonomous, human-guidance-free
approach for discovering quantum-error-correction (QEC) strategies
from scratch, in few-qubit quantum systems subject to arbitrary noise
and hardware constraints. This approach relies on a network agent
that learns feedback strategies, adapting its actions to measurement
results. As illustrated in \cref{fig:SetupAndGenerality}b-d, our
method provides a unified approach to protect a quantum memory from
noise. It covers a wide range of scenarios where one would otherwise
have to select an existing scheme (stabilizer codes, adaptive phase
estimation, etc.) and adapt it to the given situation. Our findings
are of immediate relevance to the broad field of quantum error correction
(including quantum-error-mitigation techniques)
and are best suited to be used in few-qubit quantum modules. These
could be used as stand-alone quantum memory or be part of the modular
approach to quantum computation, which has been suggested for several
leading hardware platforms\cite{monroe_scaling_2013,devoret_superconducting_2013}.

Given a collection of qubits and a set of available quantum gates,
the agent is asked to preserve an arbitrary quantum state $\alpha\left|0\right\rangle +\beta\left|1\right\rangle $
initially stored in one of the qubits. It finds complex sequences
including projective measurements and entangling gates, thereby protecting
the quantum information stored in such a few-qubit system against
decoherence. This is a very complex challenge, where both brute force
searches and even the most straightforward RL approaches fail. The
success of our approach is due to a combination of two key ideas:
(i) two-stage learning, with an RL-trained network receiving maximum
input acting as a teacher for a second network, and (ii) a measure
of the recoverable quantum information hidden inside a collection
of qubits, being used as a reward.

Recent progress in multi-qubit quantum devices \cite{chiaverini_realization_2004,schindler_experimental_2011,waldherr_quantum_2014,kelly_state_2015,veldhorst_two-qubit_2015,ofek_extending_2016,potocnik_studying_2017,debnath_demonstration_2016,takita_experimental_2017,watson_programmable_2017}
has highlighted hardware features deviating from often-assumed idealized
scenarios. These include qubit connectivity, correlated noise, restrictions
on measurements, or inhomogeneous error rates. Our approach can help finding
``hardware-adapted'' solutions. This builds on the main advantage
of RL, namely its flexibility: it can discover strategies for such
a wide range of situations with minimal domain-specific input. We
illustrate this flexibility in examples from two different domains:
in one set of examples (uncorrelated bit-flip noise), the network
is able to go beyond rediscovering the textbook stabilizer repetition
code. It finds an adaptive response to unexpected measurement results
that allows it to increase the coherence time, performing better than
any straightforward non-adaptive implementation. Simultaneously, it
automatically discovers suitable gate sequences for various types
of hardware settings. In another, very different example, the agent
learns to counter spatially correlated noise by finding non-trivial
adaptive phase-estimation strategies that quickly become intractable
by conventional numerical approaches such as brute-force search. Crucially,
all these examples can be treated by exactly the same approach, with
no fine-tuning. The only input consists in the problem specification
(hardware and noise model).

In a nutshell, our goal is to have a neural network
which can be employed in an experiment, receiving measurement results
and selecting suitable subsequent gate operations conditioned on these
results. However, in our two-stage learning approach, we do not directly
train this neural network from scratch. Rather, we first employ reinforcement
learning to train an auxiliary network that has full knowledge of
the simulated quantum evolution. Later on, the experimentally applicable
network is trained in a supervised way to mimic the behavior of this
auxiliary network.

We emphasize that feedback requires reaction towards the observations,
going beyond optimal control type challenges (like pulse shape optimization
or dynamical decoupling), and RL has been designed for exactly this
purpose. Specifically, in this work we will consider discrete-time,
digital feedback, of the type that is now starting to be implemented
experimentally \cite{riste_feedback_2012,campagne-ibarcq_persistent_2013,steffen_deterministic_2013,pfaff_unconditional_2014,riste_digital_2016},
e.\,g.\ for error correction in superconducting quantum computers.
Other wide-spread optimization techniques for quantum control, like
GRAPE, often vary evolution operators with respect to continuous parameters
\cite{khaneja_optimal_2005,machnes_comparing_2011}, but do not easily
include feedback and are most suited for optimizing the pulse shapes
of individual gates (rather than complex gate sequences acting on
many qubits). Another recent approach \cite{johnson_qvector:_2017}
to quantum error correction uses optimization of control parameters
in a pre-configured gate sequence. By contrast, RL directly explores
the space of discrete gate sequences. Moreover, it is a ``model-free''
approach \cite{russell_artificial_2016}, i.\,e.\ it does not rely
on access to the underlying dynamics. What is optimized is the network
agent. Neural-network based RL promises to complement other successful
machine-learning techniques applied to quantum control \cite{hentschel_machine_2010,hentschel_efficient_2011,palittapongarnpim_learning_2016,tiersch_adaptive_2015}.

\begin{figure}
	\includegraphics[width=0.9\columnwidth]{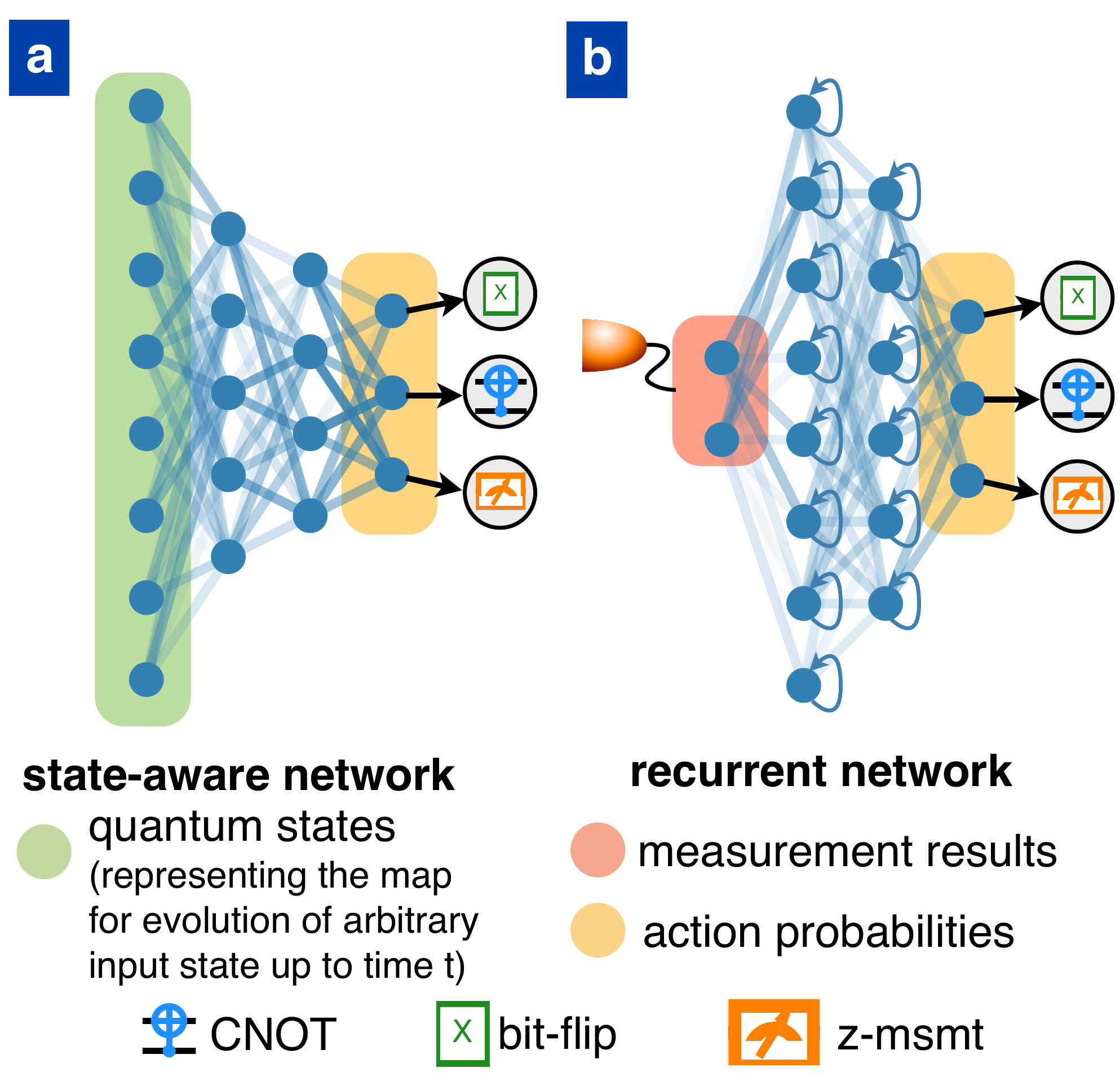}
	
	\caption{\label{fig:NetworkLayout}(color) The neural networks. (a) At each
	time $t$, the ``state-aware'' network receives a representation
	of the map $\Phi$ describing the quantum evolution of arbitrary initial
	logical qubit states up to that time (represented by four evolved
	states $\hat{\rho}$; see main text). It outputs the probabilities
	for the different actions (gates), defining the agent's policy. Like
	any neural network, it is a nonlinear function that can be decomposed
	into layer-wise linear superposition of neuron values, using trainable
	weights (visualized by connections), and the application of nonlinear
	activation functions. Examples for actions are shown here (bit-flip,
	CNOT, measurement). Each of those can be applied to various qubits
	(or qubit pairs), resulting in around 10-20 actions. (b) The recurrent
	network receives the most recent measurement result (if any) and also
	outputs action probabilities. Its long short-term memory (LSTM) neurons
	learn to keep track of information accumulated in previous time steps
	(schematically indicated by the recurrent connections here).}
\end{figure}

Conceptually, our approach aims to control a quantum system using
a classical neural network. To avoid confusion, we emphasize our approach
is distinct from future ``quantum machine learning'' devices, where
even the network will be quantum \cite{biamonte_quantum_2017,romero_quantum_2017,dunjko_machine_2017}.

\section*{Reinforcement Learning}

The purpose of RL (\cref{fig:SetupAndGenerality}a) is to find an
optimal set of actions (in our case, quantum gates and measurements)
that an ``agent'' can perform in response to the changing state
of an ``environment'' (here, the quantum memory). The objective
is to maximize the expected ``return'' $R$, i.\,e.\ a sum of
rewards.

To find optimal gate sequences, we employ a widespread version of
reinforcement learning \cite{williams_simple_1992,goodfellow_deep_2016}
where discrete actions are selected at each time step $t$ according
to a probabilistic ``policy'' $\pi_{\theta}$. Here, $\pi_{\theta}(a_{t}|s_{t})$
is the probability to apply action $a_{t}$, given the state $s_{t}$
of the RL-environment. As we will use a neural network to compute
$\pi_{\theta}$, the multi-dimensional parameter $\theta$ stands
for all the network's weights and biases. The network is fed $s_{t}$
as an input vector and outputs the probabilities $\pi_{\theta}(a_{t}|s_{t})$.
The expected return can then be maximized by applying the policy gradient
RL update rule \cite{williams_simple_1992}:
\begin{equation}
	\delta\theta_{j}=\eta\frac{\partial\mathbb{E}[R]}{\partial\theta_{j}}=\eta\,\mathbb{E}\bigg[R\sum_{t}\frac{\partial}{\partial\theta_{j}}\ln\pi_{\theta}(a_{t}|s_{t})\bigg]\,,\label{eq:simplePolicyGradient}
\end{equation}
with $\eta$ the learning rate parameter, and $\mathbb{E}$ the expectation
value over all gate sequences and measurement outcomes. These ingredients
summarize the basic policy gradient approach. In practice, improvements
of \cref{eq:simplePolicyGradient} are used; for example, we employ
a baseline, natural policy gradient, and entropy regularization (see
Appendix). Even so, several further conceptual steps are essential
to have any chance of success (see below).

\Cref{eq:simplePolicyGradient} provides the standard recipe for a
fully observed environment. This approach can be extended to a partially
observed environment, where the policy would then be a function of
the observations only, instead of the state. The observations contain
partial information on the actual state of the environment. In the
present manuscript we will encounter both cases.

\section*{Reinforcement Learning Approach to Quantum Memory\label{sec:Reinforcement-leaning-approach}}

In this work we seek to train a neural network to develop strategies
to protect the quantum information stored in a quantum memory from
decoherence. This involves both variants of stabilizer-code-based
QEC \cite{shor_scheme_1995,nielsen_quantum_2011,terhal_quantum_2015}
as well as other, more specialized (but, in their respective domain,
more resource-efficient) approaches, like decoherence-free subspaces
or phase estimation. We remind the reader that, for the particular case of stabilizer-code-based
QEC, the typical steps are: (i) the encoding, in which the logical
state initially stored in one qubit is distributed over several physical
qubits, (ii) the detection of errors via measurement of suitable multi-qubit
operators (syndromes), (iii) the subsequent correction, and (iv) the
decoding procedure that transfers the encoded state back into one
physical qubit. We stress that no such specialized knowledge will
be provided a priori to our network, thus retaining maximum flexibility
in the tasks it might be applied to and in the strategies it can
encompass (\cref{fig:SetupAndGenerality}b).

We start by storing an arbitrary quantum state $\alpha\left|0\right\rangle +\beta\left|1\right\rangle$
inside one physical qubit. The goal is to be able to retrieve this
state with optimum fidelity after a given time span. Given hardware
constraints such as the connectivity between qubits, the network agent
must develop an efficient QEC strategy from scratch solely by interacting
with the quantum memory at every time step via a set of unitary gates
(such as CNOTs and bit-flips) and measurements. They are chosen according
to the available hardware and define the action set of the agent.
Importantly, the network must react and adapt its strategy to the
binary measurement results, providing real-time quantum feedback.

This particular task seems practically unsolvable for the present
reinforcement learning techniques if no extra precautions are taken.
The basic challenge is also encountered in other difficult RL applications:
the first sequence leading to an increased return is rather long.
In our scenarios, the probability to randomly select a good sequence
is much less than $10^{-12}$. Moreover, any subsequence may be worse
than the trivial (idle) strategy: for example, performing an incomplete
encoding sequence (ending up in a fragile entangled state) can accelerate
decay. Adopting the straightforward return, namely the overlap of
the final and initial states, both the trivial strategy and the error-correction
strategy are fixed points. These are separated by a wide barrier –
all the intermediate-length sequences with lower return. In our numerical
experiments, naive RL was not successful, except for some tasks with
very few qubits and gates.

We introduce two key concepts to solve this challenge: a two-stage
learning approach with one network acting as teacher of another, and
a measure of the ``recoverable quantum information'' retained in
any quantum memory.

Before we address these, we mention that from a machine-learning point-of-view
there is another unconventional aspect: Instead of sampling initial
states of the RL-environment stochastically, we consider the evolution
under the influence of the agent's actions for \emph{all} possible
states simultaneously. This is required because the quantum memory
has to preserve arbitrary input states. Our reward will be based on
the completely positive map describing the dissipative quantum evolution
of arbitrary states. The only statistical averaging necessary is over
measurement outcomes and the probabilistic action choices. Further
below, we comment on how this is implemented in practice.

As known from other RL applications (like board games \cite{silver_mastering_2017}),
it helps to provide as much information as possible to the network.
In our case, this could mean providing the multi-qubit quantum state
at each time step. However, that information is not available in a
real experiment. In order to solve this dilemma, we train two different
networks in succession (\cref{fig:NetworkLayout}a,b): The first network
is fully \emph{state-aware}. Later on, we will use it as a teacher
for the second network which essentially only gets the measurement
results as an input (plus the information which gate or measurement
has been applied). This splits the problem into two sub-problems that
are easier to solve. In this approach, the main remaining challenge
is to train the state-aware network, while the supervised training
of the second network is fairly straightforward in our experience.
In contrast, directly training the second network via RL would be
tremendously harder, if not impossible, because the input would be
significantly less comprehensive than the completely positive map.

At this point, we see that evolving all initial states simultaneously
is not only more efficient, but even required to prevent the state-aware
network from ``cheating''. Otherwise, it might simply memorize the
initial state, wait for it to relax, and then reconstruct it – which,
of course, is not a valid strategy to preserve a principally \emph{unknown}
quantum state. Such a behavior is avoided when the network is asked
to preserve all possible logical qubit states with the same gate sequence.
It turns out that this can be implemented efficiently by evolving
just four initial quantum states $\hat{\rho}$ (for a single logical
qubit); tracking their evolution fully characterizes, at any point
in time, the completely positive map $\Phi$ of the multi-qubit system
that maps $\hat{\rho}(0)$ to $\hat{\rho}(t)$. Moreover, we have
found it useful to apply principal component analysis, i.\,e.\ to
feed only the few largest-weight eigenvectors of the evolved $\hat{\rho}$s
as input to the network (see Appendix).

We are now ready to define our problem fully from the point of view
of reinforcement learning. The state space of the RL environment is
the space of completely positive maps. This information is not accessible
in a real-world experiment, where the measurements provide partial
information about the RL-environment. This reinforcement-learning
problem is therefore classified as a partially observed Markov process.
This is what is considered in our second learning stage, and our method
to solve it relies on a recurrent network. In the modified input scheme
of the first learning stage, the agent observes the full state space
and we therefore deal with a fully observed Markov process. In both
cases, the RL environment is stochastic due to the measurements. As
described above, the action set is defined by the available hardware
instructions (unitary gates and measurements).

Two-stage learning with parallel evolution is essential, but not yet
sufficient for our challenge. We now introduce a suitable reward that
indicates the likely final success of an action sequence ahead of
time. In our case, we follow the intuitive idea that this reward should
quantify whether the original quantum information survives in the
complex entangled many-qubit state that results after application
of unitary gates and measurements, and with the system subject to
decoherence.

We note that, in the ideal case, without decoherence, two initially
orthogonal qubit states are always mapped onto orthogonal states.
Therefore, they remain 100\% distinguishable, and the original state
can always be restored. With a suitable encoding, this remains true
even after some errors have happened, if a suitable error-detection
and decoding sequence is applied (``recovery''). By contrast, irreversible
loss of quantum information means that perfect recovery becomes impossible.
In order to make these notions concrete, we start from the well-known
fact that the probability to distinguish two quantum states $\hat{\rho}_{1}$
and $\hat{\rho}_{2}$, by optimal measurements, is given by the trace
distance $\frac{1}{2}\Vert\hat{\rho}_{1}-\hat{\rho}_{2}\Vert_{1}$.
Let $\hat{\rho}_{\vec{n}}(t)$ be the quantum state into which the
multi-qubit system has evolved, given the initial logical qubit state
of Bloch vector $\vec{{n}}$. We now consider the distinguishability
of two initially orthogonal states, $\frac{1}{2}\Vert\hat{\rho}_{\vec{n}}(t)-\hat{\rho}_{-\vec{n}}(t)\Vert_{1}$.
In general, this quantity may display a non-trivial, non-analytic
dependence on $\vec{n}$. We introduce the ``\emph{recoverable quantum
information}'' as:
\begin{equation}
	\mathcal{R}_{Q}(t)=\frac{1}{2}{\rm min}_{\vec{n}}\left\Vert \hat{\rho}_{\vec{n}}(t)-\hat{\rho}_{-\vec{n}}(t)\right\Vert _{1}\,.\label{eq:RecoverableQuantumInformation}
\end{equation}
The minimum over the full Bloch sphere is taken because the logical
qubit state is unknown to the agent, so the success of an action sequence
is determined by the worst-case scenario. In other words, $\mathcal{R}_{Q}$
specifies a guaranteed value for the remaining distinguishability
for all possible logical qubit states. Thus, $\mathcal{R}_{Q}$ is
a property of the completely positive map that characterizes the dissipative
evolution.

The recoverable quantum information $\mathcal{R}_{Q}$ is much more
powerful than the overlap of initial and final states, as it can be
used to construct an \emph{immediate} reward, evaluating a strategy
even at intermediate times. In the idealized case where errors have
occured, but they could \textit{in principle} be perfectly recovered
by a suitable detection/decoding sequence, $\mathcal{R}_{Q}$ remains
1. As we will see below, this behavior steers the network towards
suitable strategies. $\mathcal{R}_{Q}$ can be extended towards multiple
logical qubits.

As far as $\mathcal{R}_{Q}$ is concerned, error correction steps
are only required to prevent the multi-qubit system from venturing
into regions of the Hilbert space where any further decoherence process
would irreversibly destroy the quantum information (and lower $\mathcal{R}_{Q}$).
If one wants the network to actually implement the final decoding
sequence, to return back an unentangled state, this can be done by
adding suitable contributions to the reward (see below).

\begin{figure}
	\includegraphics[width=1\columnwidth]{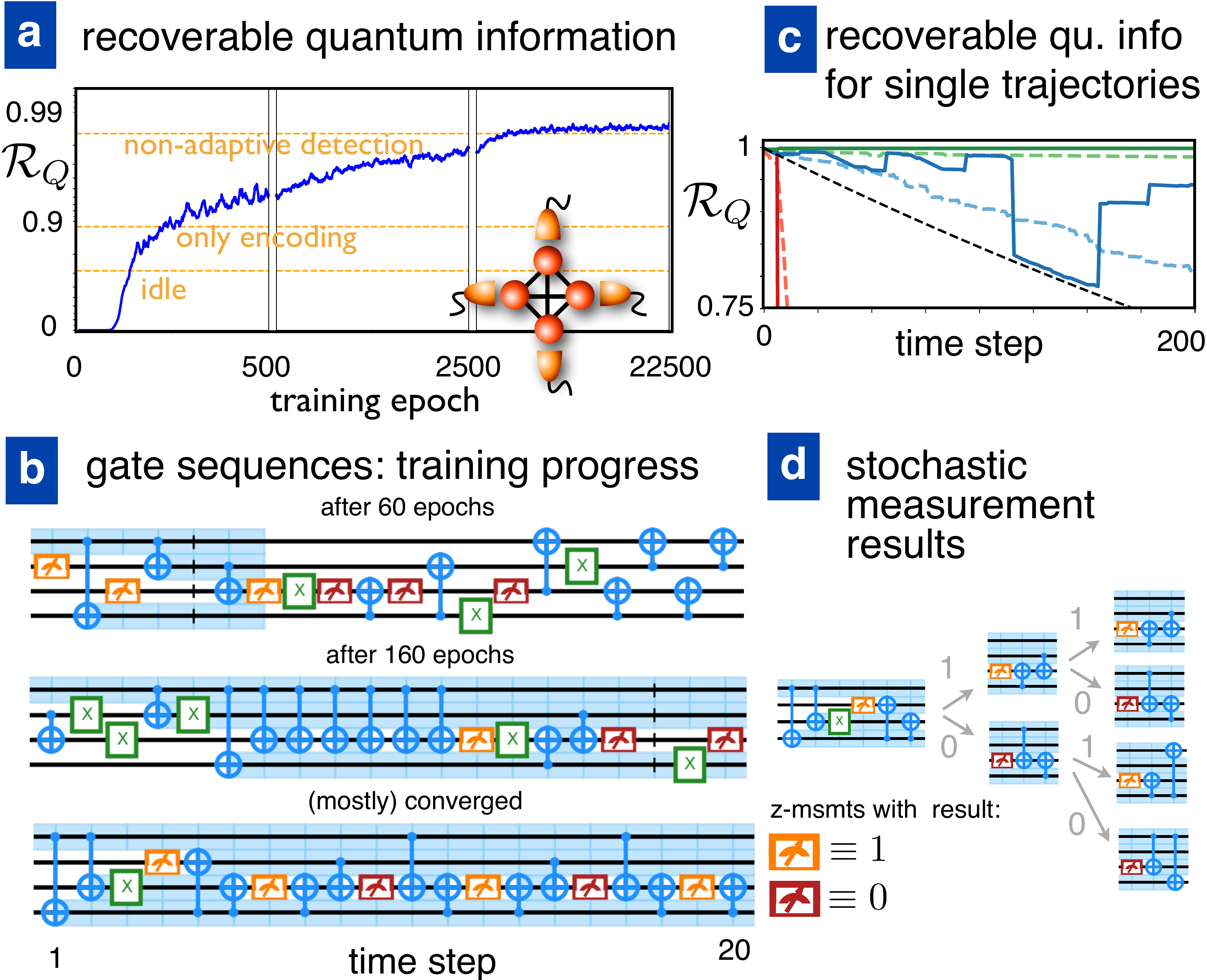}
	
	\caption{\label{fig:TrainingOverviewFigure}(color) Reinforcement learning
	for the state-aware network, with a 4-qubit setup (all qubits can
	be measured, as indicated by the detectors and the red color; all
	qubit pairs can be subject to CNOT, as indicated by the links). (a)
	Training progress in terms of the average recoverable quantum information
	$\mathcal{R}_{Q}$, evaluated at the final step of a 200-step gate
	sequence. One ``epoch'' involves training on a batch of 64 trajectories
	(gate sequences). Eventually, the network performs even better than
	a combination of encoding and periodic parity checks, due to the adaptive
	recovery sequences that are triggered by unexpected measurements (i.\,e.\ upon
	error detection). In this example, that leads to ca. 15\% increase
	of the decoherence time over a non-adaptive scheme. (b) Typical gate
	sequences at different training stages, mostly converged strategy
	at bottom. Qubits participating in encoding (holding information about
	the logical qubit state) are indicated with light blue background.
	(c) Time-evolution of $\mathcal{R}_{Q}$, at the same three different
	training stages (red,blue,green), for individual trajectories (dashed:
	averaged over many trajectories). Jumps are due to measurements. (d)
	Evolution depending on stochastic measurement results, indicated as
	``0''/``1'' and also via the color (red/orange) of the measurement
	gate symbol. The policy strongly deviates between the different branches,
	demonstrating that RL finds adaptive quantum feedback strategies,
	where the behaviour depends on the measurement outcomes. Note that
	even the mostly converged strategy is still probabilistic to some
	degree.}
\end{figure}

\section*{Results\label{sec:Results}}

We now apply the general approach to different settings, illustrating
its flexibility. The training of the state-aware network is analyzed
in \cref{fig:TrainingOverviewFigure}. In the example, the qubits
are subject to bit-flip errors uncorrelated in space and time, with
a decay term $\dot{\hat{\rho}}=T_{{\rm dec}}^{-1}\sum_{j}(\hat{\sigma}_{x}^{(j)}\hat{\rho}\hat{\sigma}_{x}^{(j)}-\hat{\rho})$
in the underlying master equation (see Appendix). All of the four
qubits may be measured, and there is full connectivity. During training
(\cref{fig:TrainingOverviewFigure}a,b), the network first learns
to avoid destructive measurements which reveal the logical qubit state.
Afterwards, it discovers a gate sequence of ${\rm CNOT}$s that creates
an entangled state, implementing some version of the 3-qubit repetition
code \cite{nielsen_quantum_2011,terhal_quantum_2015}. The particular
CNOT sequence shown in the figure generates one possible encoded state
out of several equally good ones. The symmetry between these alternative
encodings is broken spontaneously during training. The encoding already
increases the reward above the trivial level (obtained for storing
the logical qubit in one physical qubit only). Finally, the network
starts doing repeated parity measurements, of the type ${\rm CNOT}(B\mapsto A),{\rm \,CNOT}(C\mapsto A),\,{\rm M}(A)$,
flipping the state of ancilla $A$ only if the states of $B$ and
$C$ differ (here $M$ is a measurement). This implements error detection,
helping to preserve the quantum information by preventing the leakage
into states with two bit flips that cannot be corrected if undetected.
\Cref{fig:TrainingOverviewFigure}b illustrates the progression from
random quantum circuits to a nearly converged strategy. During any
single trajectory, the recoverable quantum information can have sudden
jumps when measurements are performed (\cref{fig:TrainingOverviewFigure}c),
with collapses and revivals.

\begin{figure*}
	\centering
	
	\includegraphics[width=0.8\textwidth]{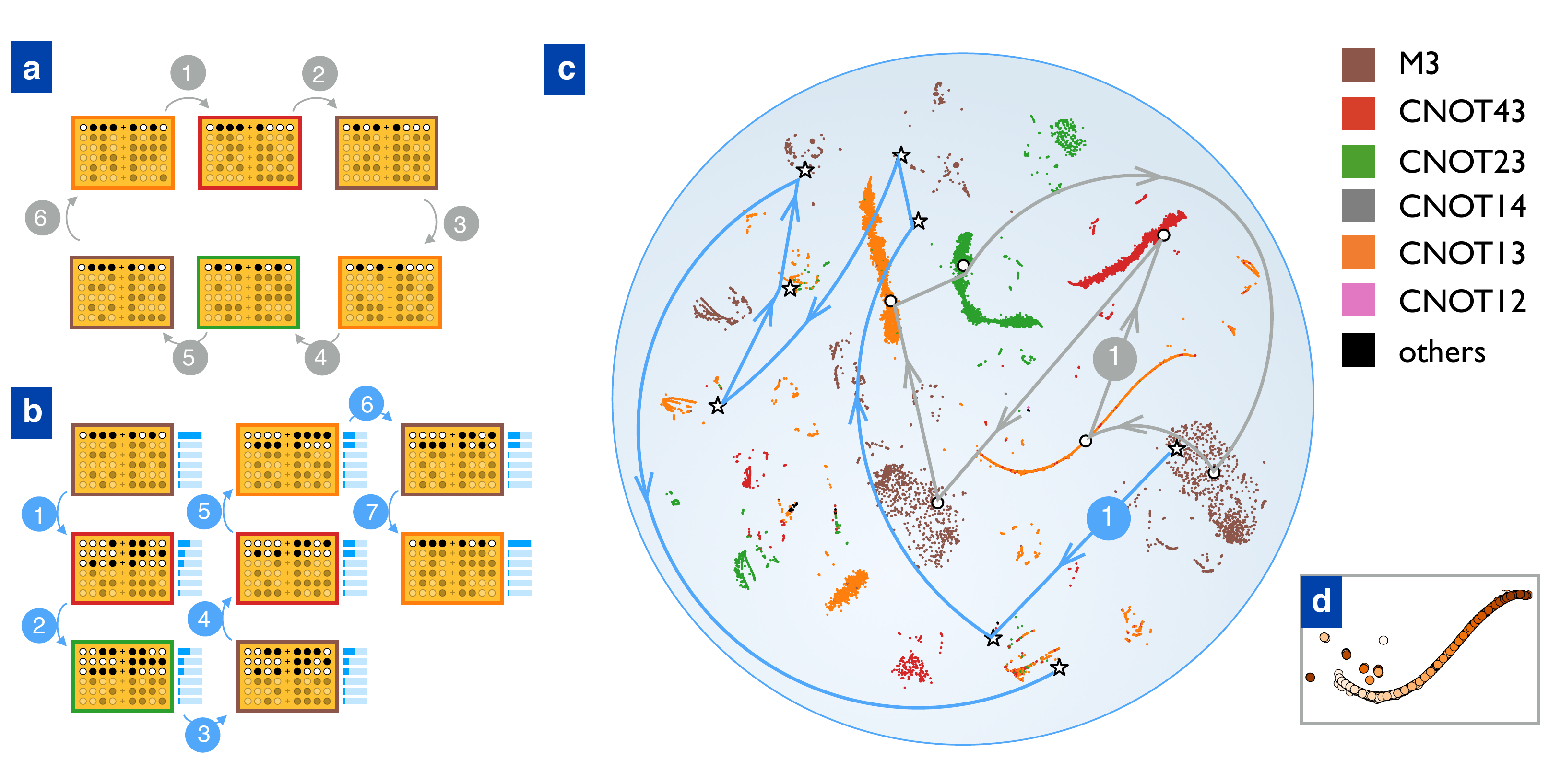}
	
	\caption{\label{fig:tSNE}(color) Visualizing the operation of the state-aware
	network. (Same scenario as in \cref{fig:TrainingOverviewFigure})
	(a) Sequence of quantum states visited during a standard repetitive
	detection cycle (after encoding), displayed for an initial logical
	qubit state $\mid+x\rangle=(\left|0\right\rangle +\left|1\right\rangle )/\sqrt{2}$,
	in the absence of unexpected measurement outcomes (no errors). Each
	state $\hat{\rho}$ is represented by its decomposition into eigenstates,
	where e.\,g.\ $\circ\bullet\bullet\bullet+\bullet\circ\bullet\circ\equiv\mid0111\rangle+\mid1010\rangle$.
	The eigenstates are sorted according to decreasing eigenvalues (probabilities).
	Eigenstates of less than 5\% weight are displayed semi-transparently.
	In this particular example, each eigenstate is a superposition of
	two basis states in the z-basis. (b) Gate sequence triggered upon
	encountering an unexpected measurement. Again, we indicate the states
	$\hat{\rho}$, with bars now showing the probabilities. This sequence
	tries to disambiguate, by further measurements, the error. (c) Visualization
	of neuron activations (300 neurons in the last hidden layer), in response
	to the quantum states (more precisely, maps $\Phi$) encountered in
	many runs. These activations are projected down to 2D using the t-SNE
	technique \cite{maaten_visualizing_2008}, which is a nonlinear mapping
	that tries to preserve neighborhood relations while reducing the dimensionality.
	Each of the $2\cdot10^{5}$ points corresponds to one activation pattern
	and is colored according to the action taken. The sequences of (a)
	and (b) are indicated by arrows, with the sequence (b) clearly proceeding
	outside the dominant clusters (which belong to the more typically
	encountered states). Qubits are numbered $1,2,3,4$, and ${\rm CNOT13}$
	has control-qubit $1$ and target $3$. (d) The zoom-in shows a set
	of states in a single cluster which is revisited periodically during
	the standard detection cycle; this means we stroboscopically observe
	the time evolution. The shading indicates the time progressing during
	the gate sequence, with a slow drift of the state due to decoherence.
	(cf.\ Supplementary for extended discussion)}
\end{figure*}

Can we understand better how the network operates? To this end, we
visualize the responses of the network responses to the input states
(\cref{fig:tSNE}c), projecting the high-dimensional neuron activation
patterns into the 2D plane using the t-SNE technique \cite{maaten_visualizing_2008}.
Similar activation patterns are mapped close to each other, forming
clearly visible clusters, each of which results in one type of action.
During a gate sequence, the network visits states in different clusters.
The sequence becomes complex if unexpected measurement results are
encountered (\cref{fig:tSNE}b). In the example shown here, the outcome
of the first parity measurement is compatible with three possibilities
(one of two qubits has been flipped, or the ancilla state is erroneous).
The network has learned to resolve the ambiguity through two further
measurements, returning to the usual detection cycle. It is remarkable
that RL finds these nontrivial sequences (which would be complicated
to construct ab initio), picking out reward differences of a few percent.

\begin{figure}
	\includegraphics[width=1\columnwidth]{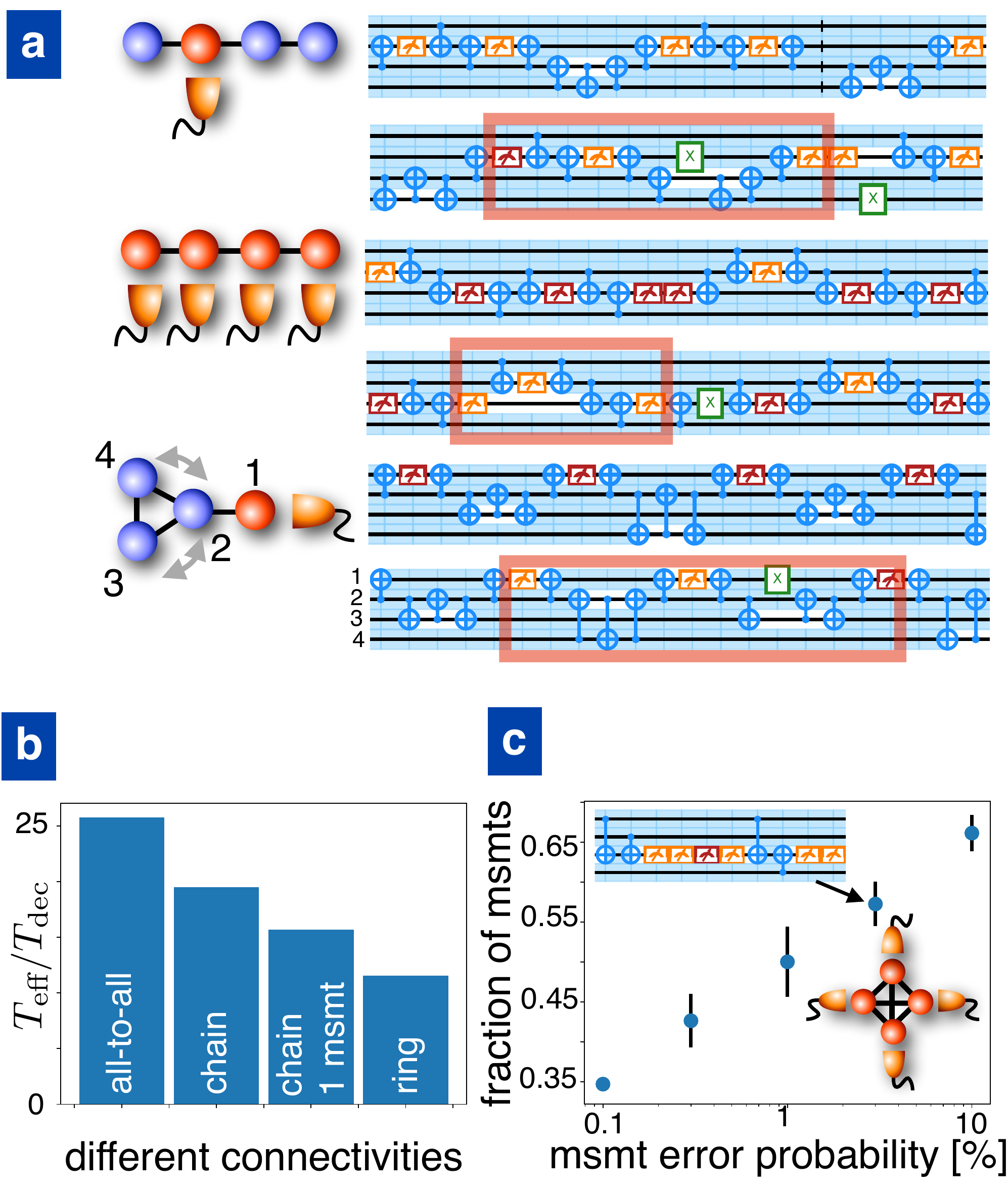}
	
	\caption{\label{fig:ManyScenarios}(color) (a) Scenarios with different qubit
	connectivity. CNOTs are allowed only between qubits connected by a
	line. In each case, we display the ``standard'' gate sequence discovered
	by the net, as well as a sequence involving corrective actions (triggered
	by an unexpected measurement). From top to bottom: chain with fixed
	measurement location, chain with arbitrary measurements, ring connected
	to an ancilla. The red rectangle highlights an interval during which
	the quantum state is not dominated by a single component, indicating
	that the precise location of the error still has to be pinpointed.
	These nontrivial detection/recovery sequences are considerably more
	complex than the periodic detection cycle. (b) The effective enhancement
	of the decoherence time via error correction, for the different scenarios.
	Here, $T_{{\rm dec}}=1200$ is the single-qubit decoherence time (in
	units of the gate time that defines the time step), and $T_{{\rm eff}}$
	was extracted from the decay of $\mathcal{R}_{Q}$ after $200$ time
	steps. The differences can be traced back to the lengths of the detection
	cycles. (c) Behaviour as a function of measurement error. The network
	discovers that redundancy is needed, i.\,e.\ the number of measurements
	in the gate sequences increases (in this plot, from $1$ per cycle
	to about $6$).}
\end{figure}

The flexibility of the approach is demonstrated by training on different
setups, where the network discovers from scratch other feedback strategies
(\cref{fig:ManyScenarios}a) adapted to the available resources. For
example, we consider a chain of qubits where CNOTs are available only
between nearest neighbours and in addition we fix a single measurement
location. Then the network learns that it may use the available CNOTs
to swap through the chain. However, if every qubit can be measured,
the net discovers a better strategy with fewer gates, where the middle
two qubits of the chain alternate in playing the role of ancilla.
We also show, specifically, the complex recovery sequences triggered
by unexpected measurements. They are a-priori unkown, and RL permits
to discover them from scratch without extra input. Generally, additional
resources (such as enhanced connectivity) are exploited to yield better
improvement of the decoherence time (\cref{fig:ManyScenarios}b).
In another scenario, \cref{fig:ManyScenarios}c, we find that the
network successfully learns to adapt to unreliable measurements by
redundancy.

\begin{figure}
	\includegraphics[width=1\columnwidth]{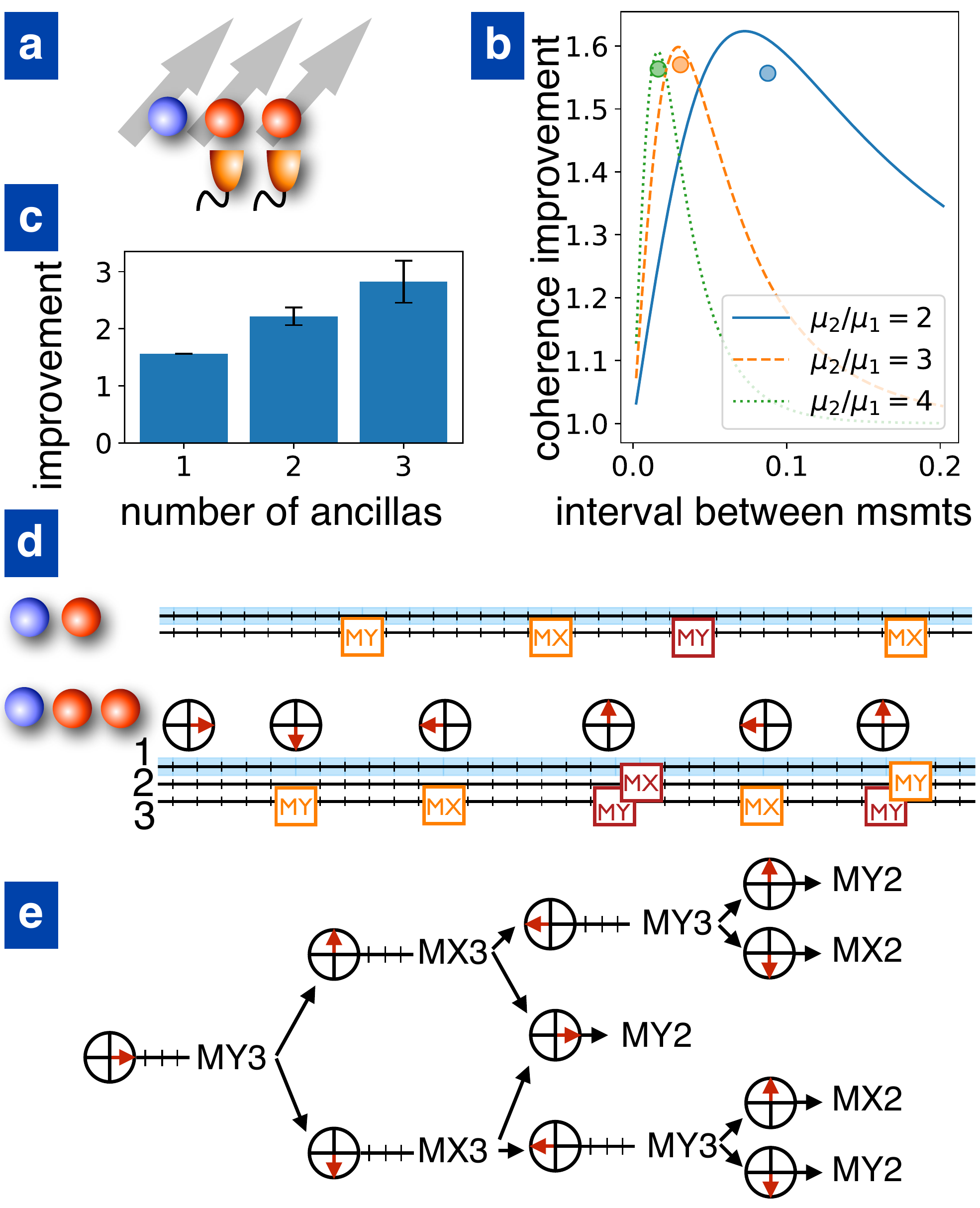}
	
	\caption{\label{fig:CorrelatedNoise}(color) Countering dephasing by measurements
	(adaptive phase estimation). (a) The setting: a data qubit, whose
	phase undergoes a random walk due to a fluctuating field. Since the
	field is spatially correlated, its fluctuations can be detected by
	measuring the evolution of nearby ancilla qubits, which can then be
	exploited for correction. In this example, the allowed actions are
	measurements along x,y (MX, MY), and the idle operation. (b) For one
	ancilla, the network performs measurements periodically, finding the
	optimal measurement interval. This can be seen by comparing the coherence
	time enhancement as a function of the interval (in units of the single-qubit
	decoherence time $T_{{\rm single}}$) for the network (circles) with
	the analytical predictions (curves; cf.\ Supplementary). Remaining
	differences are due to the discretization of the interval (not considered
	in the analytics). The coupling between noise and ancilla ($\mu_{2}$)
	is different from that between noise and data qubit ($\mu_{1}$),
	and the strategy depends on the ratio indicated here. (c) Coherence
	time enhancement, for different numbers of ancillas (here $\mu_{2}=\mu_{3}=\mu_{4}=4\mu_{1})$.
	(d) Gate sequences. For two ancillas, the network discovers an adaptive
	strategy, where measurements on qubit \#2 are rare, and the measurement
	basis is decided based on previous measurements of qubit \#3. The
	arrows show the measurement result for qubit \#3, in the equator plane
	of the Bloch sphere. (e) The 2-ancilla adaptive strategy (overview;
	see also Supplementary). Here, a brute-force search for strategies
	is still (barely) possible, becoming infeasible for higher ancilla
	numbers due to the exponentially large search space of adaptive strategies.
	[$\mu_{2}=3.8\mu_{1},\,\mu_{3}=4.1\mu_{1}$ in (d),(e)]}
\end{figure}

In a separate class of scenarios, we consider dephasing of a qubit
by a fluctuating field (\cref{fig:CorrelatedNoise}). If the field
is spatially homogeneous and also couples to nearby ancilla qubits,
then the dephasing is collective: $\hat{H}(t)=B(t)\sum_{j}\mu_{j}\hat{\sigma}_{z}^{(j)}$,
where $B(t)$ is white noise and $\mu_{j}$ are the coupling strengths
(to qubit and ancillas). Note that, in this situation, one can use
neither dynamical decoupling (since the noise is uncorrelated in time)
nor decoherence-free subspaces (since the $\mu_{j}$ can be arbitrary
in general). However, the same RL program used for the examples above
also finds solutions here (\cref{fig:CorrelatedNoise}), without any
input specific to the situation (except the available gates). It discovers
that the field fluctuations can be tracked and corrected (to some
extent) by observing the evolution of the nearby ancillas, measuring
them in suitable time intervals. For more than one ancilla, the network
discovers a strategy that is adaptive: The choice of measurement basis
depends on the history of previous observations. Brute-force searches
in this setting become quickly impossible due to the double-exponential
growth of possibilities. The computational effort involved in such
a brute-force approach is analyzed in detail in the Supplementary.

\begin{figure}
	\includegraphics[width=1\columnwidth]{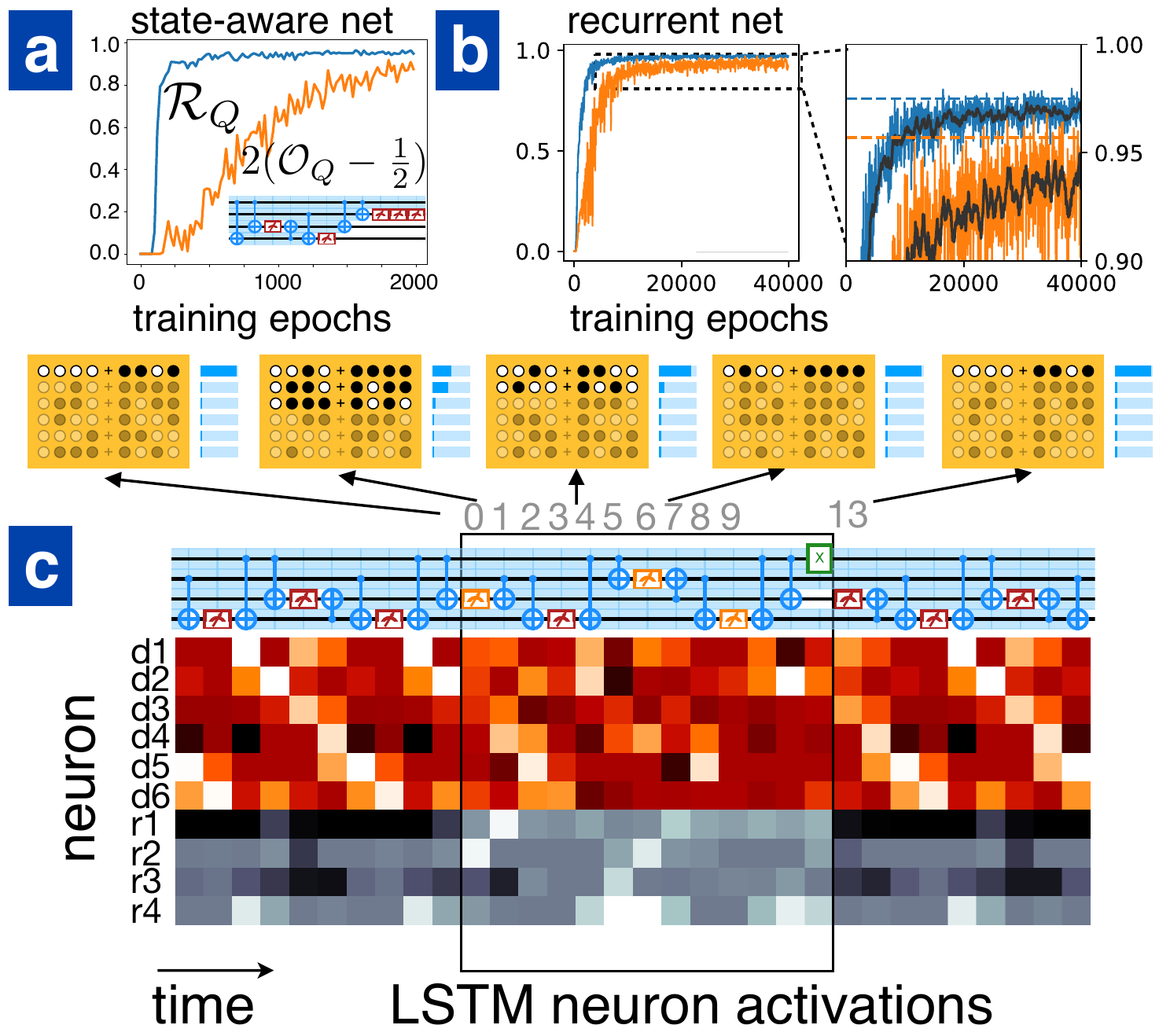}
	
	\caption{\label{fig:LSTMresults}(color) (a) Training the state-aware network
	to implement decoding back into the original data qubit, near the
	end of a 200-step gate sequence. Blue: recoverable quantum information
	$\mathcal{R}_{Q}$. Orange: (rescaled, shifted) overlap $\mathcal{O}_{Q}$
	between final and initial state of the data qubit. Inset displays
	the decoding gate sequence. (b) Training the recurrent network in
	a supervised way to imitate the state-aware network. Again, we show
	$\mathcal{R}_{Q}$ and $\mathcal{O}_{Q}$ evolving during training.
	Dashed blue/orange lines depict the performance of the state-aware
	network. (c) To investigate the workings of the recurrent network,
	we display some of the LSTM neuron activations in the second-to-last
	layer. Quantum states are illustrated even though the network is unaware
	of them. The network's input consists of the observed measurement
	result (if any) and of the actual gate choice (made in the previous
	time step, here aligned on top of the current neuron activations).
	The example gate sequence displayed here first shows the repetitive
	standard error detection pattern of repeated parity measurements that
	is interrupted once an unexpected measurement is encountered, indicating
	an error. During error recovery (boxed region), the network first
	tries to pinpoint the error and then applies corrections. The neurons
	whose behavior changes markedly during recovery are depicted in gray.
	Neuron 1r obviously keeps track of whether recovery is ongoing, while
	3r acts like a counter keeping time during the standard periodic detection
	pattern. (see Supplementary for more)}
\end{figure}

Up to now, the network only encodes and keeps track of errors by suitable
collective measurements. By revising the reward structure, we can
force it to correct errors and finally decode the quantum information
back into a single physical qubit. Our objective is to maximize the
overlap between the initial and final states, for any logical qubit
state (see Appendix). Moreover, we found that learning the decoding
during the final time steps is reinforced by punishing states where
the logical qubit information is still distributed over multiple physical
qubits. The corresponding rewards are added to the previous reward
based on the recoverable quantum information. The network now indeed
learns to decode properly (\cref{fig:LSTMresults}a). In addition,
it corrects errors. It does so typically soon after detecting an error,
instead of at the end of the gate sequence. We conjecture this is
because it tries to return as soon as possible back to the known,
familiar encoded state. For the same reason, error correction sometimes
even happens without an explicit reward.

So far, we have trained the state-aware network. However, this cannot
yet be applied to an experiment, where the quantum state is inaccessible
to us. This requires a network whose only input consists in the measurement
results (and the selected gates, since the policy is probabilistic),
requiring some sort of memory. An elegant solution consists in a \emph{recurrent}
neural network. We use the widespread long short-term memory (LSTM)
approach \cite{hochreiter_long_1997}.

Once the first, state-aware network has been trained successfully,
it is used as a teacher in supervised learning to train the second,
recurrent network (\cref{fig:LSTMresults}b). This could then be applied
as a controller to experimental runs, deciding on gate sequences depending
on measurements. It might also be refined by RL, e.\,g.\ to adapt
to changes in the parameters (decoherence rates etc.). Learning to
correct (see above) is essential for successful training of the recurrent
network, since the latter must learn to consider measurement results
distributed in time and deduce the proper corrective actions.

We have trained the recurrent network based on a fully converged state-aware
network. Inspecting the LSTM neuron activations (\cref{fig:LSTMresults}c),
we see that different neurons activate for different events and some
clearly display prolonged memory (remaining active during certain
time-intervals relevant for the strategy). For example, one neuron
switches on during the recovery sequence after an unexpected measurement,
while another seems like an internal counter operating during the
periodic detection sequence.

We now come back to the statement in the introduction that our approach
is fully autonomous and can be applied to a broad range of problems
with small human effort. In all the preceding examples, and also in
general, the only human input to our approach is the problem specification,
primarily the noise model (specifying the dissipative time evolution
governing the quantum state) and the particular action set (i.\,e.,
the available hardware instructions related to the setup and its connectivity).
Importantly, fine-tuning the hyperparameters (like learning rate,
network architecture, etc.) is not required; in the Supplementary,
we demonstrate that a common set of hyperparameters can be used for
all the scenarios.

\section*{Possible future applications}

The physical setups considered in today's quantum computing platforms
contain many components and features that go beyond the simplest scenario
of short-range coupled qubits. Conceptually, the approach developed
in the present work is general enough to find future application in
any of the following experimentally relevant domains.

An important example is cavities, which can be used as long-lived
quantum memory, especially in the microwave domain. When they are
coupled to qubits, nonlinear operations can be performed that may
aid in error correction of the cavity state, as the Yale group has
demonstrated (``kitten'' and ``cat'' codes \cite{mirrahimi_dynamically_2014,heeres_implementing_2017}).
Our approach allows to cover such situations without any changes to
the reinforcement learning method. Only the description of the physical
scenario, via the set of available actions, and of course the physics
simulation will have to be updated. Cavities also give access to unconventional
controls, e.\,g.\ naturally occurring long-distance multi-qubit
entangling gates provided by the common coupling of the qubits to
the cavity. In addition, they permit direct collective readout that
is sensitive to the joint state of multiple qubits, which may be used
to speed up error detection operations. Again, RL based quantum feedback
of the type proposed here can naturally make use of these ingredients.

Novel hardware setups, like cross-bar type geometries \cite{helmer_cavity_2009,li_crossbar_2017},
give rise to the challenge to exploit the unconventional connectivity,
for which our approach is well suited. In the future, it may even
become possible to co-optimize the hardware layout (taking into account
physical constraints) and the strategies adapted to the layout. In
the simplest case, this means discovering strategies for automatically
generated alternative layouts and comparing their performance.

The actions considered by the agent need not refer to unitary operations.
They might also perform other functions, like restructuring the connectivity
itself in real-time. This is the case for the proposed 2D ion-trap
architecture where the ions are shuffled around using electrodes \cite{monroe_scaling_2013}.
Similar ideas have been proposed for spins in quantum dots, which
can be moved around using electrodes or surface-acoustic waves. Again,
no changes to our approach would be needed. The modifications are
confined to the physics simulation. Depending on the present state
of the connectivity, the set of effective qubit gates would change.

Like any numerical approach, our method is invariably limited to modest
qubit numbers (of course, these will increase with further optimizations,
possibly up to about 10). It is important, therefore, to recall that
even an improvement of the decoherence rate in an isolated few-qubit
module can have useful applications (as a quantum memory, e.\,g.\ in
a quantum repeater). More generally, it is clear that classical simulation
of a full-scale quantum computer in the domain of quantum supremacy
is out of the question, by definition. This is a challenge widely
acknowledged by the entire community, affecting not only optimization
but also design, testing, and verification of a quantum computer.
One promising way to address this challenge at least partially, advocated
by a growing number of experimental groups, is the so-called modular
approach to quantum computation and quantum devices. This consists
in connecting small few-qubit quantum modules together via quantum
network links \cite{monroe_scaling_2013,devoret_superconducting_2013}.
The main advantage of this approach is the ability to control and
debug small quantum modules as opposed to an entire large monolithic
quantum computer. Our approach is very well suited to this strategy.
In principle one can even envision a hierarchical application of the
quantum module concept (with error correction strategies applied to
multiple modules coupled together), but for that case our approach
would need to be extended (e.\,g.\ by using RL to find one- and
two-qubit gates acting on the logical qubits stored inside the modules).

\section*{Conclusions}

We have seen how a network can discover quantum error correction techniques
from scratch. It finds a-priori unknown nontrivial detection/recovery
sequences for diverse settings without any more input than the available
gate set. The trained neural networks can in principle be used to
control experimental quantum devices. The present approach is flexible
enough to be applied directly to a range of further, qualitatively
different physical situations, like non-Markovian noise, weak measurements,
qubit-cavity systems, and error-corrected transport of quantum information
through networks. An obvious challenge for the future is to successfully
discover strategies on even more qubits, where eventually full protection
against all noise sources and multiple logical qubits could be realized.
There is still considerable leeway in improving the speed of the physics
simulation and of GPU-based training (for further
details on the current computational effort, see \cref{sec:computational-resources}).

On the machine learning side, other RL schemes can be substituted
for the natural policy gradient adopted here, like Q-learning or advantage-actor-critic
techniques, or RL with continuous controls. Recurrent networks might
be employed to discover useful subsequences. The two-stage learning
approach introduced here could also be applied in other RL scenarios,
where one would first train based on expanded state information. In
general, we have shown that neural-network based RL promises to be
a flexible and general tool of wide-ranging applicability for exploring
feedback-based control of quantum and classical systems in physics.

\textbf{Acknowledgements} We thank Hugo Ribeiro and Vittorio Peano
for fruitful comments on the manuscript.

\textbf{Author Contributions} All authors contributed to the ideas,
their implementation, and the writing of the manuscript. The numerics
was performed by T.F., P.T. and T.W.

\textbf{Data availability} The data that support the plots within
this paper and other findings of this study are available from the
corresponding author on request.

\emph{Note added}: Shortly before submission of the present manuscript
to the arXiv, a preprint \cite{august_taking_2018} appeared exploring
RL with recurrent networks for optimal quantum control (without feedback).

\appendix


\section{Physical time evolution}

To track the time evolution for an \emph{arbitrary} initial logical
qubit state (identified by its Bloch vector $\vec{n}$), we start
from $\hat{\rho}_{\vec{n}}(0)=\frac{1}{2}(\mathbb{\mathbbmss{1}}+\vec{n}\hat{\vec{\sigma}})\otimes\hat{\rho}_{{\rm Rest}}$,
factorizing out the (fixed) state of all the other qubits. Now consider
the four quantities
\begin{subequations}
	\begin{align}
		\hat{\rho}_{0}(0) & =\frac{1}{2}\big(\hat{\rho}_{\vec{\mathrm{e}}_{j}}(0)+\hat{\rho}_{-\vec{\mathrm{e}}_{j}}(0)\big)\label{eq:initialization_rho0}\\
		\delta\hat{\rho}_{j}(0) & =\frac{1}{2}\big(\hat{\rho}_{\vec{\mathrm{e}}_{j}}(0)-\hat{\rho}_{-\vec{\mathrm{e}}_{j}}(0)\big)
	\end{align}
\end{subequations} where $j\in\{\mathrm{x},\mathrm{y},\mathrm{z}\}$
and $\vec{\mathrm{e}}_{j}$ are the basis vectors; note that the right-hand
side of \cref{eq:initialization_rho0} is independent of $j$. $\hat{\rho}_{0}$
and the $\delta\hat{\rho}_{j}$ are evolved stepwise, for each time-interval
$[t_{\mathrm{i}},t_{\mathrm{f}}]$ according to the update rule
\begin{subequations}
	\begin{align}
		\hat{\rho}_{0}(t_{\mathrm{f}}) & =\frac{\phi[\hat{\rho}_{0}(t_{\mathrm{i}})]}{\tr(\phi[\hat{\rho}_{0}(t_{\mathrm{i}})])}\\
		\delta\hat{\rho}_{j}(t_{\mathrm{f}}) & =\frac{\phi[\delta\hat{\rho}_{j}(t_{\mathrm{i}})]}{\tr(\phi[\hat{\rho}_{0}(t_{\mathrm{i}})])}~.
	\end{align}
\end{subequations}
In the absence of measurements, $\phi$ is the
completely positive map for the given time-interval. In the presence
of measurements, it is an unnormalized version (see below). We explicitly
renormalize such that always $\tr(\hat{\rho}_{0}(t))=1$. $\hat{\rho}_{0}$
and the $\delta\hat{\rho}_{j}$ give us access to the density matrix
for every logical qubit state, at any time $t$:
\begin{equation}
	\hat{\rho}_{\vec{n}}(t)=\frac{\hat{\rho}_{0}(t)+\sum_{j}n_{j}\,\delta\hat{\rho}_{j}(t)}{1+\tr(\sum_{j}n_{j}\,\delta\hat{\rho}_{j}(t))}\label{eq:methods_construction_rho_xyz}
\end{equation}

\section{Physical scenarios}

We always start from the initial condition that the logical qubit
is stored in one physical qubit, and the others are prepared in the
down state ($\ket{1}$). If explicit recovery is desired, we use this
original qubit also as the target qubit for final decoding. The time
evolution is divided into discrete time steps of uniform length $\Delta t$
(set to $1$ in the main text). At the start of each of these time
slices, we perform the measurement or gate operation (which is assumed
to be quasi-instantaneous) chosen by the agent; afterwards, the system
is subject to the dissipative dynamics. Thus, the map $\phi$ for
the time interval $[t,t+\Delta t]$ is of the form $\phi[\hat{\rho}]=\mathrm{e}^{\Delta t\mathcal{D}}(\hat{U}\hat{\rho}\hat{U}^{\dagger})$
for unitary operations $\hat{{U}}$ and $\phi[\hat{\rho}]=\mathrm{e}^{\Delta t\mathcal{D}}(\hat{P}_{m}\hat{\rho}\hat{P}_{m}^{\dagger})$
for projection operators $\hat{{P}}_{m}$ ($m$ indicates the measurement
results) where $\mathcal{{D}}$ is the dissipative part of the Liouvillian
(we only consider Markovian noise). Note that the measurement results
are chosen stochastically according to their respective probability
${\rm tr(\hat{P}_{m}\hat{\rho}_{0})}$. In the examples discussed
in the figures, we use two different error models, the bit-flip error
(bf) and the correlated noise error (cn):
\begin{subequations}
	\begin{align}
		\mathcal{D}_{{\rm bf}}\hat{\rho} & =T_{{\rm dec}}^{-1}\sum_{q}\hat{\sigma}_{x}^{(q)}\hat{\rho}\hat{\sigma}_{x}^{(q)}-\hat{\rho}\\
		\mathcal{D}_{{\rm cn}}\hat{\rho} & =T_{{\rm dec}}^{-1}\left(\hat{L}_{{\rm cn}}\hat{\rho}\hat{L}_{{\rm cn}}^{\dagger}-\frac{1}{2}\left\{ \hat{L}_{{\rm cn}}^{\dagger}\hat{L}_{{\rm cn}},\hat{\rho}\right\} \right)
	\end{align}
\end{subequations}where $\hat{\sigma}_{x,y,z}^{(q)}$ applies the
corresponding Pauli operator to the $q$th qubit and $\hat{L}_{{\rm cn}}=\frac{1}{\sqrt{\sum_{q}\mu_{q}^{2}}}\sum_{q}\mu_{q}\hat{\sigma}_{z}^{(q)}$.
Here, $\mu_{q}$ denotes the coupling of qubit $q$ to the noise.
Note that in the bit-flip scenario the single qubit decay time $T_{{\rm single}}=T_{{\rm dec}}$,
whereas in the presence of correlated noise it is $T_{{\rm single}}=T_{{\rm dec}}\left(\sum_{q}\mu_{q}^{2}\right)/\mu_{1}^{2}$.

\section{Recoverable quantum information}

Based on \cref{eq:methods_construction_rho_xyz}, $\mathcal{R}_{\mathrm{Q}}$
as introduced in the main text can be written as
\begin{equation}
	\mathcal{R}_{\mathrm{Q}}(t)=\min_{\vec{n}}\Big\Vert\sum_{j}n_{j}\,\delta\hat{\rho}_{j}(t)\Big\Vert_{1}
\end{equation}
in the (for us relevant) case that $\tr(\delta\hat{\rho}_{j}(t))=0$
for all $j$.

The trace distance $\frac{1}{2}\Vert\sum_{j}n_{j}\,\delta\hat{\rho}_{j}(t)\Vert_{1}$
has often a non-trivial dependence on the logical qubit state $\vec{n}$
and finding its minimum can become nontrivial. However, the location
of the minimum can sometimes be ``guessed'' in advance. For any
CHZ quantum circuit \cite{nielsen_quantum_2011}, i.\,e.\ for all
the bit-flip examples considered here, the anti-commutator relation
$\{\delta\hat{\rho}_{j},\delta\hat{\rho}_{k}\}=0$ is satisfied for
all distinct $j\neq k$; it can be shown that this restricts the minimum
to lie along one of the coordinate axes: $\min_{\vec{n}}\Vert\sum_{j}n_{j}\,\delta\hat{\rho}_{j}(t)\Vert_{1}=\min_{j\in\{\mathrm{x},\mathrm{y},\mathrm{z}\}}\Vert\delta\hat{\rho}_{j}(t)\Vert_{1}$.
For the correlated noise, the trace distance $\frac{1}{2}\Vert\sum_{j}n_{j}\,\delta\hat{\rho}_{j}(t)\Vert_{1}$
is symmetric around the $z$-axis and takes its minimal value at the
equator.

After a measurement, the updated value of $\mathcal{R}_{\mathrm{Q}}$
may vary between the different measurement results. To obtain a measure
that does not depend on this, we introduce $\bar{\mathcal{R}}_{\mathrm{Q}}$
as the average over all possible values of $\mathcal{R}_{\mathrm{Q}}$
(after a single time step), weighted by the probability to end up
in the corresponding branch. If the action is not a measurement, there
is only one option and thus $\bar{\mathcal{R}}_{\mathrm{Q}}=\mathcal{R}_{\mathrm{Q}}$.

\section{Protection reward}

The goal of the ``protection reward'' is to maximize $\mathcal{R}_{\mathrm{Q}}$
at the end of the simulation, i.\,e.\ the ability to \textit{in
principle} recover the target state. A suitable (immediate) reward
is given by
\begin{equation}
	r_t =
	\left\{\begin{array}{c@{\hspace*{-0.5em}}cl}
		1+\frac{\mathcal{R}_\mathrm{Q}(t+1)-\mathcal{R}_\mathrm{Q}(t)}{2\Delta t/T_\mathrm{single}} &
			+\,0 &
			\text{if $\bar{\mathcal{R}}_\mathrm{Q}(t+1)>0$} \\
		\multirow{2}{*}{0} &
			\multirow{2}{*}{-\,P} &
			\text{if $\mathcal{R}_\mathrm{Q}(t)\neq0$} \\
		& & \text{$\wedge\mathcal{R}_\mathrm{Q}(t+1)=0$} \\
		\smash{\underbrace{\hspace*{3.5em}0\hspace*{3.5em}}_{\textstyle=:r_t^{(1)}}} &
			\smash{\underbrace{+\,0}_{\textstyle=:r_t^{(2)}}} &
			\text{if $\mathcal{R}_\mathrm{Q}(t)=0$}
	\end{array}\right.
	\vspace*{1.25\baselineskip}
\end{equation}
with $\mathcal{R}_{\mathrm{Q}}$ and $\bar{\mathcal{R}}_{\mathrm{Q}}$
as defined above, $T_{{\rm single}}$ the decay time for encoding
in one physical qubit only (``trivial'' encoding), $\Delta t$ the
time step, and $P$ a punishment for measurements which reveal the
logical qubit state. Based on this reward, we choose the return (the
function of the reward sequence used to compute the policy gradient)
as
\begin{equation}
	R_{t}=(1-\gamma)\Big(\sum_{k=0}^{T-t-1}\gamma^{k}r_{t+k}^{(1)}\Big)+r_{t}^{(2)}\label{eq:protection_return}
\end{equation}
where $\gamma$ is the return discount rate; for more information
on the (discounted) return, see e.\,g.\ \cite{sutton_reinforcement_1998}.

\section{Recovery reward}

The protection reward does not encourage the network to finally decode
the quantum state. If this behavior is desired, we add suitable terms
to the reward (only employed for \cref{fig:LSTMresults}):
\begin{equation}
	r_{t}^{\mathrm{(recov)}}=\beta_{\mathrm{dec}}\,(D_{t+1}-D_{t})+\beta_{\mathrm{corr}}\,C_{T}\,\delta_{t,T-1}
\end{equation}
where $D_{t}=\frac{1}{2}(I_{t}^{(q')}-\sum_{q\neq q'}I_{t}^{(q)})$,
unless $t\le T_{\mathrm{signal}}$ where we set $D_{t}=0$. This means
decoding is only rewarded after $T_{\text{signal}}$. We set $I_{t}^{(q)}=1$
if $\tr_{\bar{q}}(\delta\hat{\rho}_{{\rm j}}(t))\neq0$ for any $j$,
and otherwise $I_{t}^{(q)}=-1$. $\tr_{\bar{q}}$ denotes the partial
trace over all qubits except $q$, and $q'$ labels the target qubit.
The condition $I_{t}^{(q)}=1$ implies that the logical qubit state
is encoded in the specific qubit $q$ (this is not a necessary criterion).
$C_{T}$ is $1$ if (at the final time $T$) the logical qubit state
is encoded in the target qubit only and this qubit has the prescribed
polarization (i.\,e.\ not flipped), and otherwise $0$.

As return, we use
\begin{equation}
	R_{t}^{\mathrm{(recov)}}=(1-\gamma)\sum_{k=0}^{T-t-1}\gamma^{k}r_{t+k}^{\mathrm{(recov)}}\label{eq:recovery_return}
\end{equation}
with the same return discount rate $\gamma$ as for the protection
reward.

With this reward, we aim to optimize the minimum overlap $\mathcal{O}_{Q}=\min_{\vec{n}}\bra{\phi_{\vec{n}}}\tr_{{\rm \bar{q}'}}\left(\hat{\rho}_{\vec{n}}\right)\ket{\phi_{\vec{n}}}$
between the (pure) target state $\ket\phi_{\vec{n}}$ and the actual
final state $\hat{\rho}_{\vec{n}}$ reduced to the target qubit, given
by the partial trace $\tr_{{\rm \bar{q}'}}\left(\hat{\rho}_{\vec{n}}\right)$
over all other qubits.

\section{Input of the state-aware network}

The core of the input to the state-aware network is a representation
of the density matrices $\hat{\rho}_{0}$, $\hat{\rho}_{1}:=\hat{\rho}_{0}+\delta\hat{\rho}_{\mathrm{x}}$,
$\hat{\rho}_{2}:=\hat{\rho}_{0}+\delta\hat{\rho}_{\mathrm{y}}$, and
$\hat{\rho}_{3}:=\hat{\rho}_{0}+\delta\hat{\rho}_{\mathrm{z}}$. Together,
they represent the completely positive map of the evolution (for arbitrary
logical qubit states). For reduction of the input size (especially
in view of higher qubit numbers), we compress them via principal component
analysis (PCA), i.\,e.\ we perform an eigendecomposition $\hat{\rho}_{j}=\sum_{k}p_{k}\ket{\phi_{k}}\bra{\phi_{k}}$
and select the eigenstates $\ket{\phi_{k}}$ with the largest eigenvalues
$p_{k}$. To include also the eigenvalue in the input, we feed all
components of the scaled states $\ket{\tilde{\phi}_{k}}:=\sqrt{p_{k}}\ket{\phi_{k}}$
(which yield $\hat{\rho}_{j}=\sum_{k}\ket{\tilde{\phi}_{k}}\bra{\tilde{\phi}_{k}}$)
into the network, where the states are in addition sorted by their
eigenvalue. For our simulations, we select the 6 largest components,
so we need $768=4\cdot6\cdot16\cdot2$ input neurons (4 density matrices,
16 is the dimension of the Hilbert space, 2 for real and imaginary
part).

In addition, at each time step we indicate to the network whether
a potential measurement would destroy the quantum state by revealing
the quantum information. Explicitly, we compute for each measurement
whether $\tr(\hat{P}\delta\hat{\rho}_{j})=0$ for all $j\in\{\mathrm{x},\mathrm{y},\mathrm{z}\}$
and every possible projector $\hat{P}$ (i.\,e.\ every possible
measurement result), and feed these boolean values into the network.
Note that this information can be deduced from the density matrix
(so in principle the network could learn that deduction on its own,
but giving it directly speeds up training).

Because all relevant information for the decision about the next action
is contained in the current density matrix, knowledge about the previous
actions is not needed. However, we have found that providing this
extra information is helpful to accelerate learning. Therefore, we
provide also the last action (in a one-hot encoding). We note that
it is not necessary to feed the latest measurement result to the network,
since the updated density matrix is conditional on the measurement
outcome and therefore contains all relevant information for future
decision.

To train the state-aware network to restore the original state at
the end of a trajectory, it becomes necessary to add the time to the
input. It is fully sufficient to indicate the last few time steps
where $t>T_{{\rm signal}}$ (when decoding should be performed) in
a one-hot encoding.

\section{Layout of the state-aware network}

Our state-aware networks have a feedforward architecture. Between
the input layer and the output layer (one neuron per action), there
are two or three hidden layers (the specific numbers are summarized
in the last section of the Appendix). All neighboring layers are densely
connected, the activation function is the rectified linear unit (ReLU).
At the output layer, the softmax function $x_{j}\mapsto\mathrm{e}^{x_{j}}/\sum_{k}\mathrm{e}^{x_{k}}$
is applied such that the result can be interpreted as a probability
distribution.

\section{Reinforcement learning of the state-aware network}

Our learning scheme is based on the policy gradient algorithm \cite{williams_simple_1992}.
The full expression for our learning gradient (indicating the change
in $\theta$) reads

\begin{equation}
	\begin{aligned}
		g = & \lambda_{\mathrm{pol}}F^{-1}\cdot\mathbb{E}\Big[\sum_{t}(R_{t}-b_{t})\frac{\partial}{\partial\theta}\ln\pi_{\theta}(a_{t}|s_{t})\Big]\\
		& -\lambda_{\mathrm{entr}}\,\mathbb{E}\left[\sum_{a}\frac{\partial}{\partial\theta}\big[\pi_{\theta}(a|s)\ln\pi_{\theta}(a|s)\big]\right]
	\end{aligned}
\end{equation}

where $R_{t}$ is the (discounted) return (cmp.\ \cref{eq:protection_return,eq:recovery_return}).
This return is corrected by an (explicitly time-dependent) baseline
$b_{t}$ which we choose as exponentially decaying average of $R_{t}$,
i.\,e.\ for the training update in epoch $N$, we use $b_{t}=(1-\kappa)\sum_{n=0}^{N-1}\kappa^{n}\bar{R}_{t}^{(N-1-n)}$
where $\kappa$ is the baseline discount rate and $\bar{R}_{t}^{(n)}$
is the mean return at time step $t$ in epoch $n$. We compute the
natural gradient \cite{amari_natural_1998,kakade_natural_2002,peters_natural_2008}
by multiplying $F^{-1}$, the (Moore–Penrose) inverse of the Fisher
information matrix $F=\mathbb{E}[(\frac{\partial}{\partial\theta}\ln\pi_{\theta}(a|s))(\frac{\partial}{\partial\theta}\ln\pi_{\theta}(a|s))^{\mathsf{T}}]$.
The second term is entropy regularization \cite{williams_function_1991};
we use it only to train the state-aware network shown in \cref{fig:LSTMresults}.
As update rule, we use adaptive moment estimation (Adam \cite{kingma_method_2015})
without bias correction.

\section{Layout of the recurrent network}

The recurrent network is designed such that it can in principle operate
in a real-world experiment. This means in particular that (in contrast
to the state-aware network) its input must not contain directly the
quantum state (or the evolution map); instead, measurements are its
only way to obtain information about the quantum system. Hence, the
input to the recurrent network contains the present measurement result
(and additionally the previous action). Explicitly, we choose the
input as a one-hot encoding for the action in the last time step,
and in case of measurements, we additionally distinguish between the
different results. In addition, there is an extra input neuron to
indicate the beginning of time (where no previous action was performed).
Since this input contains only the most recent ``event'', the network
requires a memory to perform reasonable strategies, i.\,e.\ we need
a recurrent network. Therefore, the input and output layer are connected
by two successive long short-term memory (LSTM) layers \cite{hochreiter_long_1997}
with $\tanh$ inter-layer activations. After the output layer, the
softmax function is applied (like for the state-aware network).

\section{Supervised learning of the recurrent network}

The training data is generated from inference of a state-aware network
which has been trained to sufficiently good strategies (via reinforcement
learning); for every time step in each trajectory, we save the network
input and the policy, i.\,e.\ the probabilities for all the actions,
and we train on this data. It is possible to generate enough data
such that overfitting is not a concern (for the example in \cref{fig:LSTMresults},
each trajectory is reused only 5 times during the full training process).
For the actual training of the recurrent network, we use supervised
learning with categorical cross-entropy as cost function ($q$ is
the actual policy of the recurrent network to train, and $p$ the
desired policy from the state-aware network):
\begin{equation}
	C(q,p)=-\sum_{a}p(a)\ln q(a)
\end{equation}
Due to the LSTM layers, it is necessary to train on full trajectories
(in the true time sequence) instead of individual actions. Dropout
\cite{hinton_improving_2012} is used for regularization. The training
update rule is adaptive moment estimation (Adam \cite{kingma_method_2015}).

\section{Physical parameters and hyperparameters}

\global\long\def\figLearningProgress{{\rm {Fig.\,3a}}}
\global\long\def\figLearningGateSeq{{\rm {Fig.\,3b}}}
\global\long\def\figRecovQExamples{{\rm {Fig.\,3c}}}
\global\long\def\figBranching{{\rm {Fig.\,3d}}}
\global\long\def\figTSNEcircle{{\rm {Fig.\,4a}}}
\global\long\def\figTSNEmap{{\rm {Fig.\,4b}}}
\global\long\def\figTSNEzoomin{{\rm {Fig.\,4c}}}
\global\long\def\figConnectSeq{{\rm {Fig.\,5a}}}
\global\long\def\figConnectBarplot{{\rm {Fig.\,5b}}}
\global\long\def\figMsmtErrors{{\rm {Fig.\,5c}}}
\global\long\def\figStateAwareCorrectedValidation{{\rm {Fig.\,7a}}}
\global\long\def\figStateBlindValidation{{\rm {Fig.\,7b}}}
\global\long\def\figLSTMneuronActivation{{\rm {Fig.\,7c}}}
\global\long\def\figStateAwareTraining{{\rm {Fig.\,3}}}
\global\long\def\figTSNE{{\rm {Fig.\,4}}}
\global\long\def\figCorrelatedNoise{{\rm {Fig.\,6}}}
\global\long\def\figCorrNoiseTwo{{\rm {Fig.\,6b}}}
\global\long\def\figCorrNoiseBarplot{{\rm {Fig.\,6c}}}
\global\long\def\figCorrNoiseTrajs{{\rm {Fig.\,6d}}}
\global\long\def\figBitFlip{{\rm {Figs.\,3,4,5,7}}}

The physical parameters used throughout the main text are summarized
in the following table. Times are always given in units of the time
step (gate time).

\begin{tabular}{>{\raggedright}p{0.7\columnwidth}>{\raggedright}p{0.2\columnwidth}}
	\textbf{Physical parameters} & \tabularnewline
	\hline \hline
	$\figBitFlip$ decoherence time $T_{{\rm dec}}$ & $1200$\tabularnewline
	$\figCorrelatedNoise$ single qubit decoherence time $T_{{\rm single}}=T_{{\rm dec}}/\mu_{1}^{2}$ & $500$\tabularnewline
	$\figBitFlip$ number of time steps $T$ & $200$\tabularnewline
	$\figCorrelatedNoise$ number of time steps $T$ & $100$\tabularnewline
	\hline
\end{tabular}

We have used a few separately trained neural network throughout this
work which differ slightly in hyperparameters (e.\,g.\ in the number
of hidden layers and neurons per layer). This is not due to fine-tuning,
and in the Supplementary Information we demonstrate that we can successfully
train the neural networks in all scenarios with one common set of
hyperparameters. The strategies found by the neural networks are not
influenced by using different sets of hyperparameters. Different hyperparameters
may influence the training time etc. In the following table, we summarize
the architecture of the networks, i.\,e., we list the number of neurons
in each layer.

\begin{tabular}{>{\raggedright}p{0.5\columnwidth}>{\raggedright}p{0.45\columnwidth}}
	\textbf{Network architectures} & \textbf{Neurons per layer}\tabularnewline
	\hline \hline
	$\figStateAwareTraining$ & $(793,\,300,\,300,\,21)$\tabularnewline
	$\figTSNE$ & $(793,\,300,\,300,\,300,\,21)$\tabularnewline
	$\figConnectSeq$-$\figConnectBarplot$ all-to-all connected: & $(793,\,300,\,300,\,21)$\tabularnewline
	$\figConnectSeq$-$\figConnectBarplot$ chain: & $(787,\,300,\,300,\,300,\,15)$\tabularnewline
	$\figConnectSeq$-$\figConnectBarplot$ chain with one msmt qubit: & $(781,\,600,\,300,\,12)$\tabularnewline
	$\figConnectSeq$-$\figConnectBarplot$ circle: & $(783,\,300,\,300,\,300,\,14)$\tabularnewline
	$\figMsmtErrors$ (all identical) & $(793,\,300,\,300,\,300,\,21)$\tabularnewline
	$\figCorrNoiseTwo$-$\figCorrNoiseTrajs$: one ancilla & $(101,\,100,\,50,\,3)$\tabularnewline
	$\figCorrNoiseBarplot$, $\figCorrNoiseTrajs$: two ancillas & $(265,\,100,\,50,\,5)$\tabularnewline
	$\figCorrNoiseBarplot$: three ancillas & $(781,\,200,\,100,\,7)$\tabularnewline
	$\figStateAwareCorrectedValidation$ & $(803,\,300,\,300,\,21)$\tabularnewline
	$\figStateBlindValidation$-$\figLSTMneuronActivation$ (hidden layers
	with LSTM units) & $(26,\,128,\,128,\,21)$\tabularnewline
	\hline
\end{tabular}

Each output neuron represents one action that can be performed by
the agent and thus, the output layer size is equal to the number of
actions. In the bit-flip scenarios ($\figBitFlip$), the actions are
CNOTs according to connectivity, measurements along $z$ as indicated
in the corresponding sketches, deterministic bit flips on each qubit,
and idle. When dealing with correlated noise, cf.\ $\figCorrelatedNoise$,
the available actions are instead measurements along $x$ and $y$
on all ancilla qubits, and the idle operation. Note that our whole
approach is general and able in principle to deal with arbitrary quantum
gates.

The hyperparameters used for training the state-aware networks are
summarized in the following table:\linebreak{}

\begin{threeparttable}[!h]
	\begin{longtable}{>{\raggedright}p{0.55\columnwidth}>{\raggedright}p{0.45\columnwidth}}
		\textbf{Hyperparameters state-aware network} & \tabularnewline
		\hline \hline
		training batch size & $64$\tabularnewline
		Adam parameters $\eta,\,\beta_{1},\,\beta_{2}$ & $0.0003$\tnote{a}, $0.9$, $0.999$ (no bias correction)\tabularnewline
		return discount rate $\gamma$ & $0.95$\tabularnewline
		baseline discount rate $\kappa$ & $0.9$\tabularnewline
		punishment reward coefficient $P$ & $0.1$ \tabularnewline
		decoding reward coefficient $\beta_{{\rm dec}}$(only used in $\figStateAwareCorrectedValidation$) & $20$\tabularnewline
		correction reward coefficient $\beta_{{\rm corr}}$(only used in $\figStateAwareCorrectedValidation$) & $10$\tabularnewline
		decoding signal time $T_{\mathrm{signal}}$ (only used in $\figStateAwareCorrectedValidation$) & $T_{{\rm signal}}=T-10=190$\tabularnewline
		reward scale $\lambda_\mathrm{pol}$ & $4.0$ \tabularnewline
		entropy regularization $\lambda_\mathrm{entr}$(only used in $\figStateAwareCorrectedValidation$) & $\lambda_\mathrm{entr}=5\cdot10^{-3}$ for the first $12\,000$ training epoches, $\lambda_\mathrm{entr}=0$ afterwards\tabularnewline
		\hline
	\end{longtable}
	\begin{tablenotes}
		\item[a] The exact value for the learning rate used in the simulations is in fact $0.0001\sqrt{10}$; the irrational factor of $\sqrt{10}$ is caused by a slight deviation between our implementation and the standard Adam scheme which in the end resulted only in a redefinition of the learning rate.
	\end{tablenotes}
\end{threeparttable}
\vspace*{0.5\baselineskip}

The hyperparameters used for training the recurrent network (cf.\ \cref{fig:LSTMresults}b
and \cref{fig:LSTMresults}c) are summarized in the following
table:\linebreak{}
\footnotetext{The exact value for the learning rate used in the simulation is in fact $0.0001\sqrt{10}$; the irrational factor of $\sqrt{10}$ is caused by a slight deviation between our implementation and the standard Adam scheme which in the end resulted only in a redefinition of the learning rate.}

\begin{tabular}{>{\raggedright}p{0.5\columnwidth}>{\raggedright}p{0.45\columnwidth}}
	\textbf{Hyperparameters recurrent network} & \tabularnewline
	\hline
	\hline
	training batch size & $16$\tabularnewline
	Adam parameters $\eta,\,\beta_{1},\,\beta_{2}$ & $0.001,\,0.9,\,0.999$\tabularnewline
	dropout level (after each LSTM layer) & $0.5$\tabularnewline
	\hline
\end{tabular}\linebreak{}

Our RL implementation relies on the Theano framework \cite{al-rfou_et_al._theano:_2016}
(and Keras for defining networks).

\section{Computational resources\label{sec:computational-resources}}

The computationally most expensive tasks in this
paper are the training runs of the state-aware networks depicted in
\cref{fig:TrainingOverviewFigure,fig:ManyScenarios,fig:LSTMresults}a.
Full training for a given scenario can be achieved using 1.6 million
training trajectories, which can be run within 6 hours on a single
CPU+GPU node (CPU: Intel Xeon E5-1630 v4, GPU: Nvidia Quadro P5000).
Currently, more than 2/3 of the time are spent on the numerical simulation
of the physics time evolution, which is still performed on the CPU.
We expect that the total runtime can be improved significantly by
a more efficient implementation and more powerful hardware (including
also an implementation of the physics simulation on GPU). The memory
consumption is very modest (for these examples below 200\,MB),
dominated mostly by the need for storing the input to the network
for all trajectories inside a batch (here, 64 trajectories with 200
time steps each) and less by the network weights (here, up to about
600,000). For a more detailed discussion, we refer the reader to the
supplementary material (sec.\ 6).

\bibliographystyle{unsrt}
\bibliography{refs_maintext}

\begin{thebibliography}{10}

\bibitem{lecun_deep_2015}
Yann LeCun, Yoshua Bengio, and Geoffrey Hinton.
\newblock Deep learning.
\newblock {\em Nature}, 521(7553):436--444, May 2015.

\bibitem{russell_artificial_2016}
Stuart Russell and Peter Norvig.
\newblock {\em Artificial {Intelligence}: {A} {Modern} {Approach}}.
\newblock CreateSpace Independent Publishing Platform, September 2016.
\newblock Google-Books-ID: PQI7vgAACAAJ.

\bibitem{carleo_solving_2017}
Giuseppe Carleo and Matthias Troyer.
\newblock Solving the quantum many-body problem with artificial neural
  networks.
\newblock {\em Science}, 355(6325):602--606, February 2017.

\bibitem{carrasquilla_machine_2017}
Juan Carrasquilla and Roger~G. Melko.
\newblock Machine learning phases of matter.
\newblock {\em Nat Phys}, 13(5):431--434, May 2017.

\bibitem{van_nieuwenburg_learning_2017}
Evert P.~L. van Nieuwenburg, Ye-Hua Liu, and Sebastian~D. Huber.
\newblock Learning phase transitions by confusion.
\newblock {\em Nature Physics}, 13(5):435--439, May 2017.

\bibitem{torlai_neural_2017}
Giacomo Torlai and Roger~G. Melko.
\newblock Neural {Decoder} for {Topological} {Codes}.
\newblock {\em Physical Review Letters}, 119(3):030501, July 2017.

\bibitem{august_using_2017}
Moritz August and Xiaotong Ni.
\newblock Using {Recurrent} {Neural} {Networks} to {Optimize} {Dynamical}
  {Decoupling} for {Quantum} {Memory}.
\newblock {\em Physical Review A}, 95(1), January 2017.
\newblock arXiv: 1604.00279.

\bibitem{dunjko_machine_2017}
Vedran Dunjko and Hans~J. Briegel.
\newblock Machine learning and artificial intelligence in the quantum domain.
\newblock {\em arXiv:1709.02779 [quant-ph]}, September 2017.
\newblock arXiv: 1709.02779.

\bibitem{baireuther_machine-learning-assisted_2017}
P.~Baireuther, T.~E. O'Brien, B.~Tarasinski, and C.~W.~J. Beenakker.
\newblock Machine-learning-assisted correction of correlated qubit errors in a
  topological code.
\newblock {\em arXiv:1705.07855 [cond-mat, physics:quant-ph]}, May 2017.
\newblock arXiv: 1705.07855.

\bibitem{krastanov_deep_2017}
Stefan Krastanov and Liang Jiang.
\newblock Deep {Neural} {Network} {Probabilistic} {Decoder} for {Stabilizer}
  {Codes}.
\newblock {\em Scientific Reports}, 7(1):11003, September 2017.

\bibitem{mehta_high-bias_2018}
Pankaj Mehta, Marin Bukov, Ching-Hao Wang, Alexandre G.~R. Day, Clint
  Richardson, Charles~K. Fisher, and David~J. Schwab.
\newblock A high-bias, low-variance introduction to {Machine} {Learning} for
  physicists.
\newblock {\em arXiv:1803.08823 [cond-mat, physics:physics, stat]}, March 2018.
\newblock arXiv: 1803.08823.

\bibitem{king_automation_2009}
Ross~D. King, Jem Rowland, Stephen~G. Oliver, Michael Young, Wayne Aubrey, Emma
  Byrne, Maria Liakata, Magdalena Markham, Pinar Pir, Larisa~N. Soldatova,
  Andrew Sparkes, Kenneth~E. Whelan, and Amanda Clare.
\newblock The {Automation} of {Science}.
\newblock {\em Science}, 324(5923):85--89, April 2009.

\bibitem{schmidt_distilling_2009}
Michael Schmidt and Hod Lipson.
\newblock Distilling {Free}-{Form} {Natural} {Laws} from {Experimental} {Data}.
\newblock {\em Science}, 324(5923):81--85, April 2009.

\bibitem{mnih_human-level_2015}
Volodymyr Mnih, Koray Kavukcuoglu, David Silver, Andrei~A. Rusu, Joel Veness,
  Marc~G. Bellemare, Alex Graves, Martin Riedmiller, Andreas~K. Fidjeland,
  Georg Ostrovski, Stig Petersen, Charles Beattie, Amir Sadik, Ioannis
  Antonoglou, Helen King, Dharshan Kumaran, Daan Wierstra, Shane Legg, and
  Demis Hassabis.
\newblock Human-level control through deep reinforcement learning.
\newblock {\em Nature}, 518(7540):529--533, February 2015.

\bibitem{silver_mastering_2017}
David Silver, Julian Schrittwieser, Karen Simonyan, Ioannis Antonoglou, Aja
  Huang, Arthur Guez, Thomas Hubert, Lucas Baker, Matthew Lai, Adrian Bolton,
  Yutian Chen, Timothy Lillicrap, Fan Hui, Laurent Sifre, George van~den
  Driessche, Thore Graepel, and Demis Hassabis.
\newblock Mastering the game of {Go} without human knowledge.
\newblock {\em Nature}, 550(7676):354, October 2017.

\bibitem{chen_fidelity-based_2014}
C.~Chen, D.~Dong, H.~X. Li, J.~Chu, and T.~J. Tarn.
\newblock Fidelity-{Based} {Probabilistic} {Q}-{Learning} for {Control} of
  {Quantum} {Systems}.
\newblock {\em IEEE Transactions on Neural Networks and Learning Systems},
  25(5):920--933, May 2014.

\bibitem{bukov_machine_2017}
Marin Bukov, Alexandre G.~R. Day, Dries Sels, Phillip Weinberg, Anatoli
  Polkovnikov, and Pankaj Mehta.
\newblock Machine {Learning} {Meets} {Quantum} {State} {Preparation}. {The}
  {Phase} {Diagram} of {Quantum} {Control}.
\newblock {\em arXiv:1705.00565 [cond-mat, physics:quant-ph]}, May 2017.
\newblock arXiv: 1705.00565.

\bibitem{melnikov_active_2018}
Alexey~A. Melnikov, Hendrik~Poulsen Nautrup, Mario Krenn, Vedran Dunjko, Markus
  Tiersch, Anton Zeilinger, and Hans~J. Briegel.
\newblock Active learning machine learns to create new quantum experiments.
\newblock {\em Proceedings of the National Academy of Sciences}, page
  201714936, January 2018.

\bibitem{monroe_scaling_2013}
C.~Monroe and J.~Kim.
\newblock Scaling the {Ion} {Trap} {Quantum} {Processor}.
\newblock {\em Science}, 339(6124):1164--1169, March 2013.

\bibitem{devoret_superconducting_2013}
M.~H. Devoret and R.~J. Schoelkopf.
\newblock Superconducting {Circuits} for {Quantum} {Information}: {An}
  {Outlook}.
\newblock {\em Science}, 339(6124):1169--1174, March 2013.

\bibitem{chiaverini_realization_2004}
J.~Chiaverini, D.~Leibfried, T.~Schaetz, M.~D. Barrett, R.~B. Blakestad,
  J.~Britton, W.~M. Itano, J.~D. Jost, E.~Knill, C.~Langer, R.~Ozeri, and D.~J.
  Wineland.
\newblock Realization of quantum error correction.
\newblock {\em Nature}, 432(7017):602, December 2004.

\bibitem{schindler_experimental_2011}
Philipp Schindler, Julio~T. Barreiro, Thomas Monz, Volckmar Nebendahl, Daniel
  Nigg, Michael Chwalla, Markus Hennrich, and Rainer Blatt.
\newblock Experimental {Repetitive} {Quantum} {Error} {Correction}.
\newblock {\em Science}, 332(6033):1059--1061, May 2011.

\bibitem{waldherr_quantum_2014}
G.~Waldherr, Y.~Wang, S.~Zaiser, M.~Jamali, T.~Schulte-Herbr{\"u}ggen, H.~Abe,
  T.~Ohshima, J.~Isoya, J.~F. Du, P.~Neumann, and J.~Wrachtrup.
\newblock Quantum error correction in a solid-state hybrid spin register.
\newblock {\em Nature}, 506(7487):204, February 2014.

\bibitem{kelly_state_2015}
J.~Kelly, R.~Barends, A.~G. Fowler, A.~Megrant, E.~Jeffrey, T.~C. White,
  D.~Sank, J.~Y. Mutus, B.~Campbell, Yu~Chen, Z.~Chen, B.~Chiaro, A.~Dunsworth,
  I.-C. Hoi, C.~Neill, P.~J.~J. O'Malley, C.~Quintana, P.~Roushan,
  A.~Vainsencher, J.~Wenner, A.~N. Cleland, and John~M. Martinis.
\newblock State preservation by repetitive error detection in a superconducting
  quantum circuit.
\newblock {\em Nature}, 519(7541):66, March 2015.

\bibitem{veldhorst_two-qubit_2015}
M.~Veldhorst, C.~H. Yang, J.~C.~C. Hwang, W.~Huang, J.~P. Dehollain, J.~T.
  Muhonen, S.~Simmons, A.~Laucht, F.~E. Hudson, K.~M. Itoh, A.~Morello, and
  A.~S. Dzurak.
\newblock A two-qubit logic gate in silicon.
\newblock {\em Nature}, 526(7573):410--414, October 2015.

\bibitem{ofek_extending_2016}
Nissim Ofek, Andrei Petrenko, Reinier Heeres, Philip Reinhold, Zaki Leghtas,
  Brian Vlastakis, Yehan Liu, Luigi Frunzio, S.~M. Girvin, L.~Jiang, Mazyar
  Mirrahimi, M.~H. Devoret, and R.~J. Schoelkopf.
\newblock Extending the lifetime of a quantum bit with error correction in
  superconducting circuits.
\newblock {\em Nature}, 536(7617):441, July 2016.

\bibitem{potocnik_studying_2017}
Anton Potocnik, Arno Bargerbos, Florian A. Y.~N. Schr{\"o}der, Saeed~A. Khan,
  Michele~C. Collodo, Simone Gasparinetti, Yves Salath{\'e}, Celestino
  Creatore, Christopher Eichler, Hakan~E. T{\"u}reci, Alex~W. Chin, and Andreas
  Wallraff.
\newblock Studying {Light}-{Harvesting} {Models} with {Superconducting}
  {Circuits}.
\newblock {\em arXiv:1710.07466 [physics, physics:quant-ph]}, October 2017.
\newblock arXiv: 1710.07466.

\bibitem{debnath_demonstration_2016}
S.~Debnath, N.~M. Linke, C.~Figgatt, K.~A. Landsman, K.~Wright, and C.~Monroe.
\newblock Demonstration of a small programmable quantum computer with atomic
  qubits.
\newblock {\em Nature}, 536(7614):63--66, August 2016.

\bibitem{takita_experimental_2017}
Maika Takita, Andrew~W. Cross, A.~D. C{\'o}rcoles, Jerry~M. Chow, and Jay~M.
  Gambetta.
\newblock Experimental {Demonstration} of {Fault}-{Tolerant} {State}
  {Preparation} with {Superconducting} {Qubits}.
\newblock {\em Physical Review Letters}, 119(18):180501, October 2017.

\bibitem{watson_programmable_2017}
T.~F. Watson, S.~G.~J. Philips, E.~Kawakami, D.~R. Ward, P.~Scarlino,
  M.~Veldhorst, D.~E. Savage, M.~G. Lagally, Mark Friesen, S.~N. Coppersmith,
  M.~A. Eriksson, and L.~M.~K. Vandersypen.
\newblock A programmable two-qubit quantum processor in silicon.
\newblock {\em arXiv:1708.04214 [cond-mat, physics:quant-ph]}, August 2017.
\newblock arXiv: 1708.04214.

\bibitem{riste_feedback_2012}
D.~Rist{\`e}, C.~C. Bultink, K.~W. Lehnert, and L.~DiCarlo.
\newblock Feedback {Control} of a {Solid}-{State} {Qubit} {Using}
  {High}-{Fidelity} {Projective} {Measurement}.
\newblock {\em Physical Review Letters}, 109(24):240502, December 2012.

\bibitem{campagne-ibarcq_persistent_2013}
P.~Campagne-Ibarcq, E.~Flurin, N.~Roch, D.~Darson, P.~Morfin, M.~Mirrahimi,
  M.~H. Devoret, F.~Mallet, and B.~Huard.
\newblock Persistent {Control} of a {Superconducting} {Qubit} by {Stroboscopic}
  {Measurement} {Feedback}.
\newblock {\em Physical Review X}, 3(2):021008, May 2013.

\bibitem{steffen_deterministic_2013}
L.~Steffen, Y.~Salathe, M.~Oppliger, P.~Kurpiers, M.~Baur, C.~Lang, C.~Eichler,
  G.~Puebla-Hellmann, A.~Fedorov, and A.~Wallraff.
\newblock Deterministic quantum teleportation with feed-forward in a solid
  state system.
\newblock {\em Nature}, 500(7462):319--322, August 2013.

\bibitem{pfaff_unconditional_2014}
W.~Pfaff, B.~J. Hensen, H.~Bernien, S.~B.~van Dam, M.~S. Blok, T.~H. Taminiau,
  M.~J. Tiggelman, R.~N. Schouten, M.~Markham, D.~J. Twitchen, and R.~Hanson.
\newblock Unconditional quantum teleportation between distant solid-state
  quantum bits.
\newblock {\em Science}, 345(6196):532--535, August 2014.

\bibitem{riste_digital_2016}
D.~Rist{\`e} and L.~DiCarlo.
\newblock Digital feedback in superconducting quantum circuits.
\newblock In {\em Superconducting {Devices} in {Quantum} {Optics}; eds.
  {R}.{H}. {Hadfield} and {G}. {Johansson}; see {arXiv}:1508.01385}. Springer,
  February 2016.
\newblock arXiv: 1508.01385.

\bibitem{khaneja_optimal_2005}
Navin Khaneja, Timo Reiss, Cindie Kehlet, Thomas Schulte-Herbr{\"u}ggen, and
  Steffen~J. Glaser.
\newblock Optimal control of coupled spin dynamics: design of {NMR} pulse
  sequences by gradient ascent algorithms.
\newblock {\em Journal of Magnetic Resonance}, 172(2):296--305, February 2005.

\bibitem{machnes_comparing_2011}
S.~Machnes, U.~Sander, S.~J. Glaser, P.~de~Fouqui{\`e}res, A.~Gruslys,
  S.~Schirmer, and T.~Schulte-Herbr{\"u}ggen.
\newblock Comparing, optimizing, and benchmarking quantum-control algorithms in
  a unifying programming framework.
\newblock {\em Physical Review A}, 84(2):022305, August 2011.

\bibitem{johnson_qvector:_2017}
Peter~D. Johnson, Jonathan Romero, Jonathan Olson, Yudong Cao, and Al{\'a}n
  Aspuru-Guzik.
\newblock {QVECTOR}: an algorithm for device-tailored quantum error correction.
\newblock {\em arXiv:1711.02249 [quant-ph]}, November 2017.
\newblock arXiv: 1711.02249.

\bibitem{hentschel_machine_2010}
Alexander Hentschel and Barry~C. Sanders.
\newblock Machine {Learning} for {Precise} {Quantum} {Measurement}.
\newblock {\em Physical Review Letters}, 104(6):063603, February 2010.

\bibitem{hentschel_efficient_2011}
Alexander Hentschel and Barry~C. Sanders.
\newblock Efficient {Algorithm} for {Optimizing} {Adaptive} {Quantum}
  {Metrology} {Processes}.
\newblock {\em Physical Review Letters}, 107(23):233601, November 2011.

\bibitem{palittapongarnpim_learning_2016}
Pantita Palittapongarnpim, Peter Wittek, Ehsan Zahedinejad, Shakib Vedaie, and
  Barry~C. Sanders.
\newblock Learning in {Quantum} {Control}: {High}-{Dimensional} {Global}
  {Optimization} for {Noisy} {Quantum} {Dynamics}.
\newblock {\em arXiv:1607.03428 [quant-ph, stat]}, July 2016.
\newblock arXiv: 1607.03428.

\bibitem{tiersch_adaptive_2015}
M.~Tiersch, E.~J. Ganahl, and H.~J. Briegel.
\newblock Adaptive quantum computation in changing environments using
  projective simulation.
\newblock {\em Scientific Reports}, 5:12874, August 2015.

\bibitem{biamonte_quantum_2017}
Jacob Biamonte, Peter Wittek, Nicola Pancotti, Patrick Rebentrost, Nathan
  Wiebe, and Seth Lloyd.
\newblock Quantum machine learning.
\newblock {\em Nature}, 549(7671):195, September 2017.

\bibitem{romero_quantum_2017}
Jonathan Romero, Jonathan~P. Olson, and Alan Aspuru-Guzik.
\newblock Quantum autoencoders for efficient compression of quantum data.
\newblock {\em Quantum Science and Technology}, 2(4):045001, 2017.

\bibitem{williams_simple_1992}
Ronald~J. Williams.
\newblock Simple statistical gradient-following algorithms for connectionist
  reinforcement learning.
\newblock {\em Machine Learning}, 8(3-4):229--256, May 1992.

\bibitem{goodfellow_deep_2016}
Ian Goodfellow, Yoshua Bengio, and Aaron Courville.
\newblock {\em Deep {Learning}}.
\newblock The MIT Press, Cambridge, Massachusetts, November 2016.

\bibitem{shor_scheme_1995}
Peter~W. Shor.
\newblock Scheme for reducing decoherence in quantum computer memory.
\newblock {\em Physical Review A}, 52(4):R2493--R2496, October 1995.

\bibitem{nielsen_quantum_2011}
Michael~A. Nielsen and Isaac~L. Chuang.
\newblock {\em Quantum {Computation} and {Quantum} {Information}: 10th
  {Anniversary} {Edition}}.
\newblock Cambridge University Press, Cambridge ; New York, anniversary edition
  edition, January 2011.

\bibitem{terhal_quantum_2015}
Barbara~M. Terhal.
\newblock Quantum error correction for quantum memories.
\newblock {\em Reviews of Modern Physics}, 87(2):307--346, April 2015.

\bibitem{maaten_visualizing_2008}
Laurens van~der Maaten and Geoffrey Hinton.
\newblock Visualizing {Data} using t-{SNE}.
\newblock {\em Journal of Machine Learning Research}, 9(Nov):2579--2605, 2008.

\bibitem{hochreiter_long_1997}
Sepp Hochreiter and J{\"u}rgen Schmidhuber.
\newblock Long {Short}-{Term} {Memory}.
\newblock {\em Neural Computation}, 9(8):1735--1780, November 1997.

\bibitem{mirrahimi_dynamically_2014}
Mazyar Mirrahimi, Zaki Leghtas, Victor~V. Albert, Steven Touzard, Robert~J.
  Schoelkopf, Liang Jiang, and Michel~H. Devoret.
\newblock Dynamically protected cat-qubits: a new paradigm for universal
  quantum computation.
\newblock {\em New Journal of Physics}, 16(4):045014, 2014.

\bibitem{heeres_implementing_2017}
Reinier~W. Heeres, Philip Reinhold, Nissim Ofek, Luigi Frunzio, Liang Jiang,
  Michel~H. Devoret, and Robert~J. Schoelkopf.
\newblock Implementing a universal gate set on a logical qubit encoded in an
  oscillator.
\newblock {\em Nature Communications}, 8(1):94, July 2017.

\bibitem{helmer_cavity_2009}
F.~Helmer, M.~Mariantoni, A.~G. Fowler, J.~von Delft, E.~Solano, and
  F.~Marquardt.
\newblock Cavity grid for scalable quantum computation with superconducting
  circuits.
\newblock {\em EPL (Europhysics Letters)}, 85(5):50007, 2009.

\bibitem{li_crossbar_2017}
R.~Li, L.~Petit, D.~P. Franke, J.~P. Dehollain, J.~Helsen, M.~Steudtner, N.~K.
  Thomas, Z.~R. Yoscovits, K.~J. Singh, S.~Wehner, L.~M.~K. Vandersypen, J.~S.
  Clarke, and M.~Veldhorst.
\newblock A {Crossbar} {Network} for {Silicon} {Quantum} {Dot} {Qubits}.
\newblock {\em arXiv:1711.03807 [cond-mat, physics:quant-ph]}, November 2017.
\newblock arXiv: 1711.03807.

\bibitem{august_taking_2018}
Moritz August and Jos{\'e}~Miguel Hern{\'a}ndez-Lobato.
\newblock Taking gradients through experiments: {LSTMs} and memory proximal
  policy optimization for black-box quantum control.
\newblock {\em arXiv:1802.04063 [quant-ph]}, February 2018.
\newblock arXiv: 1802.04063.

\bibitem{sutton_reinforcement_1998}
Richard~S Sutton and Andrew~G Barto.
\newblock {\em Reinforcement learning: {An} introduction}, volume~1.
\newblock MIT press Cambridge, 1998.

\bibitem{amari_natural_1998}
Shun-ichi Amari.
\newblock Natural {Gradient} {Works} {Efficiently} in {Learning}.
\newblock {\em Neural Computation}, 10(2):251--276, February 1998.

\bibitem{kakade_natural_2002}
Sham~M Kakade.
\newblock A natural policy gradient.
\newblock In {\em Advances in neural information processing systems}, pages
  1531--1538, 2002.

\bibitem{peters_natural_2008}
Jan Peters and Stefan Schaal.
\newblock Natural actor-critic.
\newblock {\em Neurocomputing}, 71(7-9):1180--1190, 2008.

\bibitem{williams_function_1991}
Ronald~J Williams and Jing Peng.
\newblock Function optimization using connectionist reinforcement learning
  algorithms.
\newblock {\em Connection Science}, 3(3):241--268, 1991.

\bibitem{kingma_method_2015}
D~Kingma and Jimmy~Ba Adam.
\newblock A method for stochastic optimisation.
\newblock In {\em International {Conference} for {Learning} {Representations}},
  volume~6, 2015.

\bibitem{hinton_improving_2012}
Geoffrey~E Hinton, Nitish Srivastava, Alex Krizhevsky, Ilya Sutskever, and
  Ruslan~R Salakhutdinov.
\newblock Improving neural networks by preventing co-adaptation of feature
  detectors.
\newblock {\em arXiv preprint arXiv:1207.0580}, 2012.

\bibitem{al-rfou_et_al._theano:_2016}
Rami Al-Rfou~et al.
\newblock Theano: {A} {Python} framework for fast computation of mathematical
  expressions.
\newblock {\em arXiv e-prints}, abs/1605.02688, May 2016.

\end{thebibliography}


\begin{thebibliography}{10}

\bibitem{suppl:nielsen_quantum_2011}
Michael~A. Nielsen and Isaac~L. Chuang.
\newblock {\em Quantum {Computation} and {Quantum} {Information}: 10th
  {Anniversary} {Edition}}.
\newblock Cambridge University Press, Cambridge ; New York, anniversary edition
  edition, January 2011.

\bibitem{suppl:laflamme1996perfect}
Raymond Laflamme, Cesar Miquel, Juan~Pablo Paz, and Wojciech~Hubert Zurek.
\newblock Perfect quantum error correcting code.
\newblock {\em Physical Review Letters}, 77(1):198, 1996.

\bibitem{suppl:sutton_reinforcement_1998}
Richard~S Sutton and Andrew~G Barto.
\newblock {\em Reinforcement learning: {An} introduction}, volume~1.
\newblock MIT press Cambridge, 1998.

\bibitem{suppl:williams_simple_1992}
Ronald~J. Williams.
\newblock Simple statistical gradient-following algorithms for connectionist
  reinforcement learning.
\newblock {\em Machine Learning}, 8(3-4):229--256, May 1992.

\bibitem{suppl:kingma_method_2015}
D~Kingma and Jimmy~Ba Adam.
\newblock A method for stochastic optimisation.
\newblock In {\em International {Conference} for {Learning} {Representations}},
  volume~6, 2015.

\bibitem{suppl:peters_natural_2008}
Jan Peters and Stefan Schaal.
\newblock Natural actor-critic.
\newblock {\em Neurocomputing}, 71(7-9):1180--1190, 2008.

\bibitem{suppl:amari_natural_1998}
Shun-ichi Amari.
\newblock Natural {Gradient} {Works} {Efficiently} in {Learning}.
\newblock {\em Neural Computation}, 10(2):251--276, February 1998.

\bibitem{suppl:kakade_natural_2002}
Sham~M Kakade.
\newblock A natural policy gradient.
\newblock In {\em Advances in neural information processing systems}, pages
  1531--1538, 2002.

\bibitem{suppl:peters2003nat_grad}
Jan Peters, Sethu Vijayakumar, and Stefan Schaal.
\newblock Reinforcement learning for humanoid robotics.
\newblock In {\em Proceedings of the third IEEE-RAS international conference on
  humanoid robots}, pages 1--20, 2003.

\bibitem{suppl:pascanu2013nat_grad}
Razvan Pascanu and Yoshua Bengio.
\newblock Natural gradient revisited.
\newblock 2013.

\bibitem{suppl:williams_function_1991}
Ronald~J Williams and Jing Peng.
\newblock Function optimization using connectionist reinforcement learning
  algorithms.
\newblock {\em Connection Science}, 3(3):241--268, 1991.

\bibitem{suppl:hochreiter97}
Sepp Hochreiter and J{\"u}rgen Schmidhuber.
\newblock {L}ong-{S}hort {T}erm {M}emory.
\newblock {\em Neural Computation}, 9:1735, 1997.

\bibitem{suppl:goodfellow16}
Ian Goodfellow, Yoshua Bengio, and Aaron Courville.
\newblock {\em {D}eep {L}earning}.
\newblock MIT Press, 2016.

\bibitem{suppl:maaten08}
Laurens van~der Maaten and Geoffrey Hinton.
\newblock Visualizing data using t-{SNE}.
\newblock {\em Journal of Machine Learning Research}, 9:2579, 2008.

\bibitem{suppl:al-rfou_et_al._theano:_2016}
Rami Al-Rfou~et al.
\newblock Theano: {A} {Python} framework for fast computation of mathematical
  expressions.
\newblock {\em arXiv e-prints}, abs/1605.02688, May 2016.

\bibitem{suppl:gottesman1998heisenberg}
Daniel Gottesman.
\newblock The heisenberg representation of quantum computers.
\newblock {\em arXiv preprint quant-ph/9807006}, 1998.

\bibitem{suppl:cleve1997stabilizers}
Richard Cleve and Daniel Gottesman.
\newblock Efficient computations of encodings for quantum error correction.
\newblock {\em Physical Review A}, 56(1):76, 1997.

\bibitem{suppl:aaronson2004improved_stabilizers}
Scott Aaronson and Daniel Gottesman.
\newblock Improved simulation of stabilizer circuits.
\newblock {\em Physical Review A}, 70(5):052328, 2004.

\bibitem{suppl:pedregosa11}
Fabian Pedregosa, Ga{\"e}l Varoquaux, Alexandre Gramfort, Vincent Michel,
  Bertrand Thirion, Olivier Grisel, Mathieu Blondel, Peter Prettenhofer, Ron
  Weiss, Vincent Dubourg, Jake Vanderplas, Alexandre Passos, David Cournapeau,
  Matthieu Brucher, Matthieu Perrot, and {\'E}douard Duchesnay.
\newblock Scikit-learn: {M}achine {L}earning in {P}ython.
\newblock {\em Journal of Machine Learning Research}, 12:2825, 2011.

\end{thebibliography}

\onecolumngrid
\newpage

\begin{appendices}

\crefalias{section}{appsec}
\crefalias{subsection}{appsec}

\renewcommand\thesection{\arabic{section}}
\renewcommand\thesubsection{\arabic{section}.\arabic{subsection}}

\makeatletter
\def\p@subsection{}
\makeatother

\setcounter{equation}{0}
\renewcommand\theequation{S\arabic{equation}}

\setcounter{figure}{0}
\renewcommand\thefigure{S\arabic{figure}}

\setcounter{table}{0}
\renewcommand\thetable{S\arabic{table}}


\vspace*{5\baselineskip}
\begin{center}
	\LARGE{\textbf{Supplementary Material}}
\end{center}

\vspace*{7\baselineskip}


%
%
%
%

\section{Physical time evolution}
\label{sec:phys_time_evolution}

\subsection{Lindblad equation and completely positive map}
\label{sec:phys_time_evolution:lindblad_and_cp_map}

We start by discussing how to describe the dynamics of open quantum systems as a brief introduction for people from other fields, and to fix notation.

The state of any Markovian quantum system at time $t$ is completely characterized by the density matrix $\hat{\rho}(t)$. Its time evolution can always be described by a Lindblad equation
\begin{equation}
	\frac{\total{}}{\total{t}}\,\hat{\rho} = - \frac{\mconst{i}}{\pconst{hbar}} \comm{\hat{H}}{\hat{\rho}} + \sum_j \Gamma_j \Big(\hat{L}_j\hat{\rho}\hat{L}_j^\dagger-\frac{1}{2}\anticomm{\hat{L}_j^\dagger\hat{L}_j}{\hat{\rho}}\Big)
	\label{eq:lindblad_eqn}
\end{equation}
where the Hamiltonian $\hat{H}$ represents the coherent part of the dynamics and the Lindblad operators $\hat{L}_j$ (or jump operators) the incoherent part; $\Gamma_j$ are their corresponding decay rates. $\comm{\hat{A}}{\hat{B}}=\hat{A}\hat{B}-\hat{B}\hat{A}$ is the commutator of two operators, and $\anticomm{\hat{A}}{\hat{B}}=\hat{A}\hat{B}+\hat{B}\hat{A}$ the anti-commutator. $\hat{A}^\dagger$ denotes the Hermitian conjugate of $\hat{A}$.

For compact notation, the terms on the right-hand side can be combined into one superoperator, the Liouvillian $\mathcal{L}$:
\begin{equation}
	\frac{\total{}}{\total{t}}\,\hat{\rho} = \mathcal{L}\,\hat{\rho}
\end{equation}
We now introduce the completely positive map $\pquant{strict_cp_map_tf_ti}$ to formally write down the time evolution of all density matrices in the time interval from $\pquant{ti}$ to $\pquant{tf}$:
\begin{equation}
	\hat{\rho}(\pquant{tf}) = \pquant{strict_cp_map_tf_ti}[\hat{\rho}(\pquant{ti})]
\end{equation}
If $\mathcal{L}$ does not change over time, the completely positive map is given by $\pquant{strict_cp_map_tf_ti}=\mconst{e}^{(\pquant{tf}-\pquant{ti})\mathcal{L}}$; otherwise, it can be obtained from $\pquant{strict_cp_map_tf_ti}=\Texp(\int_{\pquant{ti}}^{\pquant{tf}}\mathcal{L}(t)\,\total{t})$ where $\Texp$ denotes the time-ordered exponential.

Measurements can be nicely integrated into this framework. For each measurement variable and obtained result, there is a projection operator $\hat{P}$, and we associate a superoperator $\mathcal{P}$ defined by its action $\mathcal{P}\hat{\rho}=\hat{P}\hat{\rho}\hat{P}^\dagger$ on all density matrices $\hat{\rho}$. If some measurements are performed at intermediate times $t_1,\hdots,t_m$ with corresponding superoperators $\mathcal{P}_1,\hdots,\mathcal{P}_m$, and otherwise $\hat{\rho}$ follows the Lindblad equation characterized by the Liouvillian $\mathcal{L}$, the completely positive map has the form $\pquant{strict_cp_map_tf_ti}[\hat{\rho}]=\pquant{cp_map_tf_ti}[\hat{\rho}]/\tr(\pquant{cp_map_tf_ti}[\hat{\rho}])$ where $\pquant{cp_map_tf_ti}=\mconst{e}^{(\pquant{tf}-t_m)\mathcal{L}}\,\mathcal{P}_m\,\mconst{e}^{(t_m-t_{m-1})\mathcal{L}}\hdots\mconst{e}^{(t_2-t_1)\mathcal{L}}\,\mathcal{P}_1\,\mconst{e}^{(t_1-\pquant{ti})\mathcal{L}}$ (note that explicit normalization is required because projections are in general not trace-preserving).

For our purposes, it is more convenient to consider $\pquant{cp_map}$ instead of $\pquant{strict_cp_map}$, so in the following we will use
\begin{equation}
	\hat{\rho}(\pquant{tf}) = \frac{\pquant{cp_map_tf_ti}[\hat{\rho}(\pquant{ti})]}{\tr(\pquant{cp_map_tf_ti}[\hat{\rho}(\pquant{ti})])}
	\label{eq:linearized_cp_map}
\end{equation}
where $\pquant{cp_map_tf_ti}$ is always a linear map (even if it includes measurements).

\subsection{Simulation of a logical qubit}

This formalism is usually applied directly to single density matrices. However, as motivated in the main text, we require an efficient scheme which is capable of processing all possible logical qubit states, \abbr{ie} the ``full Bloch sphere'', in parallel. This is possible due to two circumstances: the linearity of the completely positive map, and the fact that all initial states are arranged on an affine space.

In Methods, we specify an explicit scheme to construct four quantities $\pquant{rho_0}(t)$, $\pquant{delta_rho_x}(t)$, $\pquant{delta_rho_y}(t)$ and $\pquant{delta_rho_z}(t)$ which give access to the density matrix $\hat{\rho}_{\vec{n}}(t)$ for an arbitrary logical qubit state $\vec{n}=(x,y,z)$ at any time $t$:
\begin{align}
	\hat{\rho}_{\vec{n}}(t) & =
		\frac{\pquant{rho_0}(t)+\sum_jn_j\pquant{delta_rho_j}(t)}{1+\tr(\sum_jn_j\pquant{delta_rho_j}(t))} = \nonumber \\
	& =
		\frac{\pquant{rho_0}(t)+x\,\pquant{delta_rho_x}(t)+y\,\pquant{delta_rho_y}(t)+z\,\pquant{delta_rho_z}(t)}{1+x\tr(\pquant{delta_rho_x}(t))+y\tr(\pquant{delta_rho_y}(t))+z\tr(\pquant{delta_rho_z}(t))}
	\label{eq:construction_rho_xyz}
\end{align}

Without measurements, \abbr{ie} under trace-preserving quantum operations like unitary transformations and dissipation, the variables $\pquant{rho_0}$, $\pquant{delta_rho_x}$, $\pquant{delta_rho_y}$ and $\pquant{delta_rho_z}$ follow the same dynamics as the density matrices themselves, \abbr{ie} their time evolution is given by the completely positive map $\pquant{strict_cp_map_tf_ti}$. However, they have to be treated differently under measurements where renormalization becomes important. Reusing our notion of $\pquant{cp_map_tf_ti}$ from \cref{eq:linearized_cp_map}, we have
\begin{align}
	\pquant{rho_n}(\pquant{tf}) & =
		\frac{\pquant{cp_map_tf_ti}[\pquant{rho_n}(\pquant{ti})]}{\tr(\pquant{cp_map_tf_ti}[\pquant{rho_n}(\pquant{ti})])} = \nonumber \\
	& =
		\frac{\pquant{cp_map_tf_ti}[\pquant{rho_0}(\pquant{ti})+\sum_jn_j\pquant{delta_rho_j}(\pquant{ti})]}{\tr(\pquant{cp_map_tf_ti}[\pquant{rho_0}(\pquant{ti})+\sum_jn_j\pquant{delta_rho_j}(\pquant{ti})])} = \nonumber \\
	& =
		\frac{\pquant{cp_map_tf_ti}[\pquant{rho_0}(\pquant{ti})]+\sum_jn_j\pquant{cp_map_tf_ti}[\pquant{delta_rho_j}(\pquant{ti})]}{\tr(\pquant{cp_map_tf_ti}[\pquant{rho_0}(\pquant{ti})]+\sum_jn_j\pquant{cp_map_tf_ti}[\pquant{delta_rho_j}(\pquant{ti})])}
\end{align}
We see that it is not \textit{required} to explicitly renormalize $\pquant{rho_0}$, $\pquant{delta_rho_x}$, $\pquant{delta_rho_y}$ and $\pquant{delta_rho_z}$, and in particular that it is not correct to renormalize them separately (which would be ill-defined because the $\pquant{delta_rho_j}$ can be traceless). However, a \textit{common} prefactor is allowed as they appear both in the numerator and the denominator, and for convenience we choose the prefactor $1/\tr(\pquant{cp_map_tf_ti}[\pquant{rho_0}(\pquant{ti})])$ such that $\tr(\pquant{rho_0}(t))\stackrel{!}{=}1$ is always satisfied. This leads to the update equations given in Methods.

\section{Recoverable quantum information}
\label{sec:magic_quantity}

\subsection{Definition}

In the main text, we have defined the recoverable quantum information as
\begin{equation}
	\pquant{magic} = \frac{1}{2} \min_{\vec{n}} \norm[\big]{\hat{\rho}_{\vec{n}}-\hat{\rho}_{-\vec{n}}}_1
\end{equation}
An alternative definition would be
\begin{equation}
	\pquant{magic} = \min_{\vec{n}} \norm[\Big]{\sum_jn_j\,\pquant{delta_rho_j}}_1
\end{equation}
Both expressions coincide as long as all $\tr(\pquant{delta_rho_j})=0$, \abbr{ie} no information about the logical qubit state has been revealed. For the remaining cases, it is not clear to the authors what is the most natural generalization.

For the following considerations, it will not make a difference which definition is used. Likewise, it does not play a role in the numerical simulations discussed in the main text because, due to our operation sets, only two cases can occur: either no information about the logical qubit has been revealed ($\frac{1}{2}(\hat{\rho}_{\vec{n}}-\hat{\rho}_{-\vec{n}})=\sum_jn_j\pquant{delta_rho_j}$), or all superpositions have been destroyed (both definitions yield $\pquant{magic}=0$).

%
%
%
%
%

\subsection{Properties}
\label{sec:magic_quantity:properties}

\begin{figure}[p]
	\centering
	\includegraphics{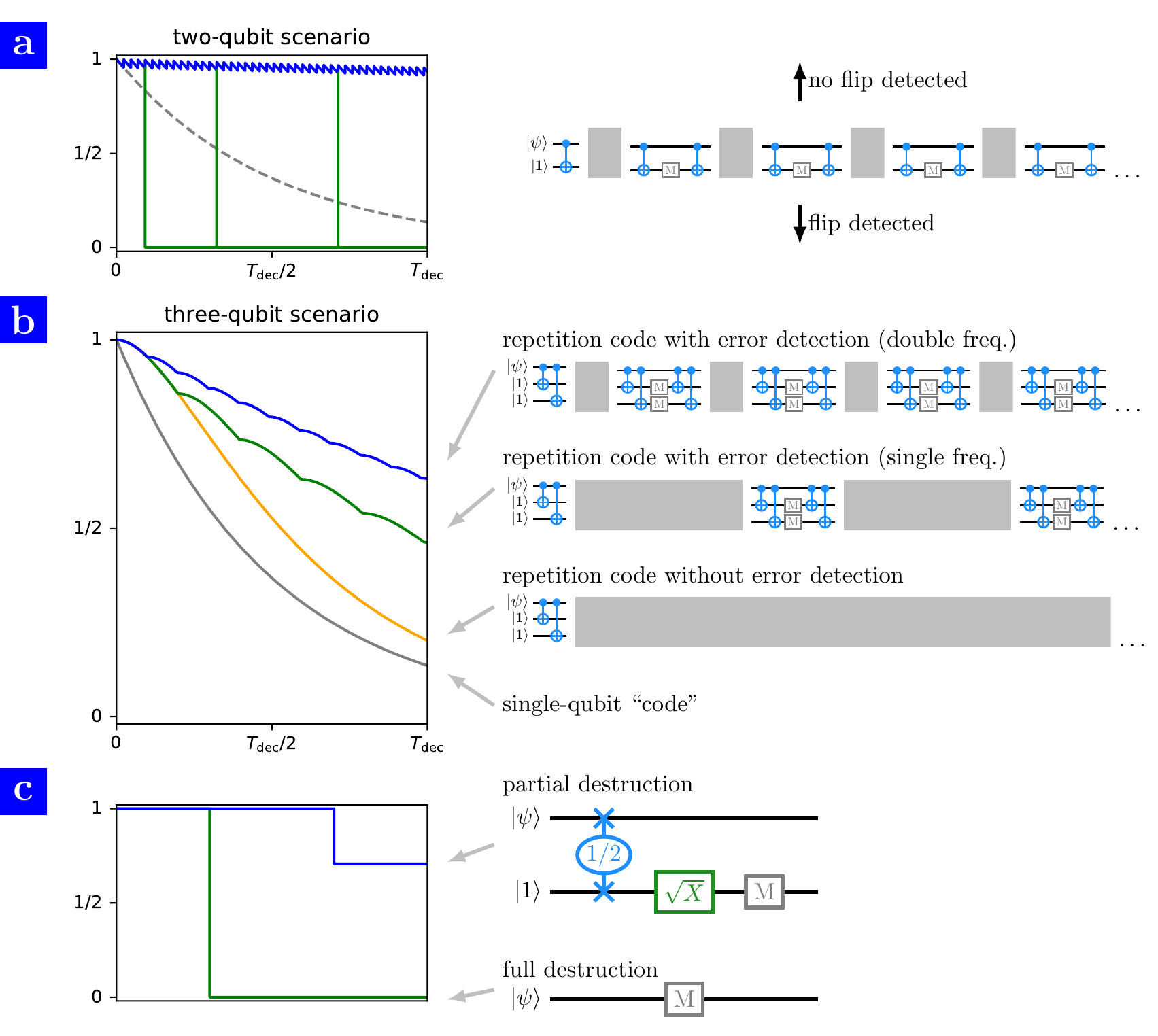}
	\caption{Behavior of $\pquant{magic}$.
		(a) Exemplary curves for $\pquant{magic}(t)$ in a two-qubit system with bit-flip errors.
		In the two-qubit encoding $\alpha\ket{\uparrow\uparrow}+\beta\ket{\downarrow\downarrow}$, parity measurements can lead to an increase of $\pquant{magic}$ if no bit flip is detected (blue), but $\pquant{magic}$ decreases to $0$ for the alternating result (green lines). The average behavior (gray) is identical to the decay for the single-qubit ``encoding'', and for the two-qubit encoding without parity measurements.
		(b) Decay of $\pquant{magic}$ in a three-qubit system with bit-flip errors. The curves show the single-qubit ``encoding'' (gray), the repetition code $\alpha\ket{\uparrow\uparrow\uparrow}+\beta\ket{\downarrow\downarrow\downarrow}$ without error detection (orange) and the repetition code with periodic parity measurements (green and blue). Note that for the two protocols with measurements, we plot the average over all possible measurement results; single trajectories will deviate from this behavior dependent on the particularly found syndroms. In addition, we neglected errors that occur during the syndrom detection sequences, \abbr{ie} we implicitly assumed the length of these sequences to be very small compared to the intermediate idle times. This assumption is seriously violated in the bit-flip scenarios on which the networks in the main text are trained (best results for measurements at the period limited by the gate operation time), such that the depicted three-qubit syndrom detection scheme would there perform much worse than the four-qubit schemes found by the neural networks (never enters a fragile state).
		(c) Drop in $\pquant{magic}$ as a consequence of total (green) or partial (blue) destruction of superpositions due to a measurement which reveals information about the logical qubit state. The gate sequence leading to the partial drop is $\sqrt{\mathrm{SWAP}}\to\text{local $90^\circ$ rotation around $x$ axis}\to\mathrm{MSMT(z)}$.
	}
	\label{fig:suppl_magic_quantity}
\end{figure}

The purpose of the following discussion is to develop an intuitive understanding for $\pquant{magic}$.

We start by giving three basic properties of the underlying trace distance $\pquant{D_n}=\norm{\sum_jn_j\,\pquant{delta_rho_j}(t)}_1$ for antipodal logical qubit states:
\begin{enumerate}
	\item $\pquant{D_n}$ is invariant under unitary transformations.
	\item $\pquant{D_n}$ can only decrease under the influence of dissipation (pure states yield the maximum value $\pquant{D_n}=1$).
	\item $\pquant{D_n}$ can be decreased and also increased by measurements, but only under the constraint that the average over all measurement results cannot exceed the prior value:
		\begin{equation}
			\sum_m p_m\,\pquant{D_n}^{(m)} \le \pquant{D_n}
		\end{equation}
		Here, $\pquant{D_n}$ is the value directly before the measurement, and $\pquant{D_n}^{(m)}$ the updated value for result $m$; $p_m$ denotes the corresponding probability to find this result. (The extreme cases $\sum_mp_m\,\pquant{D_n}^{(m)}=0$ and  $\sum_mp_m\,\pquant{D_n}^{(m)}=\pquant{D_n}$ can occur and are discussed below).
\end{enumerate}
(1) is a general property of the trace norm. (2) and (3) follow from the contractivity of the trace norm under trace-preserving quantum operations \citesuppl[\abbr{p} 406]{suppl:nielsen_quantum_2011} (here, the completely positive maps for dissipative dynamics $\hat{\rho}\mapsto\Texp(\int_{\pquant{ti}}^{\pquant{tf}}\mathcal{L}(t)\,\total{t})\hat{\rho}$ and non-selective measurements $\hat{\rho}\mapsto\sum_m\hat{P}_m\hat{\rho}\hat{P}_m^\dagger$, respectively). Trace-preserving quantum operations are those which can be written in the form $\hat{\rho}\mapsto\sum_k\hat{E}_k\hat{\rho}\hat{E}_k^\dagger$ for a complete operator set $\sum_k\hat{E}_k^\dagger\hat{E}_k=\unitmat$ \citesuppl[\abbr{p} 360]{suppl:nielsen_quantum_2011}.

To illustrate property (3) in an example, we consider the following situation: two qubits are subject to bit-flip errors, \abbr{ie} $\mathcal{D}\hat{\rho}=\pquant{T_dec}^{-1}\sum_q(\hat{\sigma}^{(q)}\hat{\rho}\hat{\sigma}^{(q)}-\hat{\rho})$. Initially, the logical qubit is stored in one physical qubit, and the ancilla is prepared in the down state ($\mconst{pauli_z}\ket{\downarrow}=-\ket{\downarrow}$). Then, the protocol depicted in \cref{fig:suppl_magic_quantity}a is applied: first, a CNOT gate entangles the data qubit with the ancilla. During the following idle time, $\pquant{magic}$ decays like $\pquant{magic}\sim\mconst{e}^{-2t/\pquant{T_dec}}$. Afterwards, the qubits are disentangled again by a second CNOT gate (which leaves $\pquant{magic}$ invariant). If in this situation a measurement on the ancilla reveals that it is still in the down state, $\pquant{magic}$ grows to $\pquant{magic}^{(\downarrow)}=2\mconst{e}^{-2t/\pquant{T_dec}}/(1+\mconst{e}^{-4t/\pquant{T_dec}})>\mconst{e}^{-2t/\pquant{T_dec}}$. However, the probability to find the ancilla in the down state is only $P_\uparrow=\frac{1}{2}(1+\mconst{e}^{-4t/\pquant{T_dec}})$, and if a flip is detected ($P_\uparrow=1-P_\downarrow$), $\pquant{magic}$ drops to $0$ as it cannot be determined which qubit has flipped. Hence, \textit{on average} nothing is won: $P_\downarrow\pquant{magic}^{(\downarrow)}+P_\uparrow\pquant{magic}^{(\uparrow)}=\pquant{magic}$. (This basically shows that the described protocol allows quite well to detect errors, but it is not possible to resolve them, and thus it cannot be used to slowdown the decay.) Exemplary $\pquant{magic}$ curves for the repeated application of this protocol are shown in \cref{fig:suppl_magic_quantity}a.

We proceed by considering the textbook three-qubit scenario with bit-flip error channels \citesuppl[\abbr{p} 427ff]{suppl:nielsen_quantum_2011} and inspecting it from the perspective of $\pquant{magic}$ (see also \cref{fig:suppl_magic_quantity}b). Compared to the decay $\pquant{magic}(t)=\mconst{e}^{-2t/\pquant{T_dec}}$ for the ``trivial'' single-qubit encoding, \abbr{eg} $(\alpha\ket{\uparrow}+\beta\ket{\downarrow})\otimes\ket{\downarrow\downarrow}$, the encoding $\alpha\ket{\uparrow\uparrow\uparrow}+\beta\ket{\downarrow\downarrow\downarrow}$ (bit-flip code, repetition code) improves the situation in the beginning: $\pquant{magic}(t)=\frac{1}{2}(3-\mconst{e}^{-4t/\pquant{T_dec}})\mconst{e}^{-2t/\pquant{T_dec}}$. However, the asymptotical behavior for $t\to\infty$ is still the same. This behavior can be explained as follows: as the populations for single bit-flips accumulate over time, the probability grows that a second bit-flip (on another qubit) occurs which cannot be resolved anymore. The only way to actually protect the logical qubit is a suitable error detection scheme, \abbr{ie} to keep track of possible bit-flips via repeated parity measurements. Such a measurement delays the decay again for some time: after a syndrom detection at $t_0$, we have $\pquant{magic}(t>t_0)=\frac{1}{4}(3-\mconst{e}^{-4t_0/\pquant{T_dec}})(3-\mconst{e}^{-4(t-t_0)/\pquant{T_dec}})\mconst{e}^{-2t/\pquant{T_dec}}$ (assuming a quasi-instantaneous syndrom detection during which no further errors can occur). Thus, even though measurements cannot (on average) improve $\pquant{magic}$ immediately (\abbr{cmp} property 3), they are useful (and necessary) to prevent future losses in $\pquant{magic}$.

Last, but not least, $\pquant{magic}$ correctly describes the loss in recoverability of the quantum state due to measurements which reveal information about the logical qubit state (see also \cref{fig:suppl_magic_quantity}c). If all superpositions are destroyed, $\pquant{magic}$ drops to $0$ (independent of the measurement result); this happens, for example, in the ``trivial encoding'' (with the logical qubit stored directly in one physical qubit) if a measurement is performed on this data qubit. In comparison, partial destruction of superpositions is reflected by an accordingly smaller decrease in $\pquant{magic}$. By contrast, measurements which reveal no information about the logical qubit state do not change $\pquant{magic}$ \textit{on average} ($\sum_mp_m\,\pquant{D_n}^{(m)}=\pquant{D_n}$).

\subsection{Computation}

The numerical evaluation of $\pquant{magic}$ is technically challenging. The computation of the trace norm $\norm{\sum_jn_j\,\pquant{delta_rho_j}(t)}_1$ for a given point $\vec{n}$ on the Bloch sphere is already expensive as it involves the eigendecomposition of its argument. Moreover, the dependence of $\norm{\sum_jn_j\,\pquant{delta_rho_j}(t)}_1$ on the Bloch vector $\vec{n}$ can be complex, which makes it non-trivial to locate the minimum that determines $\pquant{magic}$.

In our experience, it is not valid to approximate the trace-norm $\norm{\cdot}_1$ in $\pquant{magic}$ using the Hilbert-Schmidt norm $\norm{\cdot}_2$, which is numerically far easier to calculate (and minimize). Likewise, other simpler quantities (like the purity of the multi-qubit state) are not sufficient to replace $\pquant{magic}$.

In the Methods, we have claimed that for CHZ circuits (all unitary operations are combinations of CNOT, Hadamard and phase gates; measurement of variables in the Pauli group; state preparation in the computational basis \citesuppl[\abbr{p} 464]{suppl:nielsen_quantum_2011}), the anti-commutator relation $\anticomm{\pquant{delta_rho_j}}{\pquant{delta_rho_k}}=0$ is satisfied for all distinct $j,k\in\{\mathrm{x},\mathrm{y},\mathrm{z}\}$; our bit-flip scenarios fall into this category of CHZ circuits. Furthermore, we have argued that $\min_{x,y,z}\norm{x\,\pquant{delta_rho_x}+y\,\pquant{delta_rho_y}+z\,\pquant{delta_rho_z}}_1=\min_{j\in\{\mathrm{x},\mathrm{y},\mathrm{z}\}}\norm{\pquant{delta_rho_j}}_1$ (for $x^2+y^2+z^2=1$) if this anti-commutator relation is satisfied. In the following, we briefly sketch the proof for this statement. We note that ``$\le$'' is a priori clear, so just ``$\ge$'' has to be shown. The assumption $j\neq k\Rightarrow\anticomm{\pquant{delta_rho_j}}{\pquant{delta_rho_k}}=0$ directly implies that $\comm{\pquant{delta_rho_j}^2}{\pquant{delta_rho_k}^2}=0$, \abbr{ie} there is a common eigenbasis $\{\ket{\phi_n}\}$ for all three $\pquant{delta_rho_j}^2$. Further, we can conclude that
\begin{equation}
	\norm[\big]{x\,\pquant{delta_rho_x}+y\,\pquant{delta_rho_y}+z\,\pquant{delta_rho_z}}_1 =
	\sum_n\sqrt{x^2\Braket{\phi_n}{\pquant{delta_rho_x}^2}{\phi_n}+y^2\Braket{\phi_n}{\pquant{delta_rho_y}^2}{\phi_n}+z^2\Braket{\phi_n}{\pquant{delta_rho_z}^2}{\phi_n}}
\end{equation}
and for $x^2+y^2+z^2=1$, Jensen's inequality (note that square root is a concave function) yields the lower bound
\begin{align}
	\norm[\big]{x\pquant{delta_rho_x}+y\pquant{delta_rho_y}+z\pquant{delta_rho_z}}_1 & \ge
		\sum_nx^2\sqrt{\Braket{\phi_n}{\pquant{delta_rho_x}^2}{\phi_n}}+y^2\sqrt{\Braket{\phi_n}{\pquant{delta_rho_y}^2}{\phi_n}}+z^2\sqrt{\Braket{\phi_n}{\pquant{delta_rho_z}^2}{\phi_n}} = \nonumber \\
	& =
		x^2\norm{\pquant{delta_rho_x}}_1+y^2\norm{\pquant{delta_rho_y}}_1+z^2\norm{\pquant{delta_rho_z}}_1
\end{align}
Minimization over the Bloch sphere ($x^2+y^2+z^2=1$) leads to the desired result.

\section{State-Aware Network}

\subsection{Reward scheme}
\label{sec:reward_scheme}

\paragraph{Protection reward} The purpose of the protection reward is that the agent learns how to protect the quantum information against the error mechanisms of the quantum system, \abbr{ie} to preserve a state that could in principle be recovered without the need to actually perform the recovery sequence (also \abbr{cmp} Methods). For instance, in the bit-flip examples, the agent learns to go into a proper encoding and to perform afterwards repeated parity measurements. Most of the networks in \abbr{sec} ``Results'' of the maintext are trained solely with this type of reward; only for \maintextref{fig:decoding_and_event_aware} in the main text, an additional recovery reward (see below) has been considered.

For convenience, we will set the simulation time step to $\pquant{time_step}=1$ in this discussion.

As discussed in \cref{sec:magic_quantity} and the main text, $\pquant{magic}$ is a powerful measure for the capability to recover a target state (without the need to actually perform these steps), and so the objective is to maximize the value of $\pquant{magic}$ at the final time $T$. For learning efficiency, and (very important for us) to credit already partial success, we aim for an immediate reward. This reward should essentially be of the form $\tilde{r}_t=\pquant{magic}(t+1)-\pquant{magic}(t)$: as the learning algorithm optimizes the reward sum, $\sum_t\tilde{r}_t=\sum_t\pquant{magic}(t+1)-\pquant{magic}(t)=\pquant{magic}(T)-\pquant{magic}(0)=\pquant{magic}(T)-1$, the agent is guided towards the desired behavior, to maximize $\pquant{magic}(T)$.

In practice, we have found it useful to implement the reward scheme rather in the following form, which will be motivated below:
\begin{equation}
	r_t =
	\left\{\begin{array}{c@{\hspace*{-0.5em}}cl}
		1+\frac{\pquant{exp_magic}(t+1)-\pquant{magic}(t)}{2\pquant{time_step}/\pquant{T_triv}} &
			+\,0 &
			\text{if $\pquant{exp_magic}(t+1)>0$} \\
		0 &
			-\,\pquant{punish_destr} &
			\text{if $\pquant{exp_magic}(t)\neq0\wedge\pquant{exp_magic}(t+1)=0$} \\
		\smash{\underbrace{\hspace*{3.5em}0\hspace*{3.5em}}_{\textstyle=:r_t^{(1)}}} &
			\smash{\underbrace{+\,0}_{\textstyle=:r_t^{(2)}}} &
			\text{if $\pquant{exp_magic}(t)=0$}
	\end{array}\right.
	\vspace*{1.25\baselineskip}
	\label{eq:definition_magic_reward}
\end{equation}
where $\pquant{time_step}$ denotes the time per gate operation, $\pquant{T_triv}$ is the decay time of the quantum state in the trivial encoding, and $\pquant{punish_destr}$ is the punishment for a measurement which reveals the logical qubit state. $\pquant{exp_magic}$ will be defined in the next paragraph. Note that $r_t$ is still designed such that optimizing $\pquant{magic}(T)$ leads to the highest possible reward sum $\sum_tr_t$.

The updated value of $\pquant{magic}$ after a measurement can depend on its outcome. We define $\pquant{exp_magic}(t+1)$ as the expectation value for $\pquant{magic}(t+1)$. For measurements, this is the average for all possible measurement results, weighted according to their probability. For unitary gates, there is no ambiguity and thus $\pquant{exp_magic}(t+1)=\pquant{magic}(t+1)$. We will motivate below why it makes sense to consider this quantity.

When computing the (discounted) return $R_t$ from the reward, we choose to distribute only $r_t^{(1)}$ backwards in time, while $r_t^{(2)}$ is assigned directly to the respective action:
\begin{equation}
	R_t = (1-\gamma) \Big(\sum_{k=0}^{T-t-1} \gamma^k r_{t+k}^{(1)}\Big) + r_t^{(2)}
	\label{eq:definition_magic_return}
\end{equation}
$\gamma$ is the discount rate; we choose a value of $\gamma=0.95$, corresponding to a half-life in the range between 10 and 20 time steps.

The expressions in \cref{eq:definition_magic_reward,eq:definition_magic_return} are chosen for the following reasons:
\begin{itemize}
	\item For measurements, there is the special situation that, dependent on the quantum state, they might reveal information about the logical qubit state. This is always accompanied by the destruction of superpositions, and in our operation sets, it even leads to their complete destruction. Because no other action could lead to a worse state and the exceptionally strong variation in $\pquant{magic}$ would cause learning instabilities, it makes sense to treat those cases separately.
		
		These ``destructive'' measurements are characterized by the fact that $\pquant{magic}$ drops to $0$ independent of the measurement result. If this situation occurs, we deviate from a reward that is directly related to $\pquant{magic}(t+1)-\pquant{magic}(t)$, and instead set the immediate reward $r_t$ to a fixed negative value $-\pquant{punish_destr}$. Furthermore, because this situation could easily be avoided by choosing this particular action different, we do not distribute this reward over the previous time steps when we compute the return.
		
		As an alternative, those measurements could simply be excluded from being performed by an external instance. We decided against doing so because in the real-world application we aim for (control of a quantum experiment), an agent needs to be able to detect these cases on its own.
	\item Even after excluding those measurements which reveal the logical qubit state, $\pquant{magic}$ can still jump in special situations; for example, in the two-qubit scenario discussed in \cref{sec:magic_quantity:properties}, there are measurements that make $\pquant{magic}$ increase in case of the anticipated result, at the expense of getting $\pquant{magic}=0$ for the unlikely one. These strong fluctuations in $\pquant{magic}$ would lead to large variations in the reward and thereby can easily cause learning instabilities. This can be improved by using the variable $\pquant{exp_magic}$ as defined above: we can easily replace $\pquant{magic}(t+1)-\pquant{magic}(t)\to\pquant{exp_magic}(t+1)-\pquant{magic}(t)$ because in the end, the reward is used to compute an estimator (for the learning gradient, see \cref{sec:reinforcement_learning}), and on average both coincide. Since $\pquant{exp_magic}$ is much more stable than $\pquant{magic}$, we can easily get rid of the instabilities; as an intended side-effect, this also reduces the ``normal'' noise level on the learning gradient (due to the deviations between the estimator and the true value) because now each simulation averages directly over all possible measurement results, something that otherwise would have to be done over many epochs.
	\item The first case in \cref{eq:definition_magic_reward} is a scaled version of $\pquant{magic}(t+1)-\pquant{magic}(t)$ due to the following consideration: For the trivial encoding, we have (at the start of the simulation) $\pquant{magic}(t+1)-\pquant{magic}(t)\approx-2\pquant{time_step}/\pquant{T_triv}$ since in this case $\pquant{magic}(t)\approx\mconst{e}^{-2t/\pquant{T_triv}}\approx1-2t/\pquant{T_triv}$ for $t\ll\pquant{T_triv}$, whereas for perfect conservation of the quantum information, \abbr{ie} constant $\pquant{magic}(t)$, we would get $\pquant{magic}(t+1)-\pquant{magic}(t)=0$. Note that typically, even the ideal action sequence does not reach constant $\pquant{magic}(t)$, but it provides a good approximation for comparison with other strategies. From comparing the values $0$ and $-2\pquant{time_step}/\pquant{T_triv}$, we can see that the typical value range for $\tilde{r}_t=\pquant{magic}(t+1)-\pquant{magic}(t)=0$ depends on parameters of the physical model. In practice, this makes it difficult to change these properties, especially the type of error channels and the corresponding decoherence rates. For a proper normalization, we can simply divide $\pquant{magic}(t+1)-\pquant{magic}(t)$ by $2\pquant{time_step}/\pquant{T_triv}$; in addition, we can always add the constant value which we choose to be $1$, such that the trivial encoding earns $r_t\approx0$ and the ideal strategy $r_t\approx1$.
\end{itemize}

\paragraph{Recovery reward} In practice, the goal is to eventually recover the logical qubit state from the physical qubits. This means that the quantum system should finally be brought into a state where the logical qubit state can be read off easily from one target qubit, with a pre-defined interpretation for the polarization. To train the network towards this behavior, we will introduce a recovery reward $r_t^\text{(recov)}$ (see below) that is given in addition to the protection reward (which only addresses preserving the recoverable quantum information $\pquant{magic}$). Furthermore, we extend the input of the neural network to also contain the time. It is fully sufficient to add a ``countdown'' for the last time steps to signal the network when the decoding should start (when also the decoding reward can be earned). We input this countdown in a one-hot encoding. Note that it does not matter how many time steps exactly remain when giving the decoding signal (and the recovery reward), if there are sufficiently many left to fit the decoding sequence, which of course is initially unknown. For our example shown in \maintextref{fig:decoding_and_event_aware} of the main text the countdown is started when $t>T-10$, and the decoding reward is only given afterwards.

A successful recovery requires to correct for the errors (if some occurred) and to decode, such that in the end the logical qubit is stored solely on a specified target qubit. All other qubits should be disentangled and in particular the state of the target qubit should not be flipped compared to the initial state. 
To achieve this, we extend the reward scheme by two contributions: First, an additional decoding reward can be earned during the last few time steps. Second, a correction reward can be earned at the last time step (but only if the decoding has been successfully performed). We combine them into the recovery reward
\begin{equation}
	r_t^\text{(recov)} = \underbrace{\beta_\text{dec}\left(D_{t+1}-D_t\right)}_\text{decoding reward}+\underbrace{\beta_\text{corr}C_T\delta_{t,T-1}}_\text{correction reward}
	\label{eq:recovery_reward}
\end{equation}
where $\beta_\text{dec}$ and $\beta_\text{corr}$ are the scalings for the decoding and correction reward respectively. In the following, we will discuss first the explicit form of $D_t$ (for decoding) and later that of $C_t$ (for correction).

We design the decoding reward to ease learning for the network. Instead of providing a reward only if all qubits of our example are in their respective desired target state of either containing or not containing information about the logical qubit state, it helps to immediately reward also partial success, \abbr{ie} whenever one more qubit is brought into the desired state. Thereby, this disentanglement of qubits can be learned stepwise until all qubits are decoded correctly. 

In order to quantify such a reward, we aim for a criterion to determine whether a qubit (ignoring all other qubits) contains at least a small fraction of quantum information about the logical qubit state. We obtain a mathematically sufficient criterion by investigating the partial trace over all other qubits: in order to conclude that qubit $q$ definitely contains information about the logical qubit state, we evaluate whether
\begin{equation}
	\tr_{\bar{q}}(\pquant{rho_xyz}(t))
\end{equation}
depends on the logical qubit state $\vec{n}=(x,y,z)$. Expressed in terms of the variables $\pquant{delta_rho_x}$, $\pquant{delta_rho_y}$ and $\pquant{delta_rho_z}$ (\abbr{cmp} \cref{eq:construction_rho_xyz}), stating that the quantity $\tr_{\bar{q}}(\pquant{rho_xyz}(t))$ does depend on $\vec{n}$ and therefore $q$ carries information about the logical qubit state is equivalent to the condition that at least one of the corresponding partial traces is non-vanishing:
\begin{align}\label{eq:tracecrit}
	(\tr_{\bar{q}}(\pquant{delta_rho_x}(t)) \neq 0) \vee
	(\tr_{\bar{q}}(\pquant{delta_rho_y}(t)) \neq 0) \vee
	(\tr_{\bar{q}}(\pquant{delta_rho_z}(t)) \neq 0)
\end{align}
We have successfully applied this criterion to train the network analyzed in \maintextref{fig:decoding_and_event_aware} of the main text. To see why this is strictly speaking not a necessary condition (and thus not generally applicable), consider the Laflamme-Miquel-Paz-Zurek \citesuppl{suppl:laflamme1996perfect} encoding. There, this criterion would predict that there is no quantum information in any of the qubits, even though the expected result is that it is distributed over all of them. Nonetheless, for our example of the repetition code it is very useful to help the network to learn the decoding with a close to perfect success rate. One can now specify a desired target state, where \abbr{eg} the first qubit is supposed to contain the quantum information, but none of the remaining qubits. By evaluating condition \cref{eq:tracecrit} for each qubit, we identify how many qubits are in the desired target state of containing or not-containing quantum information and calculate
\begin{equation}
	D_t =
	\begin{cases}
		0 & \text{if $t<T_\text{signal}$} \\
		\frac{\text{\#correct qubits}-\#\text{wrong qubits}}{2}& \text{if $t\geq T_\text{signal}$} 
	\end{cases}
\end{equation}
where $T_\text{signal}$ denotes the time where the neural net obtains the signal to decode, \abbr{ie}, where the countdown starts. As the decoding reward is essentially $D_{t+1}-D_t$ (see \cref{eq:recovery_reward}), a positive reward is given if the number of qubits in the desired state increases from one time step to the next, and the network is punished with a negative reward if the number of qubits in the desired state decreases. The corresponding coefficient $\beta_\text{dec}$ has to be sufficiently large such that the decoding reward can compete with the protection reward (we choose $\beta_\text{dec}=20$). This is because during and after the decoding, $\pquant{magic}$ decays faster, lowering the protection reward. Note that the decoding reward is set to zero if the recoverable quantum information is already too small. In our particular example, we have chosen the threshold to be $\pquant{magic}<0.1$.

Having decoded the multi-qubit state leaves the logical qubit solely on a specified physical qubit. However, this is not necessarily the initial logical qubit state. Preserving the recoverable quantum information $\pquant{magic}$ only implies that the initial logical qubit state could in principle be recovered from the actual qubit state. To trigger the network to perform corrections, such that the final state is not flipped with respect to the initial logical qubit state, we calculate whether the Bloch vector of the specified physical qubit after the final time step is rotated with respect to the Bloch vector of the initial logical qubit state. To quantify this, we consider the overlap $\Braket{\phi_{\vec{n}}}{\tr_\mathrm{o.q.}(\hat{\rho}_{\vec{n}})}{\phi_{\vec{n}}}$ for a given logical qubit state $\vec{n}$; $\tr_\mathrm{o.q.}$ denotes the partial trace over all qubits except for the target qubit. As a successful error correction scheme needs to work for all possible logical qubit states, we focus on the worst-case, \abbr{ie} we consider
\begin{equation}
	\mathcal{O}_Q = \min_{\vec{n}} \Braket{\phi_{\vec{n}}}{\tr_\mathrm{o.q.}(\hat{\rho}_{\vec{n}})}{\phi_{\vec{n}}}
	\label{eq:worst_case_overlap}
\end{equation}
(\abbr{cmp} \cref{eq:overlap_criterion}). According to $\cref{eq:construction_rho_xyz}$, we can write $\hat{\rho}_{\vec{n}}$ in terms of $\pquant{rho_0}$, $\pquant{delta_rho_x}$, $\pquant{delta_rho_y}$ and $\pquant{delta_rho_z}$. If $\tr_\mathrm{o.q.}(\pquant{rho_0})=\frac{1}{2}\unitmat$ and no information about the logical qubit state has been revealed ($\tr_\mathrm{o.q.}(\pquant{delta_rho_j})$ for all $j\in\{\mathrm{x},\mathrm{y},\mathrm{z}\}$), $\mathcal{O}_Q$ simplifies to
\begin{equation}
	\mathcal{O}_Q =
	\frac{1+\mathrm{mineig}\,\hat{A}}{2} 
\end{equation}
with $A_{jk}=\frac{1}{4}(\Braket{\phi_{\unitvec{_j}}}{\tr_\mathrm{o.q.}(\pquant{delta_rho_k})}{\phi_{\unitvec{_j}}}+\Braket{\phi_{\unitvec{_k}}}{\tr_\mathrm{o.q.}(\pquant{delta_rho_j})}{\phi_{\unitvec{_k}}})$; $\mathrm{mineig}$ denotes the smallest eigenvalue. In our bit-flip scenarios, only the two cases $\mathrm{mineig}\,\hat{A}=\pm\pquant{magic}$ can occur after decoding into the target qubit. We set $C_T=1$ if (at the end of the simulation) the decoding has been performed completely and $\mathrm{mineig}\,\hat{A}>0$, and otherwise $C_T=0$. Like for the decoding reward, the coefficient $\beta_\text{corr}$ has to be sufficiently large such that the correction reward can compete with the protection reward (we choose $\beta_\text{corr}=10$).

In the learning gradient, we do not consider directly the reward $r_t^\text{(recov)}$, but rather the return which we choose as
\begin{equation}
	R_t^\text{(recov)}=\left(1-\gamma\right)\sum_{k=0}^{T-t-1}\gamma^k r_{t+k}^\text{(recov)}.
\end{equation}
with discount rate $\gamma$.

\subsection{Network Input}
\label{sec:state_aware:layout}

As discussed in Methods, the input to the state-aware network consists of \textit{(i)} a representation of four evolved density matrix contributions (that represent the evolution of all logical qubit states), \textit{(ii)} one neuron per measurement in the action set which predicts whether executing that particular measurement would reveal information about the logical qubit state, \textit{(iii)} the previous action, and \textit{(iv)} a counter indicating the physical time during the last few time steps; \textit{(iv)} is given only if we want the agent to perform explicit recovery. Here, we provide additional information for \textit{(i)} and \textit{(ii)}:

\begin{itemize}
	\item As the state-aware network is supposed to have perfect information, its input should make it possible to reconstruct the quantum state for any logical qubit state. 
		
		This information is contained in $\mathcal{A}=\{\pquant{rho_0},\pquant{delta_rho_x},\pquant{delta_rho_y},\pquant{delta_rho_z}\}$, or any (non-trivial) combination of these quantities. In particular, this includes $\mathcal{B}=\{\pquant{rho_0},\hat{\rho}_1:=\pquant{rho_0}+\pquant{delta_rho_x},\hat{\rho}_2:=\pquant{rho_0}+\pquant{delta_rho_y},\hat{\rho}_3:=\pquant{rho_0}+\pquant{delta_rho_z}\}$; $\mathcal{B}$ consists of the density matrices for the logical qubit states $\unitmat/2$ (fully depolarized), $\mconst{pauli_x}$, $\mconst{pauli_y}$ and $\mconst{pauli_z}$ (up to normalization which only matters if information about the logical qubit state has been revealed). We prefer $\mathcal{B}$ over $\mathcal{A}$ because it is in a more ``compact'' format: $\hat{\rho}_1$, $\hat{\rho}_2$ and $\hat{\rho}_3$ are usually dominated by one eigenstate (as long as the quantum information is well preserved), whereas $\pquant{delta_rho_x}$, $\pquant{delta_rho_y}$ and $\pquant{delta_rho_z}$ already start from two components with equally large contribution, so the double number of PCA components would be required for the same amount of information. In practice, we have found that this indeed makes a difference in learning.
		
		The straightforward approach would be to feed these density matrices directly into the neural network (one input neuron per matrix component). However, we have found it useful to pre-process the density matrices via principal component analysis (PCA). The key motivation is to reduce the input size of the neural network: compared to the $4^\mathrm{\#qubits}$ entries of a density matrix, we downsize the input to $P(\mathrm{\#qubits})\cdot2^\mathrm{\#qubits}$ components where $P(\mathrm{\#qubits})$ is the number of PCA components which we would typically choose polynomially in $\mathrm{\#qubits}$ (note that the eigenstates are of dimension $2^\mathrm{\#qubits}$, and the density matrices of dimension $2^\mathrm{\#qubits}\times2^\mathrm{\#qubits}$). So, the input size still grows exponentially with $\mathrm{\#qubits}$ even for the PCA-ed input, but compared to the non-PCA case we still win an exponential factor. A nice side-effect is that the PCA automatically decomposes density matrices into main channels and error contributions (and sorts these components by influence); we suppose that this helps the network to better recognize the input states.
	\item As already explained in \cref{sec:reward_scheme}, a particular pitfall are measurements which reveal information about the logical qubit and thereby ``destroy'' the quantum state. In the following, we describe an extension to the input which helps the network to detect these cases.
		
		From \cref{eq:construction_rho_xyz}, it can be seen that the knowledge about the logical qubit state is determined by the vector
		\begin{equation}
			\vec{b} :=
			\begin{pmatrix}
				\tr(\pquant{delta_rho_x}) \\
				\tr(\pquant{delta_rho_y}) \\
				\tr(\pquant{delta_rho_z})
			\end{pmatrix}
		\end{equation}
		The ideal case where no information has leaked out, \abbr{ie} all logical states are still equally likely, is equivalent to $\vec{b}=\vec{0}$. We now consider the effect of measurements on $\vec{b}$. For a measurement with two possible results (let $\hat{P}_+$ and $\hat{P}_-$ be the corresponding projection operators), the ``measurement bias'' vector
		\begin{equation}
			\Delta\vec{b} :=
			\begin{pmatrix}
				\tr\big((\hat{P}_+-\hat{P}_-)\,\pquant{delta_rho_x}\big) \\
				\tr\big((\hat{P}_+-\hat{P}_-)\,\pquant{delta_rho_y}\big) \\
				\tr\big((\hat{P}_+-\hat{P}_-)\,\pquant{delta_rho_z}\big)
			\end{pmatrix}
		\end{equation}
		describes the change of $\vec{b}$:
		\begin{equation}
			\vec{b} \stackrel{\pm}{\to} \frac{\vec{b}\pm\Delta\vec{b}}{2\tr(\hat{P}_\pm\pquant{rho_0})}
			\label{eq:update_knowledge_bias}
		\end{equation}
		Therefore, $\Delta\vec{b}$ indicates how much additional information about the logical qubit state is gathered by performing the corresponding measurement.
		
		For our set of operations, each measurement is either unbiased or leads to a complete ``collapse'' of the quantum state; from \cref{eq:update_knowledge_bias}, we see that the condition $\Delta\vec{b}\stackrel{?}{=}\vec{0}$ should distinguish these cases. So, we compute $\Delta\vec{b}\stackrel{?}{=}\vec{0}$ for each measurement in the action set and provide it as additional input to the network. Because this information can be extracted from $\pquant{delta_rho_x}$, $\pquant{delta_rho_y}$ and $\pquant{delta_rho_z}$, this is in the strict sense no additional input, but rather a ``rewording'' of a special property in a convenient format.
		
		We have observed that these neurons indeed accelerate the learning process, especially in the early training phase.
\end{itemize}

\subsection{Reinforcement Learning Algorithm}
\label{sec:reinforcement_learning}

Reinforcement learning\citesuppl{suppl:sutton_reinforcement_1998} is a general framework to autonomously explore strategies for ``optimum control'' problems. In the terminology of reinforcement learning, the control problem is represented by an ``environement'' and the controller by an ``agent''. This agent can successively choose between different actions; for this purpose, it typically has (at least partial) knowledge about state of the environment.

In the policy gradient\citesuppl{suppl:williams_simple_1992} approach, the agent directly computes a policy function $\pi_\theta(a_t|s_t)$ which gives the probability to choose action $a_t$ in state $s_t$; $\theta$ is a (multi-dimensional) parameter representing the internal state of the agent, for us typically the weights and biases of the neural network. Training means to search for weights $\theta$ which yield some desired behavior.

The simplest way to train such a policy is given by the ``vanilla'' policy gradient
\begin{equation}
	g_\mathrm{van} = \mathbb{E}\Big[\sum_tR_t\frac{\partial}{\partial\theta}\ln\pi_\theta(a_t|s_t)\Big]
\end{equation}
where $\mathbb{E}$ denotes the expectation value (according to the policy $\pi_\theta(a_t|s_t)$) and $R_t$ the return collected for the particular action sequence (see \cref{sec:reward_scheme} for our choice of the return). In practice, an estimate for $g_\mathrm{van}$ is computed from a finite number of simulations, and then the parameters $\theta$ are updated -- again in the simplest case -- according to the update rule
\begin{equation}
	\theta \to \theta+\eta\,g_\mathrm{van}
	\label{eq:vanilla_update_rule}
\end{equation}
(``steepest ascent'') with the learning rate $\eta$; this procedure defines one epoch, and is repeated multiple times.

There are various modifications to the learning scheme as discussed so far which often lead to significant improvements in terms of learning speed and stability. Specifically, we make use of the following techniques:
\begin{itemize}
	\item Instead of \cref{eq:vanilla_update_rule}, we actually use the adaptive moment estimation (Adam) update rule\citesuppl{suppl:kingma_method_2015}
		\begin{subequations}
			\begin{align}
				\theta & \to
					\theta + \eta \cdot B_n \cdot \frac{m}{\sqrt{v}} \\
				m & \to
					\beta_1 m + (1-\beta_1)\,g \\
				v & \to
					\beta_2 v + (1-\beta_2)\,g^{\odot2}
			\end{align}
			\label{eq:adam_update_rule}
		\end{subequations}
		with the Adam hyperparameters $\eta$, $\beta_1$ and $\beta_2$; $^{\odot2}$ denotes the element-wise square, and with $\sqrt{v}$ we mean the element-wise square root of $v$. The coefficient $B_n$ can depend on the epoch index $n$ and is proposed by the Adam inventors to counteract the zero bias for $m$ and $v$ in the early training phase ($B_n=\sqrt{1-\beta_2^n}/(1-\beta_1^n)$); in the results shown here and in the main text, we did not employ bias correction which is realized by $B_n=1$. Both variants yield comparable results for us.
	\item A particular difficulty which occurs in our challenge is that we have a quasi-continuous spectrum for the reward, and it is necessary to resolve small differences in there. Improving this aspect is one of the advantages of the natural policy gradient\citesuppl{suppl:peters_natural_2008}, defined by
		\begin{equation}
			F g_\mathrm{nat} = g_\mathrm{van}
		\end{equation}
		with the Fisher information matrix
		\begin{equation}
			F = \mathbb{E}\bigg[\Big(\frac{\partial}{\partial\theta}\ln\pi_\theta(a|s)\Big) \Big(\frac{\partial}{\partial\theta}\ln\pi_\theta(a|s)\Big)^\mathsf{T}\bigg]
		\end{equation}
		for actions $a$ and states $s$ being distributed according to the policy $\pi_\theta$; $^\mathsf{T}$ denotes vector transpose. For more information, see \citesuppl{suppl:amari_natural_1998,suppl:kakade_natural_2002,suppl:peters2003nat_grad,suppl:peters_natural_2008,suppl:pascanu2013nat_grad} (the appendix of the latter also gives a recipe for an efficient implementation based on L and R operations).
	\item Policy gradient schemes can often be enhanced by subtracting a baseline from the return, \abbr{ie}
		\begin{equation}
			g_\mathrm{van} \to \mathbb{E}\Big[\sum_t(R_t-b)\frac{\partial}{\partial\theta}\ln\pi_\theta(a_t|s_t)\Big]
		\end{equation}
		where $b$ is in the simplest case an average of past values for the return. The motivation is that for suitable $b$, the variance of $R_t-b$ is smaller than the variance of $R_t$ alone, and so the estimator for this modified learning gradient should have smaller fluctuations.
		
		There is a tailored way to choose the baseline $b$ for the natural policy gradient \citesuppl{suppl:peters2003nat_grad}, but we do not use this one because another aspect is more important for us. Due to the structure of the protection reward (see \cref{sec:reward_scheme}), the corresponding return is explicitly time-dependent: for successful strategies where $r_{t}^{(1)}\approx1$ and $r_{t}^{(2)}\approx0$, the return behaves as $R_t\approx(1-\gamma)\sum_{k=0}^{T-t-1}\gamma^k=1-\gamma^{T-t}$, \abbr{ie} $R_t$ drops from $1$ to $0$ near the end of a trajectory. Also the distribution of the recovery reward in time is highly uneven (only assigned in the final time steps). To compensate for this time dependence of the return, we use a time-dependent baseline:
		\begin{equation}
			g = \mathbb{E}\Big[\sum_t(R_t-b_t)\frac{\partial}{\partial\theta}\ln\pi_\theta(a_t|s_t)\Big]
		\end{equation}
		In practice, we choose the baseline $b_t$ as exponentially decaying average of the return $R_t$: in epoch $N$, the baseline takes the value
		\begin{equation}
			b_t = (1-\kappa) \sum_{n=0}^{N-1}\kappa^n\bar{R}_t^{(N-1-n)}
		\end{equation}
		where $\kappa=0.9$ is the discount rate and $\bar{R}_t^{(n)}$ is the mean of the return at time step $t$ found in epoch $n$.
		
		We have observed that this explicitly time-dependent baseline considerably improves learning stability and leads to much faster learning.
	\item Without any countermeasures, policies often strongly tend to prefer few actions only. In order to encourage exploration, it has been found useful to introduce entropy regularization \citesuppl{suppl:williams_function_1991}: the learning gradient $g$ used in the update equation (\abbr{cmp} \cref{eq:vanilla_update_rule,eq:adam_update_rule}) is substituted by $g+\lambda\,\partial H/\partial\theta$ where
		\begin{equation}
			H = \mathbb{E}\Big[-\sum_a \pi_\theta(a|s) \ln\pi_\theta(a|s)\Big]
		\end{equation} 
		is the Shannon entropy (note that also the alternative definition with $\log_2$ is common). For the problem we consider, however, the choice of its coefficient $\lambda$ is problematic: we observed that either there is no significant amplification of exploration, or the trade-off between maximizing the reward and maximizing the entropy is so strong that their common optimum is shifted away considerably from the optimum of the return. This means, $\pquant{magic}$ at the end of the simulations reaches a significantly lower value if entropy regularization is applied. We have found that it is the best strategy to start with entropy regularization for better exploration and later \textit{smoothly} reduce its effect (by slowly decreasing $\lambda$). Because we discovered this relatively close to the submission of this paper, only one of the networks shown in the main text (\maintextref{fig:decoding_and_event_aware:learning_curve_state_aware}) is trained with entropy regularization, suddenly switched off at epoch 12000.
\end{itemize}

\subsection{Smoothening the learning gradient}

We emphasize that, whenever possible, we try to consider as many cases as possible in one go, in order to reduce the noise on the learning gradient; this is like a golden thread running through the whole work presented here. Concretely, we take the following actions:
\begin{enumerate}
	\item Our physical simulations are based on density matrices instead of wave functions. The advantage of density matrices is that they already represent the ensemble-averaged behavior of a quantum system. Dealing with wave functions would have required to sample all possible trajectories, or at least a sufficiently large subset, in order to get comparably accurate statistics. Therefore, the fact that a single density matrix contains a complete description of the quantum system prevails over the higher consumption of computational resources (density matrices are higher dimensional objects than wave functions and therefore their time evolution is numerically more expensive and they require more memory).
	\item Furthermore, we also want the neural network to perform well on all possible logical qubit states. In principle, this can be achieved by training on randomly chosen logical qubit states (which requires a fair sampling and sufficient coverage of all possible states during training). Instead, however, it is much more efficient and also easier to consider the full Bloch sphere in each ``trajectory'': by evolving four variables with the dimension of a density matrix ($\pquant{rho_0},\pquant{delta_rho_x},\pquant{delta_rho_y},\pquant{delta_rho_z}$ as explained in \cref{sec:phys_time_evolution}) and rewarding the neural network based on its performance in the worst case, we directly train the network to find suitable strategies for all of them. Note that this kind of parallelism for the logical qubit state is also necessary to prevent the network from ``cheating'' (\abbr{cmp} \cref{sec:state_aware:layout}).
	\item Another point where we prefer to train rather on the average behavior than on specific examples is how measurements are treated in the reward scheme. The $\pquant{magic}$ value after a measurement can depend on the measurement result. If the agent (here, the neural network) has decided to perform a measurement, it has no further influence on the obtained result, and so it makes sense to make the reward for this action independent of this random decision. Therefore, we take the average over all possible $\pquant{magic}$ values (weighted by the corresponding probabilities to find them); \cref{sec:reward_scheme} describes how this can be done in a consistent way.
\end{enumerate}

\section{Recurrent network}

The state-aware network develops powerful strategies for protecting the quantum information in a given qubit architecture and for a given error model. This network has, however, complete knowledge of the quantum state (or, more precisely, the completely positive map describing the evolution, represented via four density matrices, as explained in the Methods) meaning that it cannot be employed in an experiment, where information is only gained through measurements. We therefore need a neural network that is blind to the quantum state and is only aware of the measurement results. This is inherently a non-Markovian decision process: this situation requires memory to correlate previous measurements and make a decision based on their outcome. We therefore implement this state-blind network as a recurrent neural network in the form of a long short-term memory (LSTM) network~\citesuppl{suppl:hochreiter97}. The state-aware network is used as a supervisor to teach the recurrent network the correction strategy. Here we only perform supervised training but the recurrent network could in principle be trained further with reinforcement learning such that it adapts its strategy to the concrete experimental setting with a potentially uncertain knowledge about the decoherence rates or other parameters.

In contrast to feedforward neural networks with single artificial neurons as building blocks, LSTM networks consist of neurons with internal memory and more complex structure~\citesuppl{suppl:goodfellow16}. An LSTM neuron comprises a cell, input gate, output gate, and a forget gate. Each gate is a conventional artificial neuron with a nonlinear activation function that regulates the flow of information whereas the cell is responsible to remember values over time. The ability to correlate input signals over arbitrary time intervals renders LSTM networks powerful tools for processing and predicting time series in real-world applications. A more detailed description of LSTM networks can be found in Ref.~\citesuppl{suppl:goodfellow16}.

The goal of our recurrent network is to learn the complete procedure of quantum error correction from the state-aware network: encoding the logical qubit, detection and recovery strategies, decoding the logical qubit, and finally, unitary operations that ensure that the final state is parallel to the initial state on the Bloch sphere. We consider the four-qubit system subject to bit-flip errors as explained in the main text. As a first step, the state-aware network is trained with immediate reward based on the recoverable quantum information and, in the last time steps, an additional immediate reward that enforces the decoding procedure is used. At the end of the time span, the network is given a reward to rotate the final state parallel to the initial state. The reward scheme is explained in \cref{sec:reward_scheme}. After convergence is achieved, a training dataset for the recurrent network is generated. 

\begin{figure}
	\includegraphics[width=\columnwidth]{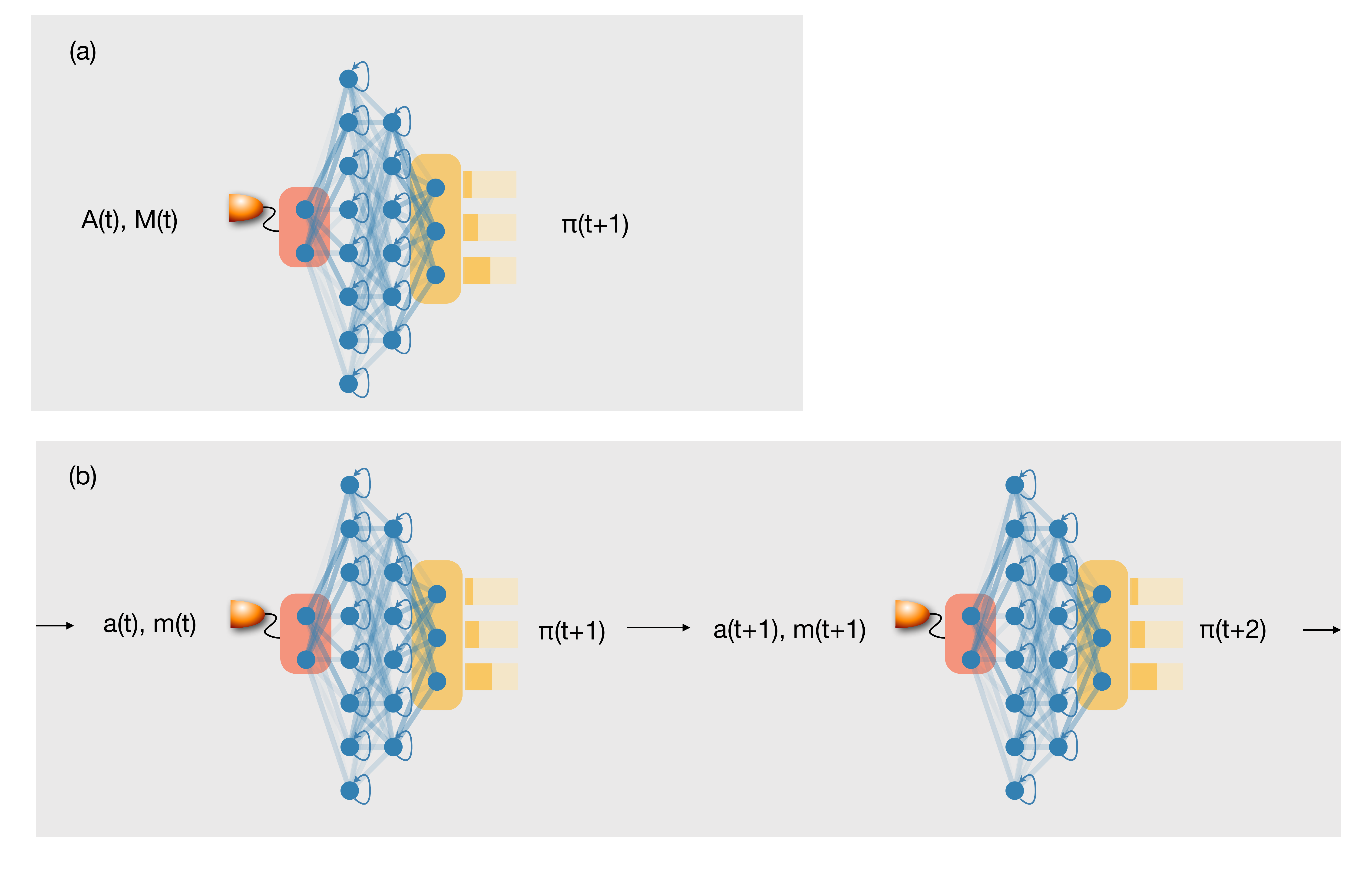}
	\caption{Visualization of the training and validation procedure of the recurrent network. (a) During training, the network is fed at the input the action $A(t)$ and measurement result $M(t)$ from the state-aware network, and outputs the policy $\pi(t+1)$ that determines the action at the next time step. This policy is trained to approximate that of the state-aware network. (b) The recurrent network is validated similarly to the state-aware network. The output policy, $\pi(t)$, results in the actions $a(t)$ and results $m(t)$, which are then fed to the input at the next time step.}
	\label{fig:lstm_training_validation}
\end{figure}

The recurrent network receives as input the action taken by the state-aware network, $A(t)$, \emph{and} the measurement result in the $z$-basis (in case a measurement was performed), $M(t)$, and outputs the policy, $\pi(t+1)$, that determines the action at the next time step, see \cref{fig:lstm_training_validation}(a). The goal of the recurrent network is therefore to find a non-trivial function, $f$, that correlates the policy at time $t$ to the input at all previous times, \abbr{ie},
\begin{equation}
    \left\{A(0), M(0), A(1), M(1), \ldots, M(t-1)\right\} \overset{f}{\rightarrow} \pi(t).
\end{equation}
In the case of quantum error correction, the long-term memory is particularly important during a recovery operation after an unexpected measurement result. This is because the network needs to carefully keep track of the error syndrome and apply a correction after the recovery operation. During validation, the performance of the recurrent network is tested on the physics environment (this is the same environment that the state-aware network is trained on). In this case, its own action at time $t$ (including the measurement outcome if the case) is fed to the input at $t+1$ as shown in \cref{fig:lstm_training_validation}(b).

\begin{table}[h!]
	\centering
	\begin{tabular}{ll}
		Hyperparameter        & Value \\ 
		\hline
		Architecture          & Two hidden layers of 128 LSTM units \\
		Training batch size   & 16    \\  
		Validation batch size & 256   \\
		Optimizer             & Adam  \\
		Learning rate         & 0.001 \\
		Activation function   & Hyperbolic tangent \\
		Loss function         & Categorical cross-entropy \\
		Dropout               & 0.5 after each LSTM layer \\
		Data set size         & \abbr{ca} 115000 gate sequences \`a 200 time steps \\
		\hline
	\end{tabular}
	\caption{List of hyperparameters used to train the recurrent network.}
	\label{tab:lstm_hyperparams}
\end{table}

The key hyperparameters of the training procedure are listed in \cref{tab:lstm_hyperparams}. The function that is optimized during training (the loss function) is taken to be the cross entropy between the policy of the recurrent and state-aware networks.

\section{Hidden representation analysis via t-SNE}

The goal of the state-aware network is to find an optimal encoding and correction scheme that protects the quantum information from decoherence. For bit-flip noise, the network finds various versions of the 3-qubit repetition code that are discussed at length in the main text. To understand how the strategy is reflected in the hidden representation of the network, we employ the t-distributed Stochastic Neighbor Embedding (t-SNE) technique~\citesuppl{suppl:maaten08} to the last hidden layer of the network. t-SNE is a nonlinear dimensionality-reduction technique that maps high-dimensional data to a two- (or three-) dimensional space and preserves the structure of the original data. In particular, nearby objects in the high-dimensional space are mapped to nearby points in the low-dimensional space.

In the following we give a brief description of the t-SNE algorithm. The similarity between the datapoints $x_i$ and $x_j$ in the high-dimensional space is modelled by a Gaussian joint probability distribution, $p_{ij}$,
\begin{equation}
	p_{ij} = \frac{\exp\left(-\norm{x_i-x_j}^2/2\sigma^2\right)}{\sum_{k\neq l}\exp\left(-\norm{x_k-x_l}^2/2\sigma^2\right)},
\end{equation}
and a Student t-distribution, $q_{ij}$, in the low-dimensional space
\begin{equation}
	q_{ij} = \frac{\left( 1+\norm{y_{i}-y_{j}}^2 \right)^{-1}}{\sum_{k\neq l}\left( 1+\norm{y_{k}-y_{l}}^2 \right)^{-1}}.
\end{equation}
The t-SNE algorithm aims at finding a low-dimensional representation of the data that minimizes the mismatch between the two probability distributions. This mismatch is quantified by the Kullback-Leibler divergence, $C=\sum_{i,j}p_{ij}\log[p_{ij}/q_{ij}]$, which the algorithm minimizes using gradient descent.

A critical parameter of the t-SNE technique is the perplexity, $p$, which is defined as
\begin{equation}
	p = 2^{-\sum_jp_{ij}\log_2p_{ij}}.
\end{equation}
The variance, $\sigma^2$, increases as the entropy (and hence the perplexity) increases. The perplexity can therefore be interpreted as a measure of the number of neighbors considered by the algorithm. With a large perplexity, the algorithm is inclined to retain the global structure of the data, whereas a low perplexity emphasizes the local structure.

\section{Implementation}
\label{sec:implementation}

Our RL implementation is based on the Theano framework \citesuppl{suppl:al-rfou_et_al._theano:_2016}, and we usually run it on GPU. Since Theano provides efficient solutions for all our neural-network related needs, we do not expect that too much can be achieved here via software optimizations. The easiest way for us to accelerate the learning part is to use one of the currently available, more powerful GPUs (currently, we have Nvidia Quadro P5000).

For the physical time evolution, we needed to implement our own simulation tools because there is no ready-to-go package meeting our specific requirements. Our current physics code is NumPy-based and runs on CPU; this is inefficient as it involves a lot of overhead and causes a memory bottleneck between physics (CPU) and network training (GPU). In addition, evolving quantum systems is a well-suited task for execution on GPU, and could therefore benefit greatly from its computing power. To improve all this, we plan a GPU-based implementation also for the physical time evolution in the future.

The numerically relevant subproblems in the time evolution scheme are \textit{(i)} updating $(\pquant{rho_0},\pquant{delta_rho_x},\pquant{delta_rho_y},\pquant{delta_rho_z})$ for the time evolution, \textit{(ii)} extracting the actual network input, and \textit{(iii)} computing $\pquant{magic}$ for the reward. Practically, \textit{(i)} means matrix-vector multiplications between a superoperator (the completely positive map) and $\pquant{rho_0}$, $\pquant{delta_rho_x}$, $\pquant{delta_rho_y}$, $\pquant{delta_rho_z}$ (interpreted as vectors); due to the sparsity of these superoperators, this operation is still relatively runtime-efficient (despite the high dimensions of the operands), and the completely positive maps describing one time step for each action of the network can be stored together in memory. For \textit{(ii)}, a principal component analysis has to be performed, and \textit{(iii)} involves the computation of trace distances. Thus, several matrices with the dimension of $\hat{\rho}$ have to be diagonalized in \textit{(ii)} and \textit{(iii)}.

For the special case of CHZ circuits (all unitary operations are combinations of CNOT, Hadamard and phase gates; measurement of variables in the Pauli group; state preparation in the computational basis \citesuppl[\abbr{p} 464]{suppl:nielsen_quantum_2011}), we can exploit the special structure of the multi-qubit quantum states that are generated. Because we still require the full density matrix, we cannot reach polynomial runtime behavior (in terms of the qubit number) like for the evolution of pure states according to the Gottesman-Knill theorem \citesuppl{suppl:gottesman1998heisenberg}; however, we can still gain an exponential factor compared to the general (non-CHZ) case. The key idea behind the efficient evolution of pure states \citesuppl{suppl:cleve1997stabilizers,suppl:aaronson2004improved_stabilizers} is to characterize such a state as the simultaneous eigenstate of several commuting stabilizers (with eigenvalue $+1$ for each of them). In our adaption, we keep track of the matrix diagonal and the stabilizers determining the eigenbasis (here: stabilizers are those which commute with the density matrix). Then, \textit{(i)} involves dense matrix-vector and sparse matrix-matrix multiplications (both for reduced dimensionality), and cheap update operations for the stabilizers. The effort for the diagonalization in \textit{(ii)} and \textit{(iii)} is completely eliminated since the matrices are already in diagonal form, and only significantly faster operations remain.

\begin{table}
	\centering
	\begin{tabular}{|l||c|c|}
		\hline
		& generic (non-CHZ) & CHZ-specific optimizations \\ \hline \hline
		time evolution &
			$\mathcal{O}(4^N)$ &
			$\mathcal{O}(4^N)$ \\
		input pre-processing &
			$\mathcal{O}(8^N)$ &
			$\mathcal{O}(P(N)\cdot2^N)$ \\
		compute $\pquant{magic}$ &
			$\mathcal{O}(8^N)$ &
			$\mathcal{O}(2^N)$ \\
		ANN training &
			\multicolumn{2}{c|}{$\mathcal{O}(P(N)\cdot2^N)$} \\ \hline
		overall &
			$\mathcal{O}(8^N)$ &
			$\mathcal{O}(4^N)$ \\ \hline
	\end{tabular}
	\caption{Runtime behavior for one epoch as a function of the qubit number $N$, separately for the different subproblems in our learning scheme (\abbr{cmp} \cref{sec:implementation}). $P(N)$ describes the dependence of PCA components on the qubit number (typically a polynomial). The training effort is assumed to be proportional to the number of input neurons.}
\end{table}

In our typical 4-qubit bit-flip examples, a reasonable level of convergence for the state-aware network is achieved within 25000 epochs. With our current implementation, the overall runtime (for the physical time evolution and the ANN training together) is around 20 hours, and the CHZ-specific optimizations reduce this time to below 6 hours. Both values are measured on the same machine (CPU: Intel Xeon E5-1630 v4, GPU: Nvidia Quadro P5000). For a well-optimized software implementation in combination with the most powerful hardware that is currently available on market, we estimate that a relative speedup of factor $10\hdots100$ is feasible.

\section{Additional information on figures}

\subsection{Training progress due to the protection reward}

In order to make the training progress in \maintextref{fig:training:learning_curve} nicely visible, we needed to rescale the figure axes. The $x$ axis is divided into three parts (to epoch 500, from epoch 500 to 2500, and from epoch 2500); inside these segments, the epoch number grows linearly. The data points are averaged over the last 10 epochs in the first segment, over the last 50 epochs in the second segment, and over the last 250 epochs in the third segments. The $y$ axis is scaled as $\pquant{magic}\mapsto-\ln(1-\pquant{magic})$, \abbr{ie} we essentially plot the negative logarithmic error probability (as $\frac{1}{2}(1+\pquant{magic})$ is the success rate for the optimal recovery sequence). The logarithmic scaling of the $y$ axis is performed after taking the average.

\begin{figure}[p]
	\centering
	\includegraphics[scale=0.7]{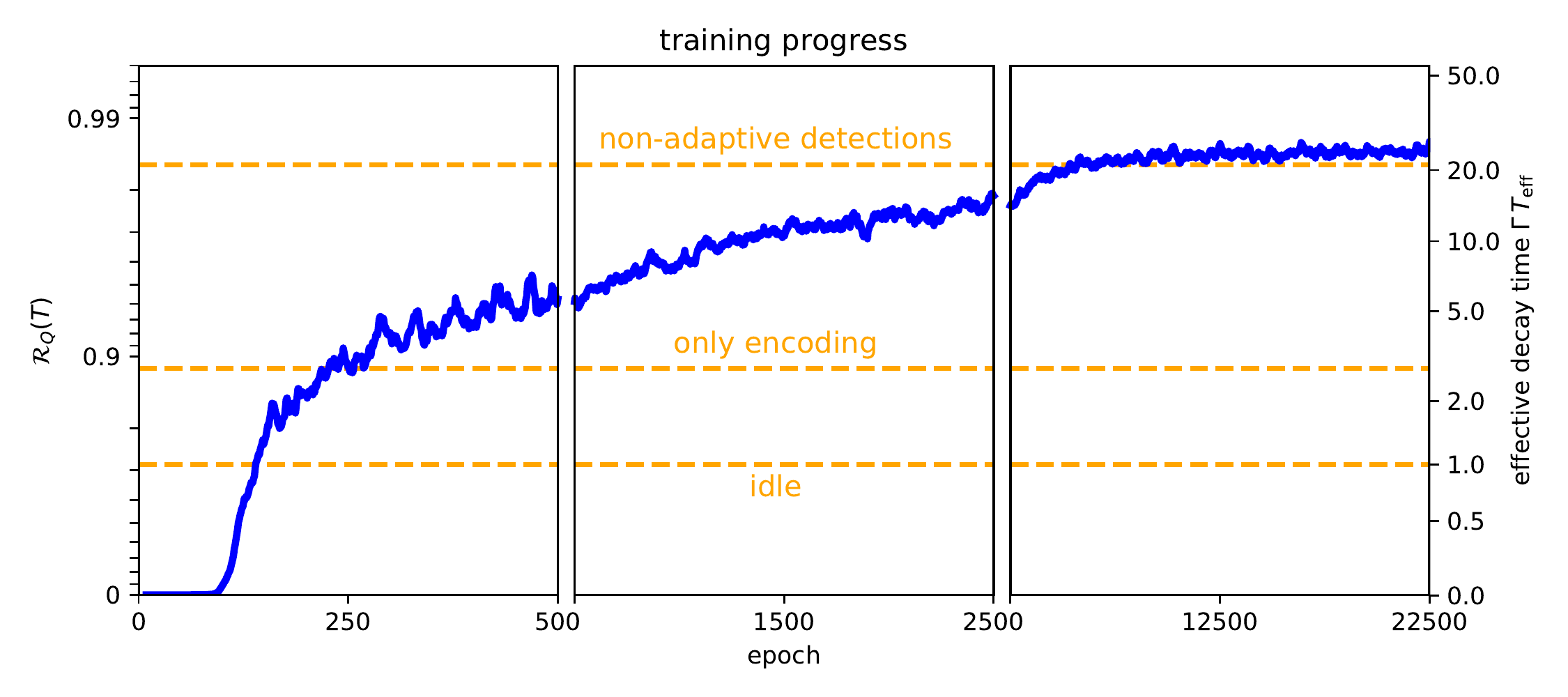}
	\caption{Replot of \maintextref{fig:training:learning_curve} with a second $y$ axis indicating the effective decay time $\pquant{T_eff}$ extrapolated from the decay of $\pquant{magic}$ during the simluation time ($T=200\pquant{time_step}$). Note that this is a well-defined quantity only for strategies where $\pquant{magic}$ decays exponentially (or at least with an exponential envelope) which is in particular not the case for ``only encoding''.}
	\label{fig:learning_curve}
\end{figure}

From the $\pquant{magic}$ value at the end of a simulation, we can extrapolate an effective decay time (\abbr{cmp} \cref{sec:more_on_figures:effective_decoherence_time}). \Cref{fig:learning_curve} replots the learning curve from \maintextref{fig:training:learning_curve} in maintext, where the ticks on the right $y$ axis indicate the effective decay time $\pquant{T_eff}$.

\begin{figure}[p]
	\includegraphics[width=\columnwidth]{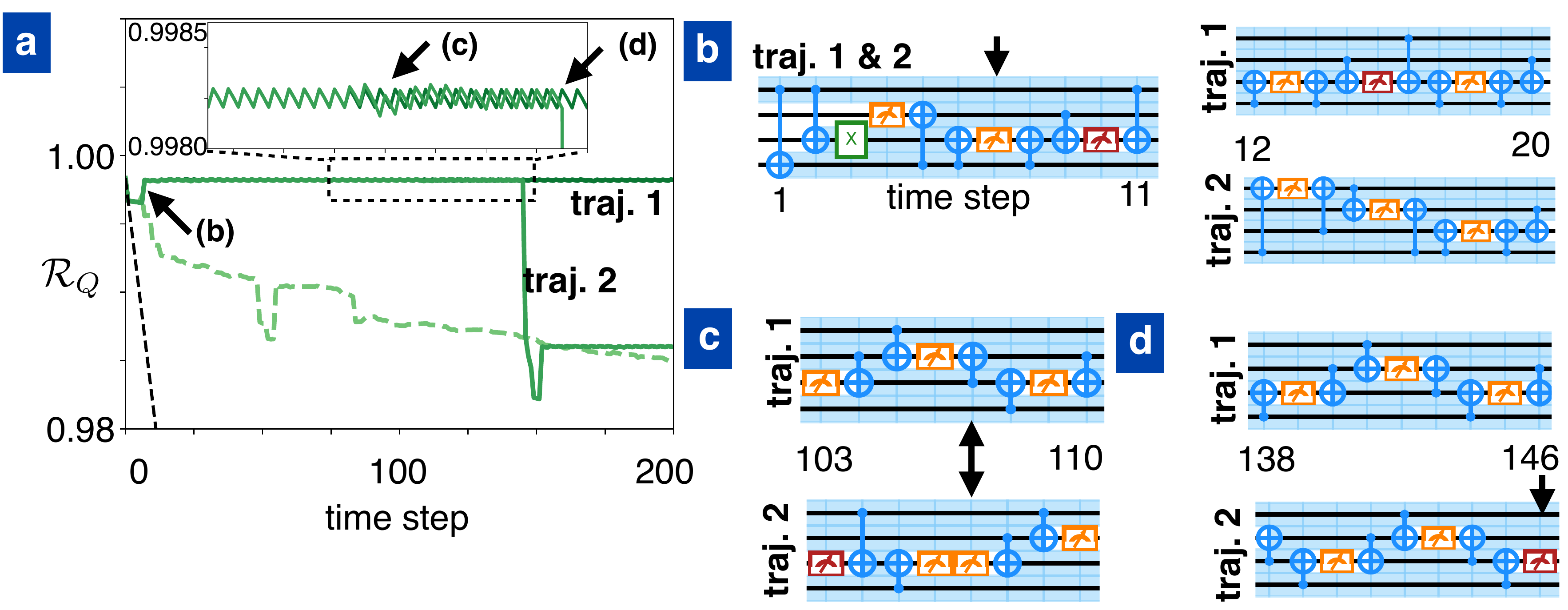}
	\caption{(a) The recoverable quantum information $\pquant{magic}$ as a function of time simulated with a converged neural network. Two typical trajectories (solid lines) and the average over a batch of $128$ trajectories (dashed) are shown. The trajectory labeled ``traj.~1'' is the same as shown in \maintextref{fig:training:magic_vs_time} of the main text. (b) The gate sequence (similar for both traj.~1 and traj.~2) responsible for the initial drop of $\pquant{magic}$. Only after the successful encoding  (time step $2$) the decay is slowed down. In time step $7$ the first parity measurement leads to an increase of $\pquant{magic}$. After time step $11$ the two trajectories show different gate sequences that, however, are equivalent in terms of preserving $\pquant{magic}$. (c) The gate sequence around the time where a shift between the respective $\pquant{magic}$ is observed. In trajectory $2$ the network repeats one measurement before continuing with repetitive parity measurements, while the other trajectory sticks to its original periodicity. (d) The gate sequences where trajectory $2$ experiences an unexpected measurement result (indicated by an arrow). This leads to a drop in $\pquant{magic}$ which is partially recovered after a few further parity measurements that pinpoint the error to one specific qubit.}
	\label{fig:trajs}
\end{figure}

\subsection{Discussion of trajectories of the converged state-aware network}

In \maintextref{fig:training}(b) and (d) we show examples of gate sequences produced by the neural network. The measurement symbols indicate measurements in the $z$-basis and the outcome encoded in the color, \abbr{ie} dark red represents ``0'' or ``up'' and orange represents ``1'' or ``down''. The light blue background of the quantum circuits indicates which qubits carry information about the logical qubit state. It is calculated from the criterion (\cref{eq:tracecrit}). In Figure 2(c) of the main text we show the recoverable quantum information $\pquant{magic}$ as a function of time at different training stages. The results of the converged network (green lines) show a barely visible decay of $\pquant{magic}$ with time. However, a close look at this trajectory reveals interesting substructure. Therefore, in \cref{fig:trajs}(a) we show this trajectory (labeled ``traj.~1'') from the main text, the corresponding sample-average, as well as another typical trajectory. Here it becomes visible, that both trajectories are subject to a fast decay in the first few time steps, where the encoding does not yet slow down the normal decoherence process. After distributing the information about the logical qubit on several physical qubits, the effective decoherence rate is reduced, but only after the first parity measurement $\pquant{magic}$ increases again. Note that this increase of the recoverable quantum information is not guaranteed, but depends on the measurement outcome: If the measurement result indicates that the two qubits are still in the same state (which is the most likely outcome after a sufficiently short time span) then the one can safely assume that no error occurred so far and a small revival of $\pquant{magic}$ is observed. However, if by chance the measurement outcome indicates a bit flip, a drop of $\pquant{magic}$ would occur instead. The corresponding gate sequences are displayed in \cref{fig:trajs}(b). The network immediately performs the CNOTs to switch to the encoded state, where the decay of $\pquant{magic}$ is slowed down. Curiously, this particular network consistently performs two unconventional additional actions (a bit flip and a measurement), before preparing the first parity measurement. (training another neural network does not reproduce these two actions). One could speculate, that the seemingly unnecessary very first measurement of the second qubit is performed in order assure the state of this qubit since it is then used as the measurement ancilla.

In the inset of \cref{fig:trajs}(a) it becomes visible, that the subsequent repeated parity measurements lead to a periodic tiny decay and recovery of the recoverable quantum information, since in these trajectories, all the measurement outcomes indicate that no error occurred. Note that, although a different sequence of parity measurements is performed for trajectory $1$ and $2$, the recoverable quantum information in both cases is the same. The slight shift of the trajectories that eventually appears is due to a repeated measurement in trajectory $2$, see \cref{fig:trajs}(c). Only when an error is detected, as in trajectory $2$ (\cref{fig:trajs}(d)), the behavior changes: The recoverable quantum information drops, until additional parity measurements determine which qubit was actually subject to the bit-flip. Then $\pquant{magic}$ partially recovers. Note that these are tiny small-scale drops compared to the large-scale drops visible for the not yet converged network in \maintextref{fig:training:magic_vs_time} of the main text (which are cause by more then single bit-flip errors or unclear situations because too much time passed between consecutive measurements).

The dashed line in \cref{fig:trajs}(a) shows the sample average over $128$ trajectories. It decays faster than trajectory $1$, where no error occurred at all due to all the trajectories with bit flips (and a corresponding drop in $\pquant{magic}$) at various different times that enter the average (note that the ``revival'' of the sample average is only an artefact due to the small number of trajectories used here).

\subsection{t-SNE}

In \maintextref{fig:analysis} and \maintextref{fig:decoding_and_event_aware} of the main text, we show graphical representations of the quantum states via the yellow rectangles. The border of the rectangles is colored according to the action the network takes for the particular input. In detail, we show the six states with the largest contribution to the density matrix $\hat{\rho}_\mathrm{x}$ ordered by their associated probability. The exact value of the probability is visualized by the blue bars next to the states. The states with a probability less than 5\% are semi-transparent. This represents as a showcase how an initial superposition state $\left(\ket{0}+\ket{1}\right)/\sqrt{2}$ evolves. The black (white) circles correspond to 1 (0) so that, \abbr{eg}, the state $\circ\bullet\circ\circ + \bullet\circ\circ\bullet$ is equivalent to $(\ket{0100}+\ket{1001})/\sqrt{2}$.

Here we apply the t-SNE technique to the hidden representation of a fully converged neural network that is trained via reinforcement learning to protect four physical qubits against bit-flip noise as explained in the main text. The network has three hidden layers of 300 neurons each. To improve speed, we first perform a principal component analysis that maps the original 300-dimensional data onto a 50-dimensional space. Aside from perplexity, all other parameters of t-SNE are kept at their default value~\citesuppl{suppl:pedregosa11}. To highlight both the global and the local structure of the hidden representation, we apply the t-SNE algorithm using three different values of perplexity (compared to the data set): small, intermediate, and large. In the following we discuss the map presented in the main text, which is computed for an intermediate value of perplexity and captures a subtle interplay between global and local features. Later we will discuss two additional maps with large and small perplexity.

\paragraph{Intermediate perplexity.} In \maintextref{fig:analysis} in the main text, we perform the t-SNE algorithm on a network validated on 1024 trajectories, each with 200 time steps, which yields a data set of 204,800 points. A perplexity of 2500 is used. A point on the map corresponds to the 2D-projected hidden representation for a particular input (the four density matrices of the RL-environment), and is colored according to the action the network takes. The map features several dense clusters as well as smaller islands and collection of points. To understand this pattern, we study a steady-state detection sequence that is visualized in \maintextref{fig:analysis:expected_msmts}. An inspection of the density matrices reveals that, in steady state, one quantum state has the largest contribution, which is the state the network aims to protect from decoherence. The other states are the leading error channels that the network aims to remove by applying the steady-state detection cycle: CNOT(1,3)CNOT(4,3)M(3)CNOT(1,3)CNOT(2,3)M(3). In this case, the network uses the second qubit as the ancilla and an encoding of the quantum information in the other three qubits of the form
\begin{equation}
	\ket{\Psi_\mathrm{enc}} = \alpha\ket{011} + \beta\ket{100}
\end{equation}
The density matrix can be uniquely identified at any time during the steady-state detection cycle. For instance: (i) the states after a measurement have a higher coherence than before the measurement, (ii) after a cnot operation the ancilla is entangled with the other qubits, and (iii) the orange and purple nuclei correspond to a different state of the ancilla qubit. These differences are clearly understood and exploited by the neural network judging from the clusters in the hidden representation.

\begin{figure}
	\includegraphics[width=\columnwidth]{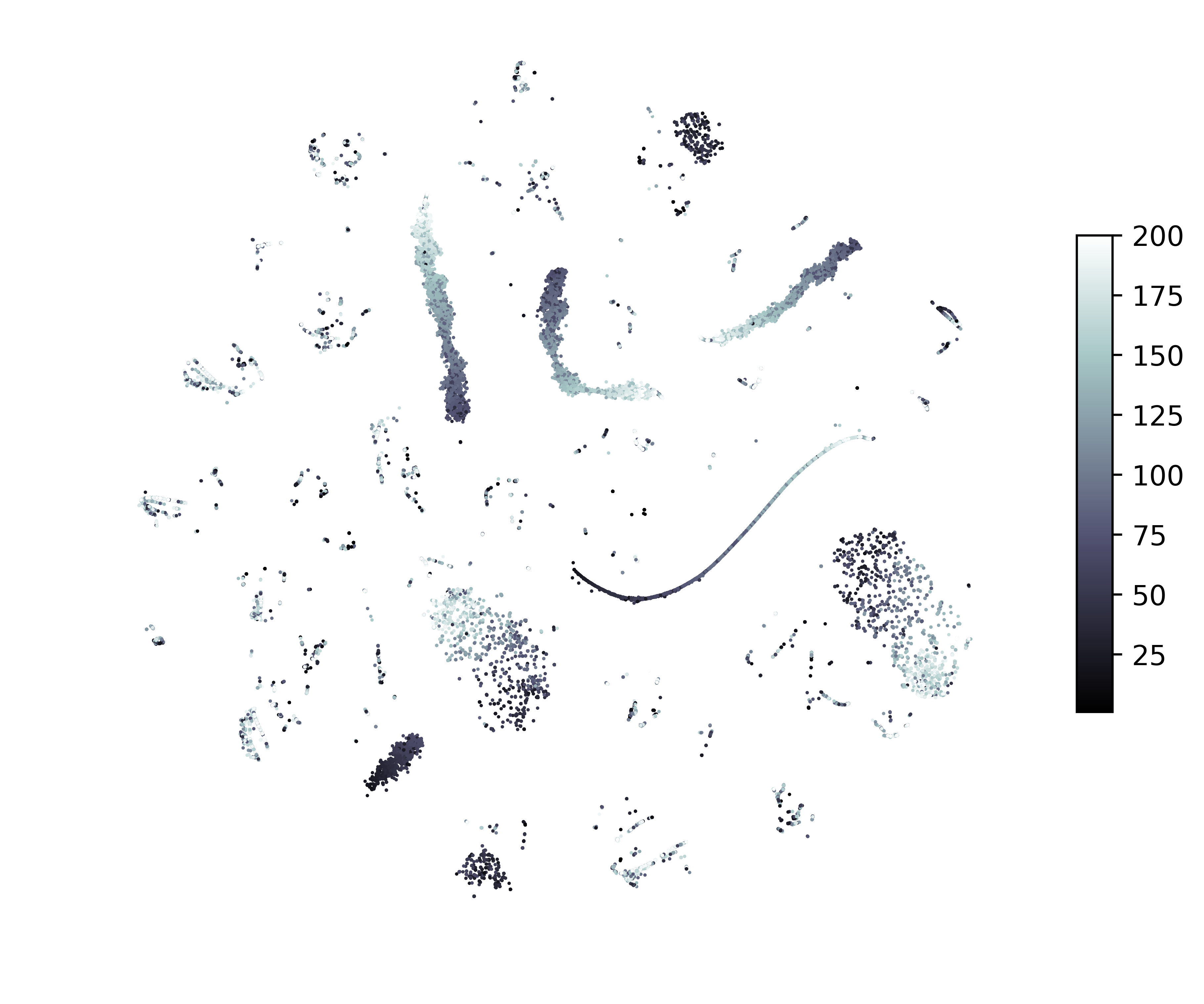}
	\caption{Same t-SNE map as in \maintextref{fig:analysis:expected_msmts} in the main text but colored according to time. The arrow of time is clearly discernible in the dominant clusters owing to the slow decoherence of the density matrix.
	\label{fig:tsne_intermperpl_time}}
\end{figure}

The detection strategy is six-fold periodic and can be visualized in the hidden representation via the gray trajectory in \maintextref{fig:analysis:tsne}. The nodes of the trajectory are displayed as white circles. The trajectory features jumps between the six largest clusters of the map, which contain most of the points of the map. This is because unexpected measurements (which interrupt the detection cycle and take the network outside the main clusters as described below) are improbable events: using the parameters from the simulation we obtain a probability of $\sim 50\%$ over the entire trajectory that all measurements on the ancilla qubit yield the expected result. A peculiar feature of the main clusters is their strongly asymmetric shape, which is due to the slow drift of the density matrix subject to decoherence. This can be visualized in the inset of \maintextref{fig:analysis:tsne}, where the orange cluster on the right-hand size of the map is colored according to a function that varies linearly with time (white corresponds to $t=1$ and dark orange to $t=200$). The detection strategy therefore forms nearly closed loops in the hidden representation that slowly drift as time progresses. Some clusters are spread out so much that they are ``cut'' by other clusters. An example can be seen in \cref{fig:tsne_intermperpl_time}, where the map is colored according to time. Comparing this map with the map in the main text one notices the dominant CNOT13 cluster on the left-hand side of the map being cut by the M3 cluster. This ``artifact'' is likely due to the fact that the perplexity is not sufficiently high and the global structure is not fully captured.

The quantum system is brought out of the steady state by an unexpected measurement result, which means that a measurement on the ancilla qubit does not project the quantum system onto the state with the largest contribution (see the discussion in the above paragraph). This interrupts the periodic detection sequence; the network employs a fully adaptive strategy in this regime that is carefully optimized during training since it plays a key role in the capability to recover the quantum information. This recovery strategy deals with markedly different quantum states and thus corresponds to different trajectories in the hidden representation. An example of a recovery trajectory is displayed by the blue line in \maintextref{fig:analysis:tsne}, where the white stars denote the nodes of the trajectory. These points belong to small clusters outside the dominant clusters.

In the following we discuss key features of this particular recovery strategy. At step 1, an unexpected measurement result projects the quantum state onto a mixture of three pure states with comparable contribution, which correspond to a bit flip of qubit 2 (largest contribution), qubit 1 (second largest), and qubit 3 (\abbr{ie}, the ancilla qubit), see \maintextref{fig:analysis:surprising_msmt}. The probability that qubit 2 flipped is roughly twice as large as the one of qubit 1 and 3 (whose probability is about the same). This can be understood as follows. Consider the state at the beginning of the detection cycle (\abbr{eg}, belonging to the blue cluster in the figure)
\begin{equation}
	\ket{\Psi} = \alpha\ket{011_\mathrm{A}1} + \beta\ket{101_\mathrm{A}0},
\end{equation}
where the subscript ``A'' denotes the ancilla. The ancilla is the physical qubit that has been designated by the network to be measured. The first part of the detection cycle, CNOT(1,3)CNOT(4,3)M(3), projects the ancilla state onto $\ket{0_\mathrm{A}}$ assuming an expected measurement result. The error channels in which qubits 1, 2, or 4 flip map the ancilla state into $\ket{1_\mathrm{A}}$, $\ket{0_\mathrm{A}}$, $\ket{1_\mathrm{A}}$, respectively, meaning that an expected measurement result $\ket{0_\mathrm{A}}$ leaves the flip of qubit 2 to be the leading error channel. The second part of the detection cycle, CNOT(1,3)CNOT(2,3)M(3), maps the ancilla state back onto $\ket{1_\mathrm{A}}$, while the error channels are mapped onto $\ket{0_\mathrm{A}}$, $\ket{0_\mathrm{A}}$, and $\ket{1_\mathrm{A}}$, respectively. Clearly, in the case of an unexpected measurement result $\ket{0_\mathrm{A}}$, the three major contributions are the flips of qubits 1, 2, and the ancilla itself. Since only qubit 2 was not corrected after the first part of the detection cycle, it carries a contribution that is twice as large. These features can be clearly seen in the density matrix.

After projection onto an unexpected state (step 1), the network begins a recovery cycle that should ideally be as short as possible. The network enforces this by targeting the state with the largest contribution in the density matrix. As a result, application of CNOT(4,3)CNOT(2,3)M(3) (steps 2--4) intentionally swaps the ancilla of the leading contribution to $\ket{1_\mathrm{A}}$. At this particular time step, the outcome of the measurement yields $\ket{0_\mathrm{A}}$ meaning that the network needs to perform another cycle (steps 5--7) to bring the quantum system back to a coherent state. After the measurement at step 7, it is finally revealed that the ancilla was flipped.

\begin{figure}[p]
	\includegraphics[width=\columnwidth]{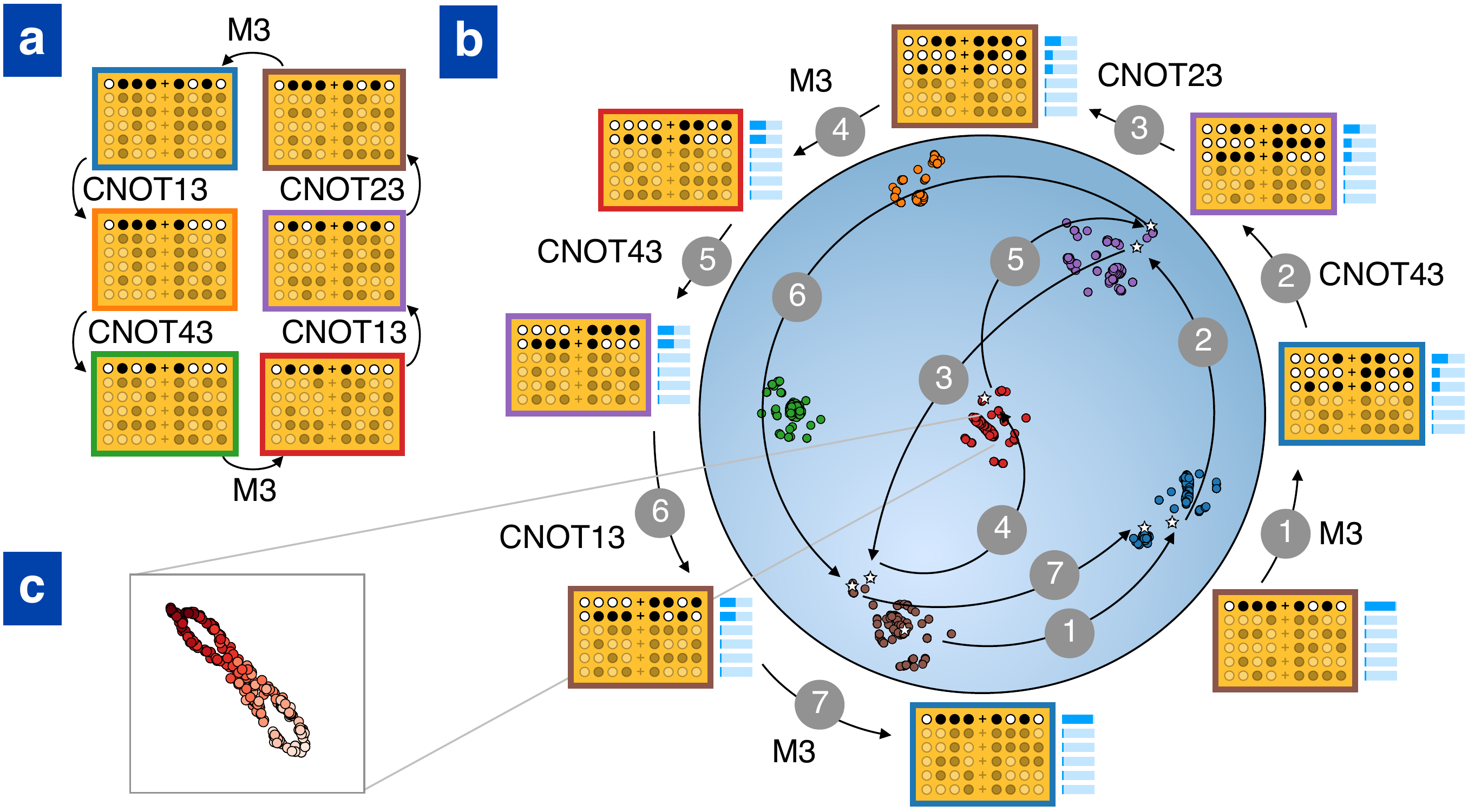}
	\caption{Visualization of the hidden representation of the neural network using the t-SNE technique in the large-perplexity regime. (a) Sequence of states visited during a standard repetitive detection cycle. (b) Visualization of neuron activations (300 neurons in the last hidden layer), sampled in several runs, projected down to 2D using the t-SNE technique. For a particular gate sequence that is triggered upon encountering unexpected measurements, we also indicate the states and the actions producing transitions between the states. Qubits are numbered 1,2,3,4, and CNOT13 has control-qubit 1 and target 3. (c) Zoom-in shows a set of states in a single cluster; the shading indicates the time progressing during the gate sequence, with a slow drift of the state due to decoherence.}
	\label{fig:tsne_largeperpl}
\end{figure}

\paragraph{Large perplexity.} In the following we perform the t-SNE algorithm on a network validated on 16 trajectories, each with 200 time steps, which yields a data set of 3200 points. To faithfully reproduce the global structure of the representation, we use a relatively large perplexity (compared to the data size) of 200. Upon a visual inspection of the map in \cref{fig:tsne_largeperpl}, six main clusters can be identified that likely reflect the six-fold periodicity in the detection scheme. Note that after the next layer (\abbr{ie}, the output layer), these clusters are converted to nearly deterministic actions. Each cluster has a non-trivial substructure with a nucleus of tightly packed points and peripheral domains with smaller density. We identify the nuclei with the steady state of the quantum system in which measurements yield the expected outcome and increase the coherence. This is the most abundant situation due to the small decoherence rate used in the simulations and, thus, a small probability that a measurement yields a result with an unexpected outcome. Consequently, the nuclei contain the majority of the points of the cluster.

\begin{figure}[p]
	\includegraphics[width=\columnwidth]{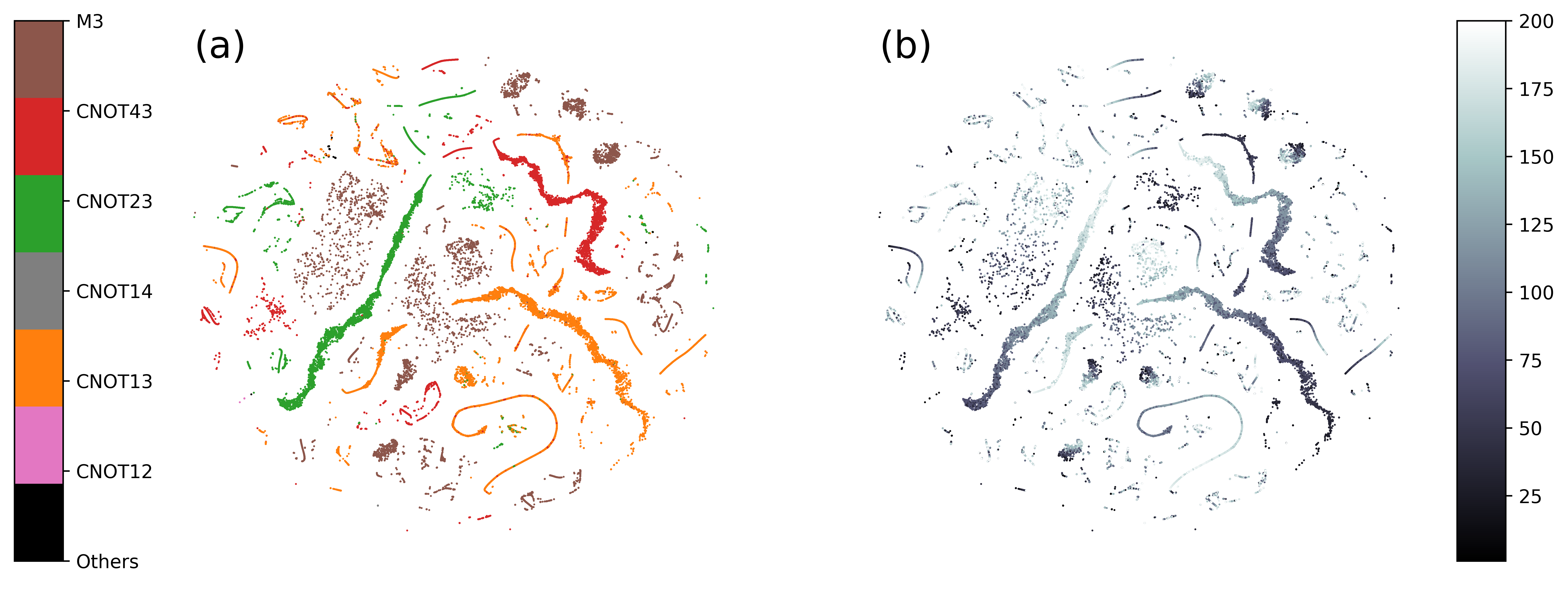}
	\caption{Visualizing the hidden representation of the state-aware neural network with the t-SNE technique in the small-perplexity regime. In (a), the map is colored according to the action that the neural network is about to take. The large circles denote the points for which the coherence of the state is low; this happens after an unexpected measurement result occurs and lasts for three or six time steps depending on the results of the next measurements, see text for details. On the right, the coloring is performed according to the time. The long and thin clusters correspond to the (slow) drift of the density matrix over time.}
	\label{fig:tsne_smallperpl}
\end{figure}

Each nucleus has a representative input (density matrix) that is shown in \cref{fig:tsne_largeperpl}(a). A steady-state detection strategy is six-fold periodic and is reflected in the hidden representation as jumps between the nuclei. An inspection of the density matrices in \maintextref{fig:analysis:surprising_msmt} reveals that, in steady state, one quantum state has the largest contribution, which is the state the network aims to protect from decoherence. The other states are the leading error channels that the network aims to remove by applying the steady-state detection cycle, see the discussion from the intermediate-perplexity section. 

Despite being tightly packed, the nuclei show some substructure as well due to the fact that the quantum system decoheres over time. This slow drift of the density matrix can be seen in \cref{fig:tsne_largeperpl}(c), where the nucleus belonging to the red cluster is colored with a function that depends linearly on time, from $t=1$ (white) to $t=200$ (black).

The peripheral domains of a cluster belong to states in which the quantum system is brought out of the steady state by an unexpected measurement result, which means that a measurement on the ancilla qubit does not project the quantum system onto the state with the largest contribution. The recovery cycle can also be visualized in the hidden representation of the network, where the white stars denote the points belonging to the cycle. Initially (before step 1) the quantum system is in steady state and the point belongs to the nucleus of the brown cluster. The points during the recovery operation belong to the peripheral domains. The corresponding density matrices have common features with the cluster they belong to, yet are sufficiently dissimilar and are clearly separated from the nuclei. After the recovery cycle (after step 7), the quantum system goes slowly back to steady state and no longer ends up in a nucleus but rather in another conglomerate of points close to it. We have observed this to be a generic feature of all trajectories and is caused by the slight decrease of coherence of the states after a recovery cycle, which the network observes through the density matrix.

\paragraph{Small perplexity.} In the following we use the t-SNE technique to visualize the hidden representation in the small-perplexity regime for getting a deeper insight into the local structure of the data. We evaluate 1024 trajectories of 200 time steps each yielding a dataset of 204,800 points with a perplexity of 400. The resulting map is displayed in \cref{fig:tsne_smallperpl} and features several long and thin clusters that would form the nuclei in the high-perplexity regime as well as smaller and less dense clusters that would likely form the peripheral domains. In \cref{fig:tsne_smallperpl}(a) the map is colored upon the action the network is about to take. There are six actions the network takes: four are related to the detection and recovery strategies mentioned above, and the other two, CNOT(1,2) and CNOT(1,4), are performed at the very beginning of the trajectory to encode the logical qubit. The slow decay of the coherence is revealed particularly well on this map, see \cref{fig:tsne_smallperpl}(b). Despite revealing the local structure of the map with great detail, the global information is somewhat distorted. For instance, note the long orange stripe on the right-hand side of the map in \cref{fig:tsne_smallperpl}(a) that is ``cut'' towards the middle by another cluster. This cut becomes particularly obvious if one follows this cluster in \cref{fig:tsne_smallperpl}(b).

\subsection{Effective decoherence time}
\label{sec:more_on_figures:effective_decoherence_time}

In \maintextref{fig:variations}(a) and (b) of the main text we show different qubit connectivities that are realized by restricting the set of available gates that the neural network can use. Only single qubit gates and CNOTs between connected qubits can be performed. We also restrict the qubits on which measurements are available. In the gate sequences shown in \maintextref{fig:variations:conn_seq} the effect of a geometrical restriction becomes clearly visible: A simple parity measurement requires now (on average) more than three operations (which would typically be two CNOT and one measurement). For example, in the case of the circular geometry, where measurements are only allowed on an ancilla qubit with a single nearest neighbour it requires at least six operations to entangle two different qubits to the ancilla. One of those can be the nearest neighbour of the ancilla, which requires a single CNOT to entangle it, but the second qubit state needs first to be swapped with the nearest neighbour state (three CNOTS), followed by the entanglement (one CNOT) and the actual measurement. Similarly also for the chains, the gate sequence for the repetition code becomes effectively longer than compared to the all-to-all connected system. This increases the average time between consecutive parity measurements which affects both the slow-down of the overall decay of $\pquant{magic}$ as well as the recovery of $\pquant{magic}$ after an error actually occurred. The physical consequences of preserving the recoverable quantum information as good as possible can be understood in terms of a prolonged effective decoherence time $\pquant{T_eff}$. Therefore, depending on the average length of the gate sequences necessary to perform parity measurements and pinpoint errors, we expect different effective decoherence times for the different configurations. In \maintextref{fig:variations:conn_bar} of the main text, we show this effect. In particular, we extract the effective decoherence time $\pquant{T_eff}$ from the average decrease of $\pquant{magic}$ after the total time $T$ and plot the ratio $\pquant{T_eff}/\pquant{T_dec}$ where $\pquant{T_dec}/\pquant{time_step}=1200$. We use that the recoverable quantum information decays approximately exponentially with time, \abbr{ie}, 
\begin{equation}
	\langle\pquant{magic}(T)\rangle=\exp{\left(\frac{-2 T}{\pquant{T_eff}}\right)},
\end{equation}
where $\langle\pquant{magic}(T)\rangle$ denotes the sample average (over $10\,000$ trajectories) of the recoverable quantum information at the final time $T$. The result show a clear trend: While all configurations increase the effective decoherence time and thus provide additional stability against bit-flip errors, the all-to-all connected systems allows for the largest improvement, followed by the chain where the neural network can use all qubits for measurements. Notably, this chain configuration performs better than the chain with one measurement qubit only (located on a central spot, at the edge would be even worse). This is, because in the first case the network switches between several qubits for measurements which leads to an advantage as compared to the fixed location. The circular configuration clearly requires the longest gate sequences and consequently it achieves a smaller increase of $\pquant{T_eff}$ than the other systems.

\subsection{Introducing measurement errors}

In \maintextref{fig:variations:msmt_errors} of the main text, we give a further example how neural networks can adapt their strategies to different environments which they are trained on. In particular, while in all other figures we have used perfectly reliable measurement outcomes, we now introduce a measurement error probability and see how a network trained with this new environment learns an adapted strategy instead of converging to the same solution as before.

We implement measurement errors by modifying the form of $\pquant{cp_map}$ (\abbr{cmp} \cref{eq:linearized_cp_map}) for measurements: for completely reliable measurement results as considered in all other simulations, measurement result $j$ is associated with $\pquant{cp_map}_j(t+\pquant{time_step},t)[\hat{\rho}]=\mconst{e}^{\pquant{time_step}\mathcal{D}}(\hat{P}_j\hat{\rho}\hat{P}_j^\dagger)$ (including the dissipative dynamics during the following idle time $\pquant{time_step}$, \abbr{cmp} Methods), and for measurement errors this expression is modified to
\begin{equation}
	\pquant{cp_map}_j(t+\pquant{time_step},t)[\hat{\rho}] = \mconst{e}^{\pquant{time_step}\mathcal{D}} \bigg(\Big(\sum_kp(R_j|A_k)\hat{P}_k\Big)~\hat{\rho}~\Big(\sum_kp(R_j|A_k)\hat{P}_k\Big)^\dagger\bigg)
\end{equation}
where $p(R|A)$ is the probability that result $R$ is reported given that the quantum system is actually projected into state $A$. Explicitly, we considered ``mixing matrices'' of the form
\begin{equation}
	\begin{pmatrix}p(0|0)&p(0|1)\\p(1|0)&p(1|1)\end{pmatrix} =
	\begin{pmatrix}1-\epsilon&\epsilon\\\epsilon&1-\epsilon\end{pmatrix}
\end{equation}
for different values of the measurement error probability $\epsilon$ between $0.001$ and $0.1$. For each value of $\epsilon$ shown in \maintextref{fig:variations:msmt_errors} we trained a separate neural network from scratch.

\begin{figure}[h]
	\centering
	\includegraphics[width=0.6\columnwidth]{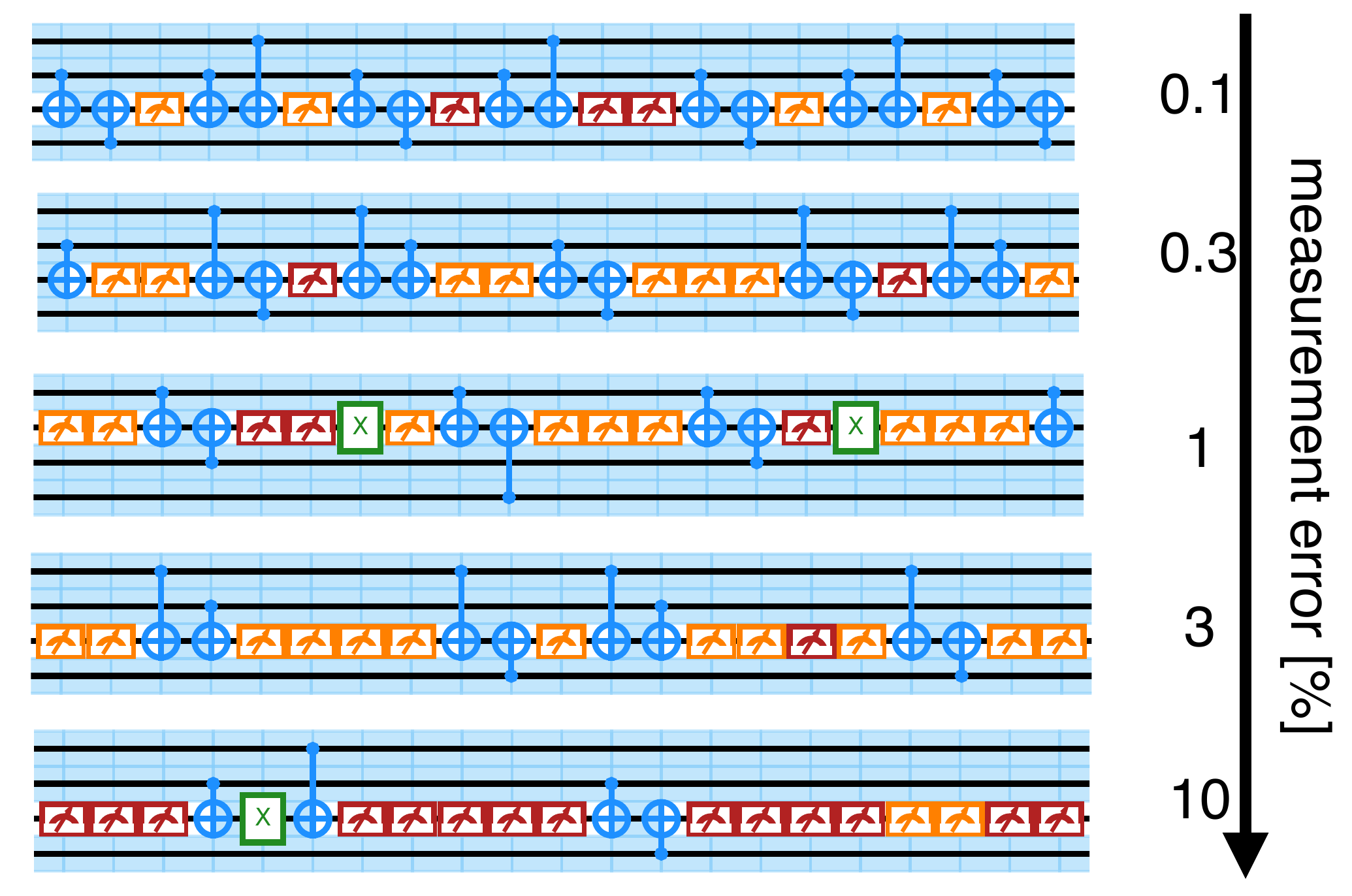}
	\caption{Typical gate sequences obtained from networks that were trained with different measurement error probabilities.}
	\label{fig:msmt_trajs}
\end{figure}

We find that the neural network counters increasing measurement errors by repeating the same measurement for several times. This helps to identify the true measurement outcome and to avoid wrong conclusions. On the other hand, repeating a measurement increases the time until the next parity measurement where new information about errors can be gained. Thus, depending on the percentage of false measurements, the neural networks learn to perform only a few or many additional measurements. We show the fraction of measurements, \abbr{ie} how many actions of all actions are measurements, in \maintextref{fig:variations:msmt_errors} of the main text. This data is obtained as an average of $2^{13}=8192$ samples and we also indicate the standard deviation (for the first data point it is so small that the error bars vanish behind the dot). We observe that the average number of measurements used increases significantly. Here, in \cref{fig:msmt_trajs} we show a typical trajectory for each measurement error probability that was used. In the lowest two trajectories, where the error probability is already very large, one can readily identify examples of repeated measurements where some show a differing, false outcome. The anomalous outcome is specific to one measurement in contrast to a bit-flip that would also affect all future measurements. 

\subsection{The behavior of the state-aware network with recovery reward}

According to the recovery reward structure described in \cref{sec:reward_scheme}, only errors of the specified target qubit need to be corrected to earn the correction reward. Errors occurring on any other qubit might or might not be corrected, as long as their status of being flipped or not is remembered for the interpretation of further parity measurements and the final decoding. Since the correction reward is only obtained after all time steps, in principle the network could decide to correct also the errors of the target qubit only at the very end (after decoding). However, we observe that the network corrects errors on the target qubit after an intermediate time - it typically performs a few further parity measurement to determine the qubit where the error occurred. Even when this is clarified with more than $90\%$ certainty, it usually performs a few more parity measurements. We believe that this is done to optimize the immediate reward earned due to preserving the recoverable quantum information. This is more pressing (since otherwise reward might be lost) than the actual correction which will only be rewarded in the very end. Furthermore, we observe that sometimes also errors on other qubits are corrected. This might be related to our observation of some ``corrections'' even in the absence of a correction reward. There we conjecture that the network performs these corrections in order to return to a well-known state.

\begin{figure}[h]
	\includegraphics[width=\columnwidth]{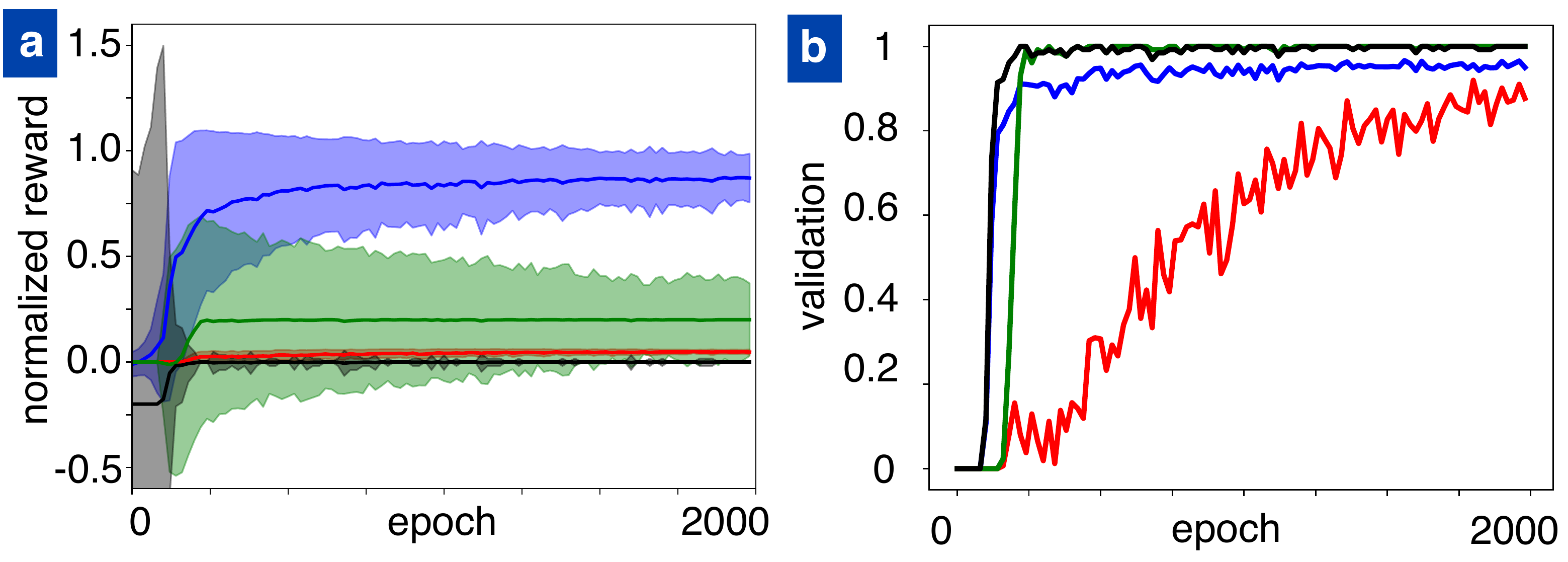}
	\caption{(a) The normalized rewards earned on the validation batch as a function of training epochs. Blue: The sum of all immediate rewards earned for preserving the recoverable quantum information. Black: The average (negative) reward obtained due to performing destructive measurements. Green: The average total decoding reward. Red: The average correction reward earned at the last time step. The shaded area indicates the standard deviation, where the  observed roughness (especially the vanishing and reappearing black shading) is due to the relatively small batch size of $128$ trajectories. (b) The training success validated on a batch of $128$ samples as a function of training epochs. Blue: The average recoverable quantum information $\pquant{magic}(T)$ after the last time step. Black: The fraction of trajectories where no destructive measurement was performed. Green: The fraction of trajectories, where the decoding to the desired target state was successful (after the last time step). Red: As measure for the overlap, $2(\mathcal{O}_Q-1/2)$ that indicates how well the logical qubit state is reproduced after the last time step.}
	\label{fig:rewards}
\end{figure}

While the correction is a rather slowly learned property (\abbr{cf} \cref{fig:rewards}), decoding is learned rapidly and with very close to perfect success. We observe that the network immediately starts decoding after receiving the signal and fills up all remaining time steps with actions that are equivalent to idle, since they do not operate on the qubit that carries the logical qubit state (\abbr{eg} measurements on already decoded qubits, \abbr{cf} inset of \maintextref{fig:decoding_and_event_aware:learning_curve_state_aware} in the main text where the error indicates the time step where the decoding signal is started). This is due to our reward structure that punishes the network if it entangles additional qubits to continue error correcting. In practice, of course, one would want to immediately read-out or proceed to perform operations on the decoded qubit, since it no longer profits from any protection.

In \cref{fig:rewards}(a) we show the mean of the rewards earned during the first $2\,000$ training epochs. We show the reward split up in its four components and averaged over the time $T$ and a batch of $128$ different trajectories: the mean of all immediate rewards earned at individual time steps for preserving the recoverable quantum information (blue); the mean reward (negative) for performing destructive measurements (black); the mean of all rewards earned for performing decoding (green); the mean correction reward (black). The shaded areas indicate the corresponding standard deviation which gives an approximate understanding of how strongly the individual parts contribute to the learning gradient.

In \cref{fig:rewards}(b) we show how successfully the different aspects of the full quantum error correction were learned: In blue we show the recoverable quantum information  $\pquant{magic}$ after the last time step averaged over the validation batch. In black we show the fraction of trajectories in the validation batch where no destructive measurement (destroying the logical qubit state) was performed. In green we show the fraction of trajectories where the decoding into the desired target state was successful (checked at the last time step). And in red we show a measure of the overlap $\mathcal{O}_Q$ to indicate how well the initial logical qubit state has been preserved overall (for the definition, see Methods). 

It is interesting to note that the overlap criterion rises much slower than all the other quantities. In particular, the network learns rather early and very fast to avoid destructive measurements and to decode properly. This is achieved with a very close to perfect success rate. There is a fast increase in preserving the recoverable quantum information at the very beginning which is associated to avoiding destructive measurements and finding a good encoding. This is followed by a slow convergence (not shown in the figure but for much longer training times of several $10\,000$ epochs) where the network optimizes the error correcting sequence by avoiding non-destructive but also not helpful actions. Also the overlap measure converges much slower, indicating that correcting actual bit flips consequently is learned only at later training times. This is partially related to the fact that it can only be learned if the recoverable quantum information is already decently preserved and the decoding is performed properly, as well as to the fact that important bit flips (on the target qubit for the logical qubit state) that really need to be corrected occur not that often. The decoding is very early on learned with a high success rate, \abbr{ie}, typically all trajectories of a batch (with very few exceptions that become even rarer during later training stages) end up correctly decoded. The large standard deviation indicated around the decoding reward despite its already high overall success is due to the fact that the decoding steps are not necessarily performed at the same time steps for all trajectories (even if at the end of the time all trajectories end up in the correct state). 

\subsection{Validating the recovery success}

We need a suitable measure to judge the success of the recovery sequence. In our (bit-flip) applications, the remains of the logical qubit state (after successful decoding) should be stored only in one target qubit with maximum overlap to the (pure) target state $\ket{\phi_{\vec{n}}}$. This is quantified by $\bra{\phi_{\vec{n}}}\tr_\mathrm{o.q.}(\hat{\rho}_{\vec{n}})\ket{\phi_{\vec{n}}}$ ($\tr_\mathrm{o.q.}$ denotes the partial trace over all qubits except for the target qubit), and to describe the worst-case scenario over the whole Bloch sphere, we consider
\begin{equation}
	\mathcal{O}_Q = \min_{\vec{n}}\bra{\phi_{\vec{n}}}\tr_\mathrm{o.q.}(\hat{\rho}_{\vec{n}})\ket{\phi_{\vec{n}}}
	\label{eq:overlap_criterion}
\end{equation}
(\abbr{cmp} \cref{eq:worst_case_overlap}). This defines how well the final state preserves coherence \emph{and} how well it matches the target state in terms of direction. Since ``random guessing'' already leads to a value of $\mathcal{O}_Q=\frac{1}{2}$, we rather plot $2(\mathcal{O}_Q-\frac{1}{2})$ in \maintextref{fig:decoding_and_event_aware}(a) and (b) of the main text.

\subsection{Performance of the recurrent network}

In \maintextref{fig:decoding_and_event_aware:learning_curve_event_aware} of the main text we plot the relevant validation quantities for the recurrent network during training. Those are the recoverable quantum information at the final time, $\pquant{magic}(T)$, and a measure of the overlap, $\mathcal{O}_Q$. Both quantities rise monotonically during training and converge towards the level of the state-aware network. Interestingly, preserving the quantum information (captured by $\pquant{magic}$) is learned faster than rotating the logical qubit (captured by $\mathcal{O}_Q$). This is likely because no long-term memory is required to preserve $\pquant{magic}$ (the periodicity of the error detection sequence is six time steps only) whereas rotating the logical qubit requires memorizing the entire recovery procedure and therefore demands longer-term memory (in \maintextref{fig:decoding_and_event_aware:neuron_act} the correction is applied thirteen time steps after the unexpected measurement result).

The key aspects of the strategy learned by the recurrent network are visualized in the quantum circuit of \maintextref{fig:decoding_and_event_aware:neuron_act} during validation. The usual detection strategy is a modified version of the repetition code in which the network alternates the measurement qubit. The encoded state of the logical qubit is of the form $\Psi_\mathrm{enc}=\alpha\ket{000}+\beta\ket{111}$. After the unexpected measurement result visualized at the left of the black rectangle at $t=1$ in the figure, the network initiates the recovery procedure. As usual, mainly three states contribute to the density matrix at this time as seen in the figure: the state in which only qubit 3 flipped (largest contribution), qubits 2 and 3 flipped, and qubits 1 and 3 flipped (smallest contribution). Note that the density matrix is only plotted for convenience --- the recurrent network does not have access to it. After $t=2$, the network copies the quantum information from qubit 2 to 3, and after $t=3$ checks whether qubit 2 flipped. At $t=5$, the state in which qubits 2 and 3 flipped is ruled out by measurement. Afterwards, the two remaining states are compared and, at $t=7$, the measurement confirms that qubit 1 flipped. In principle, the network has sufficient information to apply a bit flip to qubit 1 at this point. However, it turns out that after this six-step recovery, the quantum state has a slightly lower coherence than in steady state. Between $t=7$ and $t=10$, the network therefore conducts a further purification procedure. The recovery procedure ends at $t=13$ when a correction to qubit 1 is applied.

To understand how the strategy is reflected in the internal operation of the recurrent network, we study the hidden representation of the last hidden layer. In \maintextref{fig:decoding_and_event_aware:neuron_act} we plot the output of the selected LSTM neurons during the above-described detection and recovery procedure. The hidden representation and quantum circuit are aligned such that the neuron activations result from feeding the corresponding action to the input of the network. The first six neurons (1c--6c) are mainly involved in the steady-state detection procedure judging from their periodic activations before the unexpected measurement result. Each of these neurons fire whenever a specific gate of the detection sequence is applied. The neurons involved in the recovery process (1r--4r) display correlations over longer times:
\begin{enumerate}
	\item Neuron 1r starts firing after an unexpected measurement outcome and keeps firing until the final correction to qubit 1 is applied. This neuron presumably tells the network whether a recovery procedure is currently underway.
	\item Neuron 2r fires particularly strongly if a measurement on qubit 3 yields an unexpected result. In contrast, neuron 3c fires if the same measurement yields an \emph{expected} result.
	\item A more peculiar firing pattern can be observed for neuron 3r: during steady-state detection the neuron displays a smooth and periodic activation, which the network likely uses as an internal clock during the standard sequence. However, during recovery the pattern changes and the neuron fires when the gate CNOT(1,2) is applied. Note that this gate is only applied during recovery (it may also be applied during encoding and decoding but not during detection) and signals that qubit 2 is now disentangled and should be measured at the next time step.
	\item A dual role is played by neuron 4r as well. During steady-state detection it fires if CNOT(1,4) is applied. During recovery, however, it fires whenever qubit 2 is used as ancilla. This provides relevant information to the network regarding the encoding of the logical qubit.
\end{enumerate}

\subsection{Correlated Noise Scenario}

To further verify the flexibility of our approach, we consider in \maintextref{fig:corr_noise} of the main text a different error model than in the rest of the result discussion: in \maintextref{fig:corr_noise}, we have correlated noise described by the Lindblad equation (\abbr{cmp} Methods)
\begin{equation}
	\frac{\total{}}{\total{t}} \hat{\rho} =
	\frac{1}{\pquant{T_dec}} \bigg(\hat{L}\hat{\rho}\hat{L}^\dagger-\frac{1}{2}\Big\{\hat{L}^\dagger\hat{L},\hat{\rho}\Big\}\bigg)
\end{equation}
with decoherence time $\pquant{T_dec}$ and the Lindblad operator
\begin{equation}
	\hat{L} = \frac{1}{\sqrt{\sum_q\mu_q^2}} \sum_q \mu_q \mconst{pauli_z}^{(q)}
\end{equation}
where $q=1,2,3,\hdots$ labels the qubit, $\mu_q$ is the corresponding magnetic moment, and $\mconst{pauli_z}^{(q)}$ is the Pauli $z$ operator for the respective qubit. In this discussion, we will always assume that qubit 1 is the data qubit; the remaining ones are called the ancilla qubits.

The action set consists of measurements on these ancillas along the $x$ and $y$ axis, plus the idle operation; note that we do not allow CNOT gates here. The described setup allows to extract information about the noise on the data qubit from measurements on the ancillas. The agent can decide in which order these measurements (regarding both the qubit and the axis) are performed, and how many idle steps are chosen in between; a priori, it is not clear which strategy is the best. As we will see, this question becomes very complex for more than one ancilla, especially because improvements can be achieved by employing adaptive feedback schemes (further action sequence depending on previous measurement results).

For a benchmark of the behavior of our neural network, we extrapolate an effective decay time from the averaged value $\langle\pquant{magic}(T)\rangle$ at the end of the simulations, assuming an approximately exponential decay. These values are directly comparable to the theoretical results that we will compute below. To make their interpretation easier, we always normalize by the ``trivial'' decay time $\pquant{T_triv^cn}$ for a single qubit (see \cref{eq:corr_noise_triv_decay_time}), and call this ratio $\pquant{T_eff}/\pquant{T_triv^cn}$ the (coherence) improvement.

In order to actually judge how well our networks perform on this problem, we will in the following describe alternative approaches to find suitable measurement schemes in the two- and three-qubit scenario. We emphasize that already for this problem, the effort in terms of manpower for these alternative ways exceeds considerably that for training a neural network.

\paragraph{Single qubit scenario} If all ancillary qubits are ignored (mathematically described by a partial trace over all qubits except for the data qubit), or equivalently if there are no ancillas, $\pquant{magic}$ decays as $\pquant{magic}(t)\sim\exp\big(-2t/\pquant{T_triv^cn}\big)$ with the ``trivial'' decay time
\begin{equation}
	\pquant{T_triv^cn} = \frac{\sum_q\mu_q^2}{\mu_1^2} \cdot \pquant{T_dec}
	\label{eq:corr_noise_triv_decay_time}
\end{equation}
(called $T_\mathrm{single}$ in the Methods).

\paragraph{1+1 qubit scenario} With one ancilla qubit, it is already possible to achieve a slowdown in the decay. This two-qubit case can be treated fully analytically.

The best strategy is to measure the $x$ and $y$ axis of the ancilla qubit in an alternating manner. To get insights into the effect of the intermediate idle time $\tau$, we will derive a closed form for the effective decay time $\pquant{T_eff}(\tau)$ in the following; the result, \cref{eq:corr_noise_one_ancilla_eff_decay_time}, is used to plot the analytical predictions in \maintextref{fig:corr_noise:improvement_two_ancillas} of the main text.

We start by investigating the dynamics of the quantum system if this protocol is applied. Immediately after each measurement at time $t_m$, the quantum system (for logical qubit state $\vec{n}=(x,y,z)$) is in a product state of the two qubits:
\begin{equation}
	\hat{\rho}_{\vec{n}}(t_m) =
	\frac{1}{2}
	\begin{pmatrix}
		1+z_m&x_m-\mconst{i}y_m\\
		x_m+\mconst{i}y_m&1-z_m
	\end{pmatrix}
	\otimes
	\hat{\rho}_2(t_m)
\end{equation}
Starting from a pure state at $t_0=0$, we have $(x_0,y_0,z_0)=(x,y,z)$. The values $(x_{m+1},y_{m+1},z_{m+1})$ depend only on $(x_m,y_m,z_m)$ and the intermediate idle time $\tau_m=t_{m+1}-t_m$: the $z$ component is perfectly conserved ($z_{m+1}=z_m$), and for the $x$-$y$-components we have
\begin{equation}
	x_{m+1}+\mconst{i}y_{m+1} = g(\tau_m)\cdot\mconst{e}^{\pm\mconst{i}\vartheta(\tau_m)}\cdot(x_m+\mconst{i}y_m)
\end{equation}
(the ``$\pm$'' in front of $\vartheta$ depends on whether the measurement result indicates a rotation into clockwise or counter-clockwise direction) with
\begin{subequations}
	\begin{align}
		\tan(\vartheta(\tau)) & =
			\mconst{e}^{-2\tau/\pquant{T_triv^cn}\cdot\mu_2^2/\mu_1^2} \cdot \sinh\bigg(4\cdot\frac{\tau}{\pquant{T_triv^cn}}\cdot\frac{\mu_2}{\mu_1}\bigg)
			\label{eq:tick_tock_theta} \\
		g(\tau) & =
			\frac{1}{\cos\vartheta(\tau)}\,\mconst{e}^{-2\tau/\pquant{T_triv^cn}}
			\label{eq:tick_tock_g}
	\end{align}
\end{subequations}
$g(\tau)$ describes the loss in coherence, and $\vartheta(\tau)$ is the Bayesian guess for the acquired angle. For $\pquant{magic}$, we can conclude that
\begin{equation*}
	\pquant{magic}(t_{m+1}) = g(t_{m+1}-t_m)\,\pquant{magic}(t_m)
\end{equation*}

In order to determine the optimum value for the idle time $\tau$, we compute the effective decay time which is given by $\pquant{T_eff}(\tau)=-2\tau/\ln(g(\tau))$; this form can be obtained easiest by observing the long-term decay for repeated measurements with the same idle time $\tau$. Inserting \cref{eq:tick_tock_g}, we get
\begin{equation}
	\pquant{T_eff}(\tau) =
	\frac{\pquant{T_triv^cn}}{1-\frac{1}{4}\cdot\pquant{T_triv^cn}/\tau\cdot\ln(1+\tan^2(\vartheta(\tau)))}
	\label{eq:corr_noise_one_ancilla_eff_decay_time}
\end{equation} 
where the explicit form for $\tan\vartheta(\tau)$ is given in \cref{eq:tick_tock_theta}.

\begin{figure}
	\centering
	\includegraphics[scale=0.6]{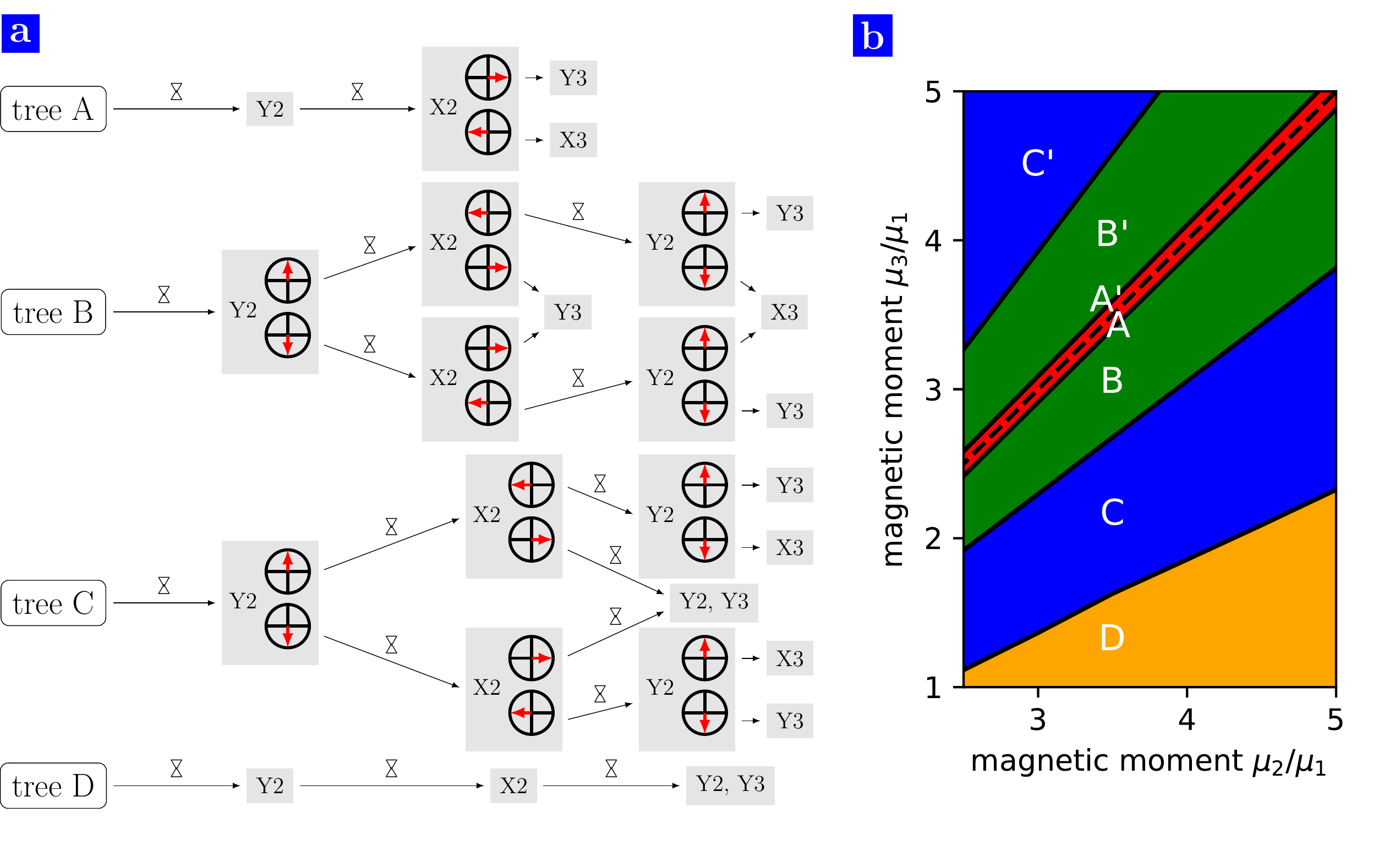}
	\caption{Strategies for the 1+2 qubit correlated noise scenario. (a) Definition of the strategies via decision trees. Idle times are marked with an hourglass symbol. We assume that we always start from a $+x$ polarization in both qubits. (b) Phase diagram indicating the prevailing strategy for different values for the magnetic moment. X' means that the roles of the qubits 2 and 3 are switched. Note that better strategies might exist which are located outside our restricted search space.}
	\label{fig:corr_noise_decision_trees}
\end{figure}

\paragraph{1+2 qubit scenario} For one data qubit plus two ancillas, the authors are not aware how to perform an analytical study like for the two-qubit scenario with reasonable effort. Instead, we follow a different strategy: we choose a maximum number of successive measurements within one cycle (we will call this the search depth) to select a subset of all possible feedback schemes, and perform a brute-force search over them.

For a proper definition of the search depth, we start from the realization that if both ancillas are measured without time delay, the quantum system is in a product state of the data qubit and the two ancillas. This means that all correlations are lifted in that process, and thus no further decision can benefit from adaptive response to measurement results before this point in time. Hence, (quasi-)simultaneous measurements split long action sequences into smaller subcycles. For any decision tree (which determines the response to the probabilistic measurement results), we define its depth as the maximum number of idle periods before the cycle is terminated, or equivalently the number of measurements excluding the very last one (that is launched without time delay).

For all strategies that are represented by a finite-depth decision tree, we can compute exact values for effective decay times (in a numerical way, see below). However, there is a considerable limitation for the depth of the brute-force search as discussed in the next paragraph.

The brute-force search does not only have to scan over all (adaptive) decision trees for measurements, but it also needs to find suitable lengths of the intermediate idle times. We discretize the (continuous) parameter range for these idle times by fixing $n$ different values. For search depth $d$, this leads to a total number $N_d$ of possible feedback schemes following the recursive expression
\begin{subequations}
	\begin{align}
		N_1 & = 16n \\
		N_{d+1} & = 4n\cdot(N_d+2)^2
	\end{align}
	\label{eq:corr_noise_bfts_complexity}
\end{subequations}
(derivation see below). From $N_d\ge2\cdot(8n)^{2^d-1}$, we can see that there is a double-exponential growth of possibilities with the search depth:
\begin{equation}
	N_d \sim \mathcal{O}(\exp(\exp(d)))
\end{equation}

$N_d$ grows so fast that in practice only very small search depths $d\le3$ are accessible in reasonable time (but $d=3$ only with very high effort). We restrict ourselves to $d=2$, but to still obtain insightful results, we fix the first measurement to the intuitively most reasonable choice (for the ancilla with the largest $\mu_q$, the axis orthogonal to its last known position), and due to the symmetry of this situation, we can analyze the two subtrees independent of each other such that the search can effectively be extended by one level. Following this approach, we could identify four different strategies (see \cref{fig:corr_noise_decision_trees}a) which prevail for different combinations $(\mu_1,\mu_2,\mu_3)$ of the magnetic moments (see \cref{fig:corr_noise_decision_trees}b). For these four fixed strategies, we in turn run a fine-grid search to further optimize the idle times. Note that all these results might be non-exhaustive due to the various restrictions of the search space.

We proceed by describing the remaining technical aspects of the search technique. First, we will discuss the rationale behind the recursive relation in \cref{eq:corr_noise_bfts_complexity}:
\begin{itemize}
	\item $d=1$: We can first choose between $n$ different idle times, and then between $4$ different measurement variables (which qubit, $x$ \abbr{vs} $y$). For $d=1$, the cycle must be terminated at this point, so an immediate measurement must follow (the $x$ or $y$ axis of the other qubit); this decision may be depend on the previous measurement result. In total, this makes $n\cdot4\cdot2^2$ possible combinations.
	\item $d\to d+1$: Again, we have the choice between $n$ different idle times and $4$ different measurement variables. Then, we can -- dependent on the measurement result -- choose between all $N_d$ combinations in the search tree with depth reduced by $1$, or the two instantaneous measurements on the other qubit to terminate the cycle immediately. In total, this makes $n\cdot4\cdot(N_d+2)^2$ possible combinations for $N_{d+1}$.
\end{itemize}

In order to judge a particular feedback scheme, we have to determine an effective decay time. For each branch $j$ in the decision tree, we can find out the probability $p_j$ to end up in this branch, the total time span $\mathcal{T}_j$ (sum of all idle times), and the loss $G_j$ of quantum information ($\pquant{magic}(t+\mathcal{T}_j)=G_j\pquant{magic}(t)$). From this information, we can compute an (averaged) effective decay time:
\begin{equation}
	\pquant{T_eff} = -\frac{2\sum_jp_j\mathcal{T}_j}{\sum_jp_j\ln G_j} = -\frac{2\langle\mathcal{T}\rangle}{\langle\ln G\rangle}
\end{equation} 
This dependency on $\mathcal{T}_j$ and $G_j$ makes it hard to narrow down the search. Suppose $A=\{j_1^\mathrm{(A)},j_2^\mathrm{(A)},\hdots\}$ and $B=\{j_1^\mathrm{(B)},j_2^\mathrm{(B)},\hdots\}$ represent two alternatives for the same subtree. If $\langle\mathcal{T}\rangle_A>\langle\mathcal{T}\rangle_B$ and $\langle\ln G\rangle_A>\langle\ln G\rangle_B$, then $A$ is clearly the better option (one ``$>$'' and one ``$\ge$'' would be enough). However, if $\langle\mathcal{T}\rangle_A>\langle\mathcal{T}\rangle_B$ and $\langle\ln G\rangle_A<\langle\ln G\rangle_B$, then it depends on the other branches of the decision tree whether $A$ or $B$ should be preferred.

\subsection{Hyperparameter analysis}
\label{sec:more_on_figures:modified_hyperparameter_sets}

\begin{table}[p]
	\begin{tabular}[t]{ll}
		\textbf{scenarios} \\
		\hline
		all-to-all & \maintextref{fig:training} and \maintextref{fig:analysis} \\
		chain (1 msmt) & \maintextref{fig:variations:conn_seq} (top) \\
		chain (all msmts) & \maintextref{fig:variations:conn_seq} (center) \\
		triangle & \maintextref{fig:variations:conn_seq} (bottom) \\
		all-to-all with small msmt noise & \maintextref{fig:variations:msmt_errors} (msmt error probability $\epsilon=0.01$) \\
		all-to-all with large msmt noise & \maintextref{fig:variations:msmt_errors} (msmt error probability $\epsilon=0.1$) \\
		\hline
	\end{tabular} \vspace*{1.5em} \\
	\begin{tabular}[t]{llll}
		\textbf{physical parameters} \\
		\hline
		parameter & a, d) main text values & b, e) faster decay & c, f) less time steps \\
		\hline
		a-c) decoherence time $\pquant{T_dec}$ for bit-flip noise &
			1200 &
			500 &
			1200 \\
		d-f) trivial decay time $\pquant{T_triv^cn}$ for correlated noise &
			500 &
			200 &
			500 \\
		d-f) magnetic moment ratios &
			\multicolumn{3}{c}{
				\begin{tabular}[t]{@{}l@{\hspace*{1em}}r@{ }l@{\hspace*{1.5em}}}
					2 qubits: & $\mu_1:\mu_2=$&$1:4$ \\
					3 qubits: & $\mu_1:\mu_2:\mu_3=$&$1:3.7:4$ \\
					4 qubits: & $\mu_1:\mu_2:\mu_3:\mu_4=$&$1:3.7:4:4.2$
				\end{tabular}
			} \\
		number of time steps $T$ &
			200 &
			200 &
			50 \\
		\hline
	\end{tabular} \vspace*{1.5em} \\
	\begin{tabular}[t]{ll}
		\textbf{hyperparameters} \\
		\hline
		ANN architecture              & (\# input neurons, 300, 300, \# actions) \\
		ANN activation function       & Softmax in output layer, elsewhere ReLU \\
		batch size                    & in a, b, d, e: 64; in c, f: 256 \\
		learning algorithm            & Adam ($\eta=0.0003$, $\beta_1=0.9$, $\beta_2=0.999$, no bias correction) \\
		\# PCA comps in input         & 6 (in d-f for two-qubit scenarios: 4 because $\dim\mathcal{H}=4$) \\
		return discount rate          & $\gamma=0.95$ \\
		baseline discount rate        & $\kappa=0.9$ \\
		reward scale                  & $\lambda_\mathrm{pol}=4.0$ \\
		punishment reward coefficient & $P=0.1$ \\
		\hline
	\end{tabular}
	\caption{Summary of the physical scenarios, their parameters and the learning hyperparameters as used in \cref{fig:common_hyperparameter_set}. Note that the exact value for the learning rate used in the simulation is in fact $0.0001\sqrt{10}$; the irrational factor of $\sqrt{10}$ is caused by a slight deviation between our implementation and the standard Adam scheme which in the end resulted only in a redefinition of the learning rate.}
	\label{tab:common_hyperparameter_set}
\end{table}

\begin{figure}[p]
	\centering
	\includegraphics[scale=0.75]{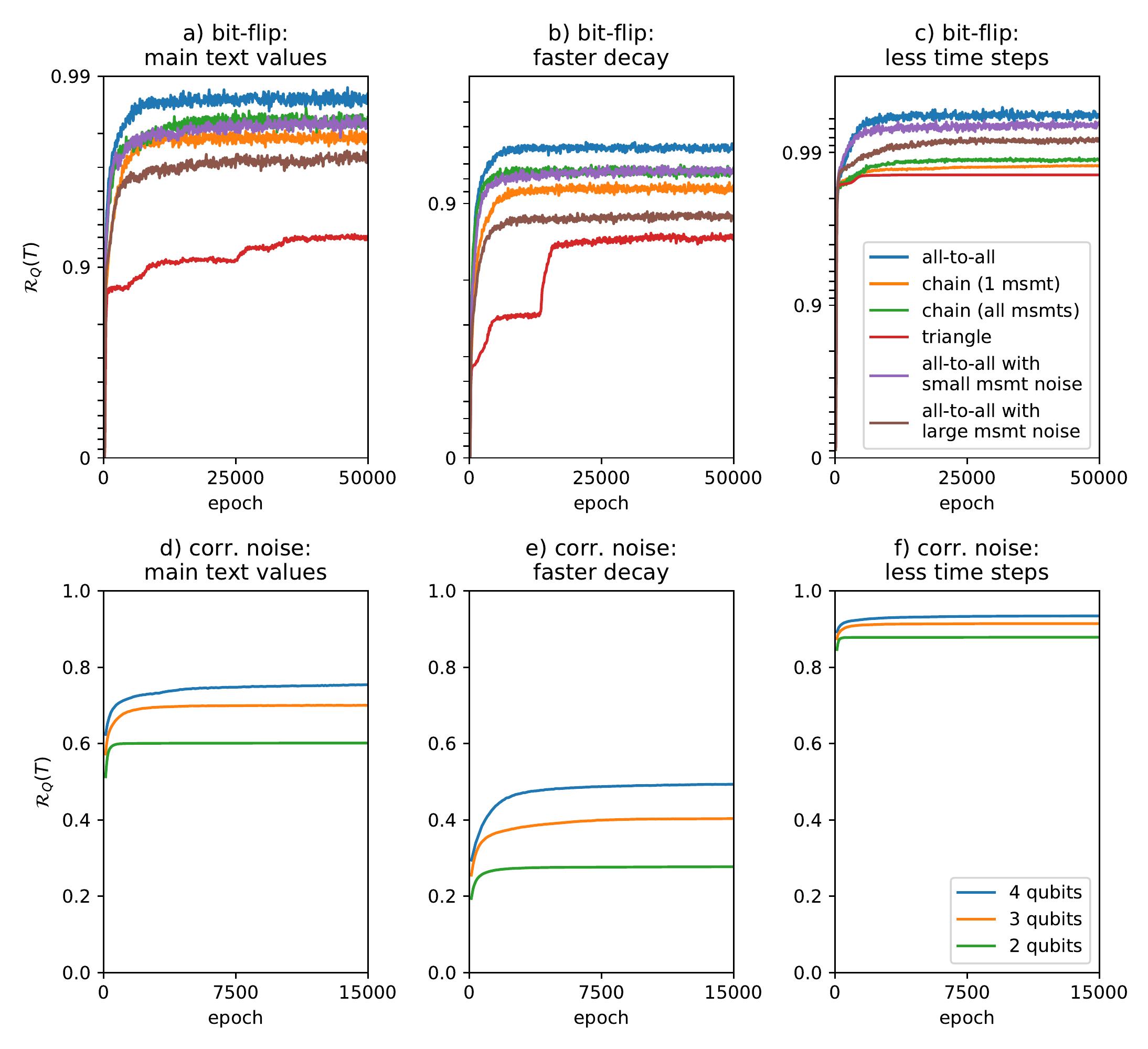}
	\caption{Performance of a common hyperparameter set for different physical systems, defined in \cref{tab:common_hyperparameter_set}. The learning curves show the value of the recoverable quantum information $\pquant{magic}(T)$ at the final time step of the episodes, averaged over the last 100 epochs. We observe successful learning for all the combinations (for the ``triangle'' scenario, there is a limited success rate, so we trained 5 networks each and show the best run here). Note that the performance for the optimum strategy depends both on the scenario and the physical parameters, so the direct comparison of the saturation levels in these plots against each other does not make a statement about a quality difference in the learning process.}
	\label{fig:common_hyperparameter_set}
\end{figure}

\begin{table}[p]
	\centering
	\begin{tabular}{|l|l|}
		\hline
		name & overridden hyperparameters \\ \hline \hline
		low learning rate &
			Adam learning rate $\eta=0.00003$ \\ \hline
		high learning rate &
			Adam learning rate $\eta=0.003$ \\ \hline
		small batch size &
			batch size 16 \\
			& Adam hyperparameters $\eta=0.00008$, $\beta_1=0.975$, $\beta_2=0.99975$ \\ \hline
		additional layer &
			ANN architecture (\# input neurons, 300, 300, 300, \# actions) \\ \hline
		pyramid &
			ANN architecture (\# input neurons, 500, 300, 100, \# actions) \\
		\hline
	\end{tabular}
	\caption{Modified hyperparameter sets as used in \cref{fig:modified_hyperparameter_set}. The values given in \cref{tab:common_hyperparameter_set} are referred to as ``standard'', and the right-hand side specifies which values are changed \abbr{wrt} ``standard'' in the corresponding variation. Note that the Adam hyperparameters are implicitly correlated with the batch size; in ``small batch size'', this makes it necessary to adjust $\eta$, $\beta_1$, $\beta_2$ for compensation to see the direct influence of the batch size (\abbr{cmp} \cref{sec:more_on_figures:modified_hyperparameter_sets}.2).}
	\label{tab:modified_hyperparameter_set}
\end{table}

\begin{figure}[p]
	\centering
	\includegraphics[scale=0.7]{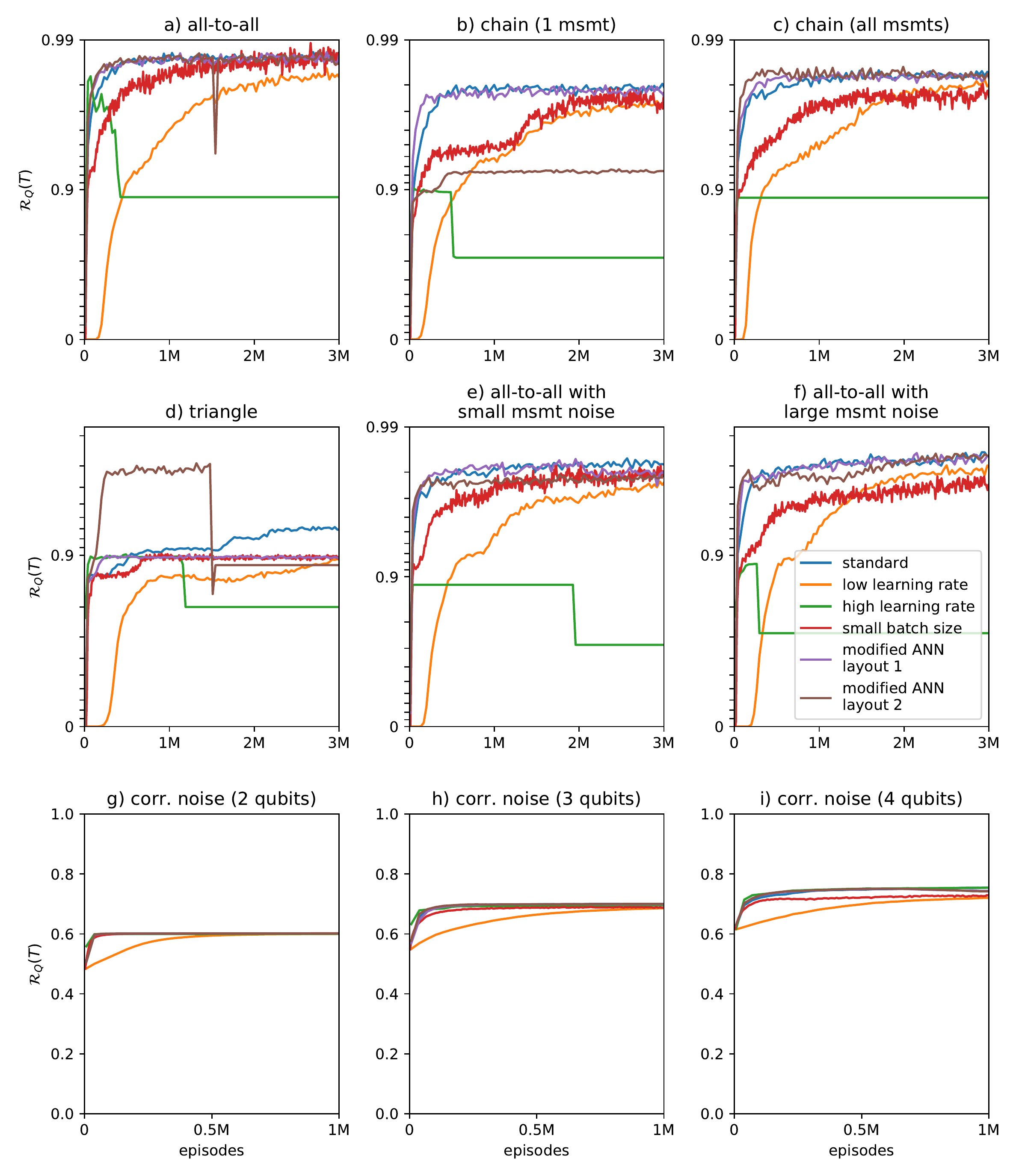}
	\caption{Influence of the hyperparameters on the training behavior. The physical properties are taken from \cref{tab:common_hyperparameter_set} (main text values), and the different hyperparameter sets are given in \cref{tab:modified_hyperparameter_set}. The learning curves show the value of the recoverable quantum information $\pquant{magic}(T)$ at the final time step of the episodes, averaged over the last 100 epochs. The only hyperparameter set which fails (for the bit-flip examples) consistently is ``high learning rate'': it starts to learn (faster than the others), but then instabilities show up and finally the training process collapses. All other hyperparameter sets work reliably and reach similar final performance, but not always converge with the same speed. The only exception is the ``triangle'' scenario: multiple attempts are necessary (independent of the hyperparameter set), so in each case we trained 5 independent networks and show the best run here. In addition, ``modified ANN layout 2'' for ``chain (1 msmt)'' (brown curve in b) failed; however, this is an outlier as the majority of the runs performed much better when retraining under the same conditions. The results are discussed in \cref{sec:more_on_figures:modified_hyperparameter_sets}.2. Note that the $x$ axis displays episodes (epoch times batch size) to allow for direct comparison between runs with different batch sizes.}
	\label{fig:modified_hyperparameter_set}
\end{figure}

In this section, we analyse the influence of hyperparameters in our learning scheme. First, we fix one common set of hyperparameters and apply it to the physical scenarios shown in the main text. In a second step, we modify this hyperparameter set in various ways and discuss the effect of these changes on the learning behavior.

The central results are:
\begin{itemize}
	\item The learning rate is the only crucial hyperparameter, but there is a clear signature which tells whether it has to be increased or decreased.
	\item All the other hyperparameters do not decide about the success of the learning process (as long as they stay in reasonable bounds).
	\item However, some of these hyperparameters have an influence on how fast the learning progresses, and thus can be tuned to optimize the training time.
\end{itemize}

\subsubsection{Common hyperparameter set}

We fix one common set of hyperparameters (values are listed in \cref{tab:common_hyperparameter_set}) and apply it to the physical scenarios shown in the main text. The resulting learning curves are plotted in \cref{fig:common_hyperparameter_set} a. The training works straightforward for all cases except for the ``triangle'' setup; there, several attempts are required, and we show the best one in each case. However, the limited success rate is not a property of the specific hyperparameter set used here, but also occurs for the hyperparameters used in \maintextref{fig:variations:conn_seq}.

To demonstrate robustness against variations of the physical parameters, we change two of them, the decoherence time (\cref{fig:common_hyperparameter_set} b) and the number of time steps (\cref{fig:common_hyperparameter_set} c), but still apply the same hyperparameter set as above. Again, we run training jobs for all the scenarios. Note that in the example with $50$ time steps (\cref{fig:common_hyperparameter_set} c) to keep the number of data points per learning update constant, we have to increase the batch size. Again, learning works smooth for all scenarios, except for the ``triangle'' setup where multiple attempts are needed.

\subsubsection{Modified hyperparameter sets}

To analyse how the hyperparameters influence the training progress, we consider several modified hyperparameter sets (\cref{tab:modified_hyperparameter_set}) and again apply them to all scenarios (with the original values for decoherence time and number of time steps). Because some of the hyperparameters are implicitly coupled, we have to properly compensate for this. For example, just decreasing the batch size effectively raises the learning rate, and so one would observe the combined effects of a smaller batch size and a higher learning rate. We counter these dependencies by suitably adjusting the correlated variables (\abbr{eg} here, decreasing the learning rate accordingly), independent of whether this would make a difference or not.

The results are shown in \cref{fig:modified_hyperparameter_set}. We make the following observations:
\begin{itemize}
	\item The learning rate has the largest impact on the learning progress. If it is chosen too small, the convergence is slowed down significantly, whereas if the high learning rate is too high, the situation is even worse as it leads to instabilities which finally lead to a collapse in the learning process. Hence, there is a simple rule of thumb how to find a suitable value for the learning rate: if instabilities show up, the learning rate has to be decreased; as long as the learning curve is smooth, the learning rate can be increased until the point where instabilities occur.
	\item The results indicate that a larger batch size seems to be favorable. The most plausible explanation for this behavior is the fact that we use the natural gradient (see \cref{sec:reinforcement_learning}) which involves the Fisher information matrix. Since the exact value of the Fisher information matrix is not accessible, we need to compute an estimate for it (in each epoch), and the quality of this estimate increases with the number of data points, \abbr{ie} the batch size. The resulting statistical noise might be amplified by the fact that it is actually the inverse of the Fisher information matrix which enters the calculation of the learning gradient.
	\item The network architecture seems to play a minor role.
\end{itemize}

\bibliographystylesuppl{unsrt}
\bibliographysuppl{suppl}

\end{appendices}

\end{document}